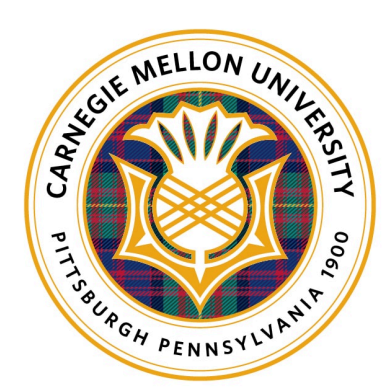

DOCTORAL DISSERTATION

*Submitted in partial satisfaction of the requirements for the degree of*
***Doctor of Philosophy*** *in* ***Human-Computer Interaction***
*at* ***Carnegie Mellon University****, Pittsburgh, PA*

# Virtual Simulation for Mental Health

**Anna M. Fang**

December 2025

**Thesis Committee:**

| | |
|---|---|
| Haiyi Zhu (Chair), | Carnegie Mellon University |
| Robert E. Kraut, | Carnegie Mellon University |
| Mayank Goel, | Carnegie Mellon University |
| Munmun De Choudhury, | Georgia Institute of Technology |
| Diyi Yang | Stanford University |

# Abstract

Poorly designed interventions or those deployed without adequate safeguards can harm the communities they aim to serve, thus exacerbating existing vulnerabilities and leaving individuals unsupported. This is especially the case for the mental health context, where there is a growing trend of relying on technological interventions due to their accessibility and ability to deliver large-scale support. However, the mental health context is also particularly sensitive to change and risks of failure are dire – at their worst, failures in mental health interventions can result in lasting negative outcomes for individuals and tragic losses as people fall through the cracks. Thus, enabling safe ways to experiment in the mental health context is vital to allow both individuals and communities to engage with new interventions without risk of their real-world consequences.

Virtual simulation – which uses virtual environments to replicate real-world interactions, processes, and behaviors – offers a promising opportunity for enabling safe, controlled experimentation with its ability to accurately replicate social situations, fears, stressors, and the potential outcomes of specific interactions. This dissertation explores how simulation approaches can support emerging mental health processes through two key ways: (1) applying agent-based modeling to create a sandbox for designing new algorithms for online mental health communities, and (2) using the simulation capabilities of VR/AR and large language models to provide individuals a low-risk, controlled space for practicing new coping techniques. I demonstrate this use of virtual simulation systems through a grounded human-centered approach, where system design is guided by empirical understanding of current real-world needs and challenges. By leveraging simulation to create environments where mental health strategies can be safely tested and practiced, this work aims to open new possibilities for designing scalable, user-centered systems that are effective and safe.

# Acknowledgements

Firstly, thank you to my family.

Thank you to Mom, Dad, and Katy – I am (and will forever be) grateful and lucky to have you. Thank you for always unconditionally supporting me. I also want to acknowledge my extended family who are across the U.S., Canada, and China, and give particular gratitude for my 爷爷, 姑父, and 姨父 who passed during the last part of my PhD.

Thank you to my dearly beloved canine companion, Milo. Over five years ago, we took a week-long cross-country road trip together to start my (our) PhD and continued to be inseparable for several years. Milo taught me patience, pure joy, resilience, and being ourselves unapologetically (such as Milo himself who was definitely a cat stuck in a gangly greyhound body). I always wanted a dog and Milo was the best one I could have ever asked for! Thanks bud, we miss you.

Thank you to my cat Pearl, who is always upset about the perpetual diet that she has to be on and swept in at the last minute to claim some credit for this PhD. I went to the shelter to find the oldest, chubbiest, sickest, laziest, cuddliest cat – and there she was in all her glory. Keep doing you, Pearl.

Thank you to Nate Klein. You have cared for me through thick & thin, and continue to be my biggest role model. Thank you for unequivocally supporting my dessert cravings on bad days, giving reassurance and push exactly when I need them, and being a remarkably kind and intelligent person.

Second, thank you to my advisor Haiyi.

Haiyi has been my advisor from Day 1, and thus has not only seen the good & bad of my research but also – even more importantly – the good and bad of my last five years. I had several ups-and-downs during my Ph.D., and I am extremely lucky to have had a mentor who never pushed me to be anything but who I am and stood as a never-ending source of stability. I had the (maybe unusual level of) freedom in this Ph.D. to pursue my ideas and the things I found important; I got very lucky to work with someone who let me grow at my own pace, stayed patient and assured, and supported my ideas while guiding them towards success. Thank you, Haiyi!

Third, thank you to my friends.

To my hometown friends who are like family to me: Stanley, Richard, Matthew, Jessica, Greg, Naheel, Lucus, and more. I consider you my brothers and sisters. Even after over a decade together, you make me laugh more than anyone.

To my undergraduate friends from Cornell (Alice, Chaska, Sitar, Quinn, and many others): thank you for being simply some of the smartest and most passionate people. At a time when I was very unsure whether I was doing the right thing, your pure love for science and engineering motivated me to pursue passion.

To my friends who I've met at CMU and beyond: I have been inspired by you every day – from your courage to pursue this path, to your humility, to your care for the communities that you work with. Thank you to many labmates (Jordan, Seyun, Jane, Anna, Tzu-Sheng, Angela, Hao-Fei, Logan, Zheng) for your continued feedback and being an inspiration to me throughout these years as we start our careers together. Thank you to the people who have brought immense joy and peace to my life: Yi-Fei, Sanika, Hugo, Yi-Hao, Sam, Ningjing, Erica, Mrinal, Cella, Shivani, Aakash, Chris, Wesley, Nayana, Tony, Yuwen, and many more. A special shoutout to Jordan Taylor and Rose Chang for caring for me like family, to Franklin Li for cooking great food that could power me through any bad day, to Seyun Kim for being my coworking and break buddy, and to Meryl Ye for your vibrant and honest spirit.

Finally, thank you to my mentors, collaborators, and committee.

To my early research inspirations — Zina Ben Miled, who was my first research mentor at IUPUI and supported me from the start until long after I finished my internship; Michael Clarkson and Nate Foster who endlessly inspired, encouraged, and instilled confidence in me to pursue academia when I was at Cornell; and Jon Kleinberg at Cornell for inspiring my interest in social computing and being my early role model for what an excellent professor looks like.

To my mentees and collaborators — Hriday, Alekhya, Jasmine, and Bosi for being the best RAs I could have ever asked for. Thank you to Yuhan, Chris, Raj, Haard, PosiTech, and Glen & Cris at 7Cups for showing me how collaboration grows these ideas beyond anything I could ever imagine myself.

To every participant and community member who was part of my work over these last 5 years – your vulnerability and courage is what has kept any of this going. Thank you.

Last but certainly not least, to the committee: Mayank, Bob, Diyi, and Munmun. I asked you to be on this thesis committee simply because I have a deep admiration for your work & brilliance. I want to be like you as I grow up. It is the best feeling to feel humbled and constantly learn in each conversation I have had with each of you. Thank you for your feedback, patience, and time as you've helped me complete this!

# Table of Contents

**– Part 2 –**

***Simulation for Individual Training**: Embodied Practice for Self-Care Skills*

The contributions of this dissertation are based upon the following works:

**Fang, A.**, *& Zhu, H.* (2022). *Matching for Peer Support: Exploring Algorithmic Matching for Online Mental Health Communities. Proceedings of the ACM on Human-Computer Interaction,* 6(*CSCW*2), 1-37.

**Fang, A.*** *& Yang, W.*, Shah, R. S., Mathur, Y., Yang, D., Zhu, H., & Kraut, R. E.* (2024). *What Makes Digital Support Effective? How Therapeutic Skills Affect Clinical Well-Being. Proceedings of the ACM on Human-Computer Interaction,* 8(*CSCW*1), 1-29.

**Fang, A.*** *& Liu, Y.*, Moriarty, G., Firman, C., Kraut, R. E., & Zhu, H.* (2024). *Exploring Trade-Offs for Online Mental Health Matching: Agent-Based Modeling Study. JMIR Formative Research,* 8, *e*58241.

**Fang, A.**, *Chhabria, H., Maram, A., & Zhu, H.* (2025). *Social Simulation for Everyday Self-Care: Design Insights from Leveraging VR, AR, and LLMs for Practicing Stress Relief. In Proceedings of the* 2025 *CHI Conference on Human Factors in Computing Systems* (*CHI* '25). *Association for Computing Machinery, New York, NY, USA, Article* 716, 1–23. *Honorable Mention.*

**Fang, A.**, *Shi, J., Chhabria, H., Li, B., & Zhu, H.* (2026). *Social Simulation for Integrating Self-Care: Measuring the Effects of Contextual Environments in Augmented Reality for Mental Health Practice.*

*Chapter 1*

# Introduction

Mental health is a highly sensitive context where real-world testing can be dangerous. Although many people struggle with their mental health and do not have the proper mechanisms for dealing with their stressors and fears (Demyttenaere et al., 2004a; Folkman, 2020), learning to engage in new ways of coping, soothing, and confronting is intimidating and risky. Attempting new personal coping techniques – for example, learning how to engage in conflict resolution more calmly or overcoming social anxiety through new experiences – may be daunting as failed attempts risk setbacks and real-world consequences.

Similarly, the vulnerability inherent in the mental health context translates into risks when developing and refining technological interventions. While tools like online support communities and mobile applications have expanded access to mental health resources, many of these technologies still struggle to be meaningfully effective or address real-world needs and contexts. As a result, HCI researchers have focused on improving these technologies in recent years. However, the process of iterating and introducing new algorithms into established mental health systems can disrupt people's primary resource of care and potentially have severe consequences for the communities they aim to support. Failures or interruptions in these systems can have significant, harmful impacts. Therefore, while technology has shown great potential in expanding access to mental health care, ongoing development is necessary to ensure that interventions can be deployed safely into real-world settings.

In general, we then see that the mental health context (including the digital one) requires a way to anticipate effects and experiment with different mechanisms for safe execution in the real world. Simulation technology offers an opportunity to address these challenges by creating safe and virtual, yet realistic settings where users can engage with mental health resources in a low-risk way. Through applying the concept of simulation, we can develop a type of "practice" space for designers to try out new algorithms before deployment in the sensitive context of mental health (and beyond) as well as for individuals to try new methods for improving their well-being. From this, we create an environment still accurate to real life to improve mental health techniques in an effective but low-risk setting.

As a result, my dissertation argues that simulating human interactions can help to improve existing mental health support mechanisms and create innovative, interactive environments where users can safely develop self-care strategies. In Part 1, I begin by reviewing my published research on first understanding the limitations in online mental health communities that provide non-professional, peer support. Although I found that these communities can be effective for clinical outcomes like depression and anxiety, I also found that they lack effective mechanisms for matching users compared to traditional therapeutic settings. To address this gap, I apply agent-based modeling to build a simulation for experimenting and understanding trade-offs among different matching protocols for connecting users within these communities. In Part 2, I explore the role of simulation for individuals to practice self-care techniques. Since real-world practice of these techniques can be daunting when it comes to high-stress situations, I first probe users for their perspectives on applying VR/AR and LLM integrated simulation to replicate everyday stress situations for rehearsing calming skills like deep breathing and mindfulness exercises. Building on insights first gained from a design study gathering user needs for simulation for self-care, I then evaluate the quantitative effectiveness of how people apply and transfer learned mental health skills from learning in an AR social simulation. By combining human-centered research and empirical analysis, this work provides actionable insights into user needs that then guide the design of simulations tailored to mental health, and advocates for the role of safe experimentation in mental health.

## 1.1 Thesis Overview

In my thesis, I demonstrate how using simulation technology can help build environments that can support (1) controlled experiments to help improve mental health technology and (2) creating lower risk settings for people to practice new mental health skills. To do so, I use human-centered, empirical analysis to understand user needs, then use those insights to guide the development of novel simulation.

**Chapter 2** synthesizes the relevant literature on human-computer interaction (HCI) for mental health and existing work on simulation for health applications.

**Chapter 3** begins Part 1 of my dissertation. This chapter is an introduction to online peer support communities, including foundational work in understanding the efficacy of these communities.

**Chapter 4** then describes our study using both qualitative interviews with 12 users of one of the largest peer support communities along with quantitative data analysis finding that matching based on demographic has the potential to increase users' experiences.

**Chapter 5** uses results from our empirical findings to create an agent-based simulation to test matching policies on an online mental health community platform, revealing trade-offs with different matching policies.

**Chapter 6** begins Part 2 of my dissertation. I detail my research involving 19 prototype-driven interviews to uncover how people think about self-care technology. I build multiple iterations of low to mid fidelity design prototypes in VR, AR, and text-based chatbots to probe how people envision immersive technology powered by large language models to help with practicing stressful situations.

**Chapter** 7 outlines my final work that designs a full fidelity simulation system developed using the insights from Chapter 6, which I then deploy over a two-week, 44-person user study in order to understand quantitatively how training mental health skills using an AR social simulation affects people's application and transfer of these skills.

**Chapter 8** provides a concluding chapter, in which I discuss the contributions, reflection, and implications from my work on why "*virtual simulation for mental health*".

## 1.2 Summary of Contributions

Altogether, my dissertation contributes a novel application of simulation to the domain of mental well-being. My work shows how simulation, particularly when its design is guided by human-centered empirical understanding, offers a safe environment for experimenting with new interventions for the sensitive context of mental health. More specifically, my work has made the following contributions:

- A review of background and related research on mental health technology and simulation approaches in HCI for mental health
- Empirical study of the efficacy of online peer support communities
- Synthesized findings regarding user needs, opportunities, and challenges for designing algorithmic matching in online peer support communities
- A novel agent-based modeling system for simulating different matching protocols in online support communities
- An introduction into using simulation for self-care practice
- Empirical understanding of user needs and opportunities for simulation for practicing mental health skills using XR and LLM capabilities
- A novel immersive, AR simulation for practicing different types of self-care techniques to address everyday stress
- An evaluation study of the simulation system that evaluates its usefulness and effectiveness at reducing stress levels and applying mental health skills for everyday experiences

Overall, my research thesis advocates for the importance and usefulness of safe experimentation in mental health contexts, and introduces how human-centered design with empirical evidence can guide the creation of these simulations for supporting well-being across multiple domains. Together, these contributions demonstrate the potential of simulation to advance mental health support through both community-driven and self-directed approaches, creating safe spaces for exploration, skill-building, and experimentation.

*Chapter 2*

# Related Work

In this chapter, I review the relevant literature on mental health as a domain, the role of human-computer interaction (HCI) in addressing mental health needs, and the current state of simulation approaches that form the foundation to this dissertation.

## 2.1 The Mental Health Crisis and HCI

The mental health crisis is an urgent and complex issue affecting millions worldwide. Mental health disorders such as anxiety, depression, psychosis, and others impact not only individual well-being but also that of families, communities, and societies at large. Furthermore, poor mental health is indeed an active crisis – today, more people than ever struggle with mental health disorders like depression, anxiety, PTSD, and substance use disorders (Y. Wu et al., 2023), and this number continues to rise each year alongside greater pressures of modern life such as economic instability and social isolation further exacerbating these struggles (Coley & Baum, 2021; Knapp & Wong, 2020; Steel et al., 2014).

However, mental health concerns are not unique to our present. Mental health issues can be traced back centuries, and mental health care as we currently define it has been consistently met with stigmatization, neglect, and insufficient resources (Kohn et al., 2004; Suite et al., 2007). While there may never be a time where mental health issues do not exist at all, and systemic causes such as economic strain and social isolation are important to address in the long-term, the pervasive nature of mental health challenges underscores the need for immediate and accessible solutions that can reach people in the here and now. With the advancement of technology through apps, online platforms, virtual reality, AI, and more, there are growing attempts to make more powerful, effective, and accessible tools that address mental health today.

As mental health struggles continue to not only exist but to rise globally, a primary (and still growing) area of research among HCI scholars has been on understanding and developing digital interventions that support mental and emotional well-being. These technological solutions have demonstrated potential to provide accessible and greater variety of mental healthcare options, such as virtual therapy sessions (Liu et al., 2021; O'leary et al., 2018; Thieme et al., 2023; Zamanifard & Robb, 2023) and self-tracking applications (Kruzan et al., 2023; R. Wang et al., 2018).

The Diagnostic and Statistical Manual of Mental Disorders (DSM-5) (Georgiopoulos et al., 2025) defines Major Depressive Disorder (MDD) as encompassing symptoms such as persistent low mood, diminished interest in activities, sleep disruptions, appetite changes, and feelings of worthlessness. Depression's high prevalence, affecting approximately 280 million people worldwide (*Depressive Disorder* (*Depression*), n.d.), underscores the importance of technological solutions that can facilitate early detection and timely intervention. In response, HCI has made significant strides in automatic detection and treatment of depressive disorders. For instance, advancements in predictive algorithms analyze behavioral data from social media, wearables, and other digital phenotyping sources to recognize early symptoms of depression (Balcombe & De Leo, 2022; M. Choudhury et al., 2013; Hickey et al., 2021; Xu et al., 2019). These tools aim to support proactive care by alerting users and caregivers to potential mental health issues before symptoms intensify.

Anxiety and stress are also two major topics of HCI research, which are also discussed in this dissertation. The DSM-5 defines anxiety disorders as a group of conditions characterized by excessive fear and anxiety, along with related behavioral disturbances. Generalized Anxiety Disorder (GAD) is the most common anxiety disorder, which is characterized by excessive and persistent anxiety and worry that is difficult to control and causes significant distress. Work on anxiety and stress is generally a broad spectrum, including how digital health interventions can alleviate symptoms but also how technology like social media can exacerbate distress (Keles et al., 2020; Thieme et al., 2023). In general, the most recent work has largely been focused on interventions with AI (e.g. chatbots) that can deliver professional therapy or non-professional support (Bae Brandtzæg et al., 2021; D'Alfonso et al., 2017; Kian et al., 2024). Ultimately, technological approaches in HCI hold significant promise for improving mental health outcomes by broadening access to mental health resources, facilitating early intervention, and supporting long-term personal growth and resilience. However, past work in HCI also underscores the importance of well-intentioned and well-designed technology for mental health, given evidence that technology optimized for other means (e.g. profit), poorly designed, or with unanticipated effects can have significant harmful effects (Braghieri et al., 2022; D. Choudhury et al., 2023; Twenge, 2020).

## 2.2 Simulation Approaches to Well-Being

Simulation has a long history in the social sciences for modeling and testing different aspects of human behavior, providing valuable insights into complex social phenomena and enabling safe experimentation (Gilbert & Troitzsch, 2005; Tesfatsion & Judd, 2006a). In the field of HCI and computational social sciences, simulations have traditionally examined virtual behaviors within online networks (Alvarez-Galvez, 2016a; Plikynas et al., 2015a; Schweitzer & Garcia, 2010a) or been used to guide design decisions for online communities (Ren & Kraut, 2014). Simulation from an embodiment standpoint rather than an agent-based standpoint has been used in HCI for a variety of applications, usually related to skill- and experience-building. For example, HCI researchers have used VR simulation to explore skills anywhere from driving (Matviienko et al., 2023; Walch et al., 2017; Yeo et al., 2020), to providing care in a variety of contexts (Hilty et al., 2020; Hodge et al., 2018; Tang et al., 2022). Importantly, embodied simulation has also been used for probing initial designs and user needs before the stage of building new systems (Goedicke et al., 2022; Luchak et al., 2024; Tigwell, 2021).

### *Agent-Based Modeling*

Agent-based modeling, which is the simulation of the actions of agents to understand their behaviors and interactions under different conditions, has been applied to a wide range of research domains including economics, urban studies, and social sciences (Gilbert & Troitzsch, 2005; Huang et al., 2014; Tesfatsion & Judd, 2006b). Relevant to my work, agent-based modeling has been historically useful for informing the design of online communities, given its ability to simulate how different design choices impact outcomes (Alvarez-Galvez, 2016b; Ren & Kraut, 2014; Z. Tan et al., 2011; van Maanen & van der Vecht, 2013). One other primary benefit of agent-based modeling, as opposed to direct experimentation, is that effects can be observed over long periods as one can run the agent-based model repeatedly and for lengthy time cycles; thus, downstream and even unintended effects (rather than just first-order effects) can be identified (Kraut & Ren, n.d.). Importantly, agent-based modeling serves as a testing ground for community designers and researchers to surface how different theories affect the community broadly but also allow them to isolate the factors leading to particular outcomes. For example, Ren and Kraut (Ren & Kraut, 2014) have shown how agent-based modeling can be used to apply social science theories and understand trade-offs in design decisions; they applied these methods to explore motivations for online community participation and how different moderation methods affect discussion in online communities. Beyond that, agent-based models have been used to design algorithms for connecting individuals and communities in online communities and to capture the emotional behaviors of agents (Mitrović & Tadić, 2012; Schweitzer & Garcia, 2010b; Sibley & Crooks, 2020).

### *Extended Reality (XR)*

Extended reality (XR) is another simulation method that can serve as an emulation of the real-world, including human behaviors and interactions similarly to agent-based modeling. However, as opposed to agent-based modeling that is usually used to simulate actions for observation of patterns and interactions by the simulators (*e.g.*, researchers), XR simulates environments often for direct interaction and immersion for users. Extended reality (XR), including virtual reality (VR) and augmented reality (AR), has shown immense promise in HCI for immersing users into visual, auditory, and sensory environments. XR encompasses both virtual and mixed real-and-virtual environments that use computer-generated elements to immerse users into a digital environment. VR is defined by fully immersing users in a digital environment, while AR overlays content onto the real world. XR technology has made significant strides in recent years within healthcare to provide sensory and immersive experiences that can help provide treatment and exposure for health purposes (Freeman, 2008; Meyerbröker & Emmelkamp, 2010; Reger et al., 2016). My dissertation work includes using the simulation and immersion capabilities of XR to help users' exposure to environments that are mental health concerns. Below, I review relevant literature on XR in HCI for mental health purposes.

XR interventions have demonstrated efficacy in fostering sensory awareness, facilitating mindfulness, and aiding users in managing a variety of mental health challenges. Studies have shown that XR can support practices like meditation, mindfulness, sensory awareness, and emotional regulation (Bruggeman & Wurster, 2018; Kaplan-Rakowski et al., 2021; Seabrook et al., 2020; Waller et al., 2021; Yildirim & O'Grady, 2020) through immersing users into sensory environments that allow more tactile interaction with visual, auditory, and other sensory triggers or visualizations. Relevant to my dissertation research, VR can act as an immersive environment for exposure therapy, allowing people to become familiar with simulated realistic environments that have previously caused them anxiety and fear responses. By immersing users in therapeutic environments, VR has proven effective for addressing specific mental health conditions as well as more general well-being.

Beyond condition-specific treatments, some XR research has also begun focusing on supporting everyday users in cultivating mental health skills. This is exemplified by HCI efforts in delivering mindfulness and meditation practices through VR and AR platforms, such as teaching breathwork techniques (Prpa et al., 2018), and guiding walking meditations in multimodal settings (Chung et al., 2016; F. F.-Y. Tan et al., 2023a). The embodied, interactive nature of XR also allows users to engage physically with abstract concepts, for instance, by "punching away" negative thoughts in VR environments (Grieger et al., 2021). While much of the work in XR for mental health has been oriented toward immersive, otherworldly settings — such as calming natural landscapes or tranquil, abstract environments (Chandrasiri et al., 2020a; Miller, Stepanova, Desnoyers-Stewart, Adhikari, Kitson, Pennefather, Quesnel, Brauns, Friedl-Werner, Stahn, et al., 2023)— the focus of my research introduced in Part 2 of this dissertation diverges by simulating realistic, socially and environmentally grounded scenarios, such as public speaking or social anxiety in everyday settings, where participants can actively practice in-the-moment mental health skills.

### *Large Language Models (LLMs)*

Large language models (LLMs) have been shown to be incredibly effective at simulating human conversation. Now more than ever, we are able to simulate realistic human dialogue and intricate social dynamics. For example, recent notable work by Park et al. demonstrated the use of LLM-powered "generative agents" that behave as human-like avatars in a simulated virtual town, engaging in social interactions, forming relationships, and coordinating activities in lifelike ways (Park et al., 2023). Given its potential to thus emulate human companionship, care conversations, and even therapeutic or professional-like advice, there has been a rapid expansion of large language models in the mental health domain in both clinical settings and broader non-professional contexts. LLMs have been used to offer therapeutic support (Fu et al., 2023; Lai et al., 2023; Loh & Raamkumar, 2023), classify mental health experiences (Xu et al., 2024), and teach coping skills (Hu et al., 2024; Shaikh et al., 2024; Yang et al., 2024). Emerging research also explores LLMs as tools for individuals to gain mental health insights, such as by helping users reflect on journaling entries and summarizing key themes for caregivers, as well as assisting in restructuring negative thought patterns (T. Kim et al., 2024; Sharma et al., 2024).

LLMs power conversational agents and mental health chatbots that deliver support to both care-seekers and providers. LLM-based agents can generate contextually aware responses for direct mental health support, helping users reframe thoughts and manage emotions (Maddela et al., 2023). However, LLM-based interventions are not without risks, as these models may propagate misinformation or exhibit biases that could be detrimental in sensitive mental health contexts (D. Choudhury et al., 2023; Zhou et al., 2023). Consequently, our approach prioritizes a user-centered design, emphasizing the importance of understanding human preferences, needs, and boundaries when engaging with these technologies. Our study does not employ LLMs to deliver direct mental health guidance; instead, we use LLMs to simulate realistic dialogues and gauge user responses to indirect, system-generated recommendations, such as breathing guidance timing.

Overall, my dissertation uses a combination of simulation-based approaches – including agent-based modeling, XR, and LLMs – to showcase their usefulness in the mental health domain. Through showcasing the capabilities of these simulation methods to be realistic and informative, my work first and foremost shows that simulation's features of *safety* and *control* are valuable and effective at helping to improve mental health technology and provide an environment for individuals to improve their own coping skills.

# Part 1

*Simulation for Designing Online Mental Health Communities*

---

*The following chapters are based upon the following previously published works*:

> ***Fang, A.****, Yang, W.*, Shah, R. S., Mathur, Y., Yang, D., Zhu, H., & Kraut, R. E.* (2024). *What Makes Digital Support Effective? How Therapeutic Skills Affect Clinical Well-Being. Proceedings of the ACM on Human-Computer Interaction,* 8(*CSCW*1), 1-29.
>
> ***Fang, A.****, & Zhu, H.* (2022). *Matching for Peer Support*: *Exploring Algorithmic Matching for Online Mental Health Communities. Proceedings of the ACM on Human-Computer Interaction,* 6(*CSCW*2), 1-37.
>
> ***Fang, A.****, Liu, Y.*, Moriarty, G., Firman, C., Kraut, R. E., & Zhu, H.* (2024). *Exploring Trade-Offs for Online Mental Health Matching*: *Agent-Based Modeling Study. JMIR Formative Research,* 8, *e*58241.
>
> **indicates equal contribution*

---

In *Part One of this thesis proposal, I review my work in online mental health communities, and how simulation can aid in safe experimentation for new algorithms in evolving, virtual support technology.*

*Online peer support communities have become a crucial supplement to traditional mental health care, offering accessible, anonymous, and immediate support. However, their effectiveness is constrained by algorithmic limitations such as how users are connected to one another. While interventions – such as improved matching algorithms – hold the potential to enhance these communities, real-world experimentation comes with immense ethical risks; altering matching systems without understanding their impact could disrupt users' primary source of mental health support. Simulation offers a powerful way to address these challenges by creating a controlled environment where different matching strategies can be tested without real-world consequences. In this section, I apply a human-centered approach to designing agent-based modeling to explore how we may safely evaluate effects of different designs for these virtual support communities, including demonstration on how computational models can provide insights into optimizing these platforms while prioritizing user well-being. This work lays the foundation for designing safer, more effective online mental health interventions and sets the stage for broader applications of simulation in mental health technology.*

# *Chapter 3*
# Understanding Online Mental Health Communities

In my first demonstration of how simulation can provide a safe environment in the mental health context, I conducted three pieces of work that yield understanding of OMHCs as a whole and build a simulation system to address unmet needs. In Chapter 3, I lay the groundwork for understanding OMHCs, providing the essential context needed for detailed discussions of research done in Chapters 4 and 5. This chapter includes an overview of the research platform of interest and a brief summary of a published study that contributed to our foundational understanding in the efficacy of OMHCs.

## 3.1 Overview of OMHCs

Mental health problems continue to rise globally and under-treatment of serious mental health problems remains a major problem. Although there is significant evidence supporting the effectiveness of professional treatments for mental health, such as therapy, a substantial proportion of people with mental health problems fail to receive any treatment because they lack access to services or have needs unmet by health services (H. S. Kim et al., 2008; Mojtabai et al., 2011). As a result, peer-to-peer support through online mental health communities (OMHCs) has emerged as an accessible tool for achieving mental and emotional support. OMHCs include sites like 7 Cups and TalkLife, which usually provide free 24/7 peer-to-peer support from volunteers; online support communities are thus able to provide support at scale for a wide variety of mental and emotional problems (Dowling & Rickwood, 2013; Fukkink, 2011). OMHCs may refer to support groups within general-purpose social networks, such as Facebook groups dedicated to specific health issues (Bender et al., 2011; Greene et al., 2011), and platforms aimed entirely at health support, such as CrisisTextLine, 7Cups.com, and TalkLife that connect support seekers with volunteer peer counselors for supportive chats. Peer support through online communities is able to fill gaps in accessing health services for many, helping those whose needs are unmet by traditional resources (Gidugu et al., 2015; McBeath et al., 2017; Rassau & Arco, 2003) or individuals who lack adequate peer networks in their daily lives to achieve needed support (Kawachi & Berkman, 2001; Kessler et al., 2005; Solano, 1986). Support through OMHCs has been found to have numerous benefits, such as yielding meaningful relationships and increasing trust in getting mental health treatment (Fan et al., 2014; Pfeiffer et al., 2011; Rogers et al., 2015). Additionally, OMHCs circumvent many key barriers that prevent help-seeking and allow immediate addressing of mental health needs (Demyttenaere et al., 2004b). Both general purpose social networks and ones focused exclusively on mental health are important in spreading health information, reducing harmful thoughts, empowering help-seeking for stigmatized populations, reducing suicidal ideation, and fundraising or spreading awareness for health issues (De Choudhury & De, 2014; De Choudhury & Kıcıman, 2017; Gui et al., 2017; Huh, 2015; Prescott et al., 2020). There is evidence, though, to support that online mental health platforms may deviate from conventional practices in traditional, offline therapy contexts; for example, there are less distinct boundaries between support-seeker and support-provider, and volunteer counselor behaviors are influenced by training that takes place through platform-provided resources (Yao et al., 2022).

### *Research Site*

For this series of research (Chapters 3, 4, and 5), I collaborated with one of the world's largest online peer support platforms, which I will refer to anonymously in this dissertation as the "online support platform" or "research site". The online support platform is a pseudonymous support service where individuals ("support-seekers") can engage in text-based chats with volunteer support-providers ("counselors", or "support-providers") on a variety of mental health problems.

The online support platform allows support-seekers to take free clinical mental health exams. Support-seekers can complete questionnaires called the PHQ-9 and PHQ-2 to measure their depression and GAD-7 and GAD-2 to measure anxiety. These assessment instruments are routinely used in both the research and medical setting for diagnostic and severity measurement (Behar et al., 2009, 2009; Löwe et al., 2006; Ruiz et al., 2011; K. Wang et al., 2018). PHQ-2 includes the first two questions of PHQ-9 (Kroenke Kurt & Spitzer Robert L, 2002), which is the 9-item questionnaire measuring depression presence and severity. Similarly, GAD-2 includes the first two questions of GAD-7 (Spitzer et al., 2006), the 7-item questionnaire measuring anxiety severity. Support-seekers can take the PHQ and GAD exams every two weeks to measure any changes in their depression and anxiety symptoms. Although the 2-item versions for PHQ and GAD are not as accurate as the longer questionnaires, PHQ-2 and GAD-2 have been shown in prior work to maintain relatively high accuracy including both considerable sensitivity and specificity for measuring depression and anxiety, respectively (Arroll et al., 2010; Plummer et al., 2016). Sensitivity and specificity for PHQ-2 are 86% and 78%, respectively; GAD-2 has a sensitivity of 76% and specificity of 81% (Arroll et al., 2010; Plummer et al., 2016). A complete list of items in PHQ-2 and GAD-2 can be found in the appendix. Because these scales ask about the frequency of symptoms of depression and anxiety in the past two weeks, a *higher* score on PHQ and GAD indicates *worse* mental health, while a lower score indicates better mental health. For both PHQ-2 and GAD-2, scores range from 0 (better mental health) to 6 points (worse mental health).

Also included in the dataset is "chat ratings", which are a rating out of an equidistant scale of 1 to 5 stars that support-seekers can give once a chat has continued for a certain length of time or after the chat ends. Support-providers are not able to rate a chat. Below, we review the quantitative description of our dataset's chat rating and clinical questionnaire data.

Each chapter will include a reminder for the descriptive statistics of the study.

## 3.2 Do OMHCs work?

Despite the many benefits available through OMHCs, many aspects of their effectiveness such as clinical improvement was not yet understood. A foundational aspect to the first leg of my research analyzing OMHC users' needs and improving these communities relies on that these communities *should* be improved in the first place. After all, a fundamental element to human-centered technology design is also considering whether technology should have a place at all. That is – are they useful, and do they work?

Online support providers on online platforms receive relatively little training (Yao et al., 2022) in contrast to the extensive training for mental health professionals and even volunteers for crisis intervention programs (Boswell & Castonguay, 2007; Gould et al., 2022; Mufson et al., 2006; Paukert et al., 2004). OMHCs often require volunteer support providers to complete short training sessions along with optional specialized training; for example, 7Cups.com provides short training sessions based on skills established in face-to-face mental health support such as active listening, empathy, and motivational interviewing (MI) techniques. Another online support platform, TalkLife, provides small-group and module-led training for peer support techniques including asking questions, showing empathy, and reflection. However, there is little known about whether therapeutic skills and other "common factors", which are effective for improving people's mental health when used by professionals in offline settings (Jani et al., 2012; Lundahl et al., 2010; Norcross & Lambert, 2019), are effective when used by peer supporters online. Resolving this uncertainty is especially important, as doing so can inform the design of online support providers' training and build understanding of whether OMHCs are effective tools for delivering support.

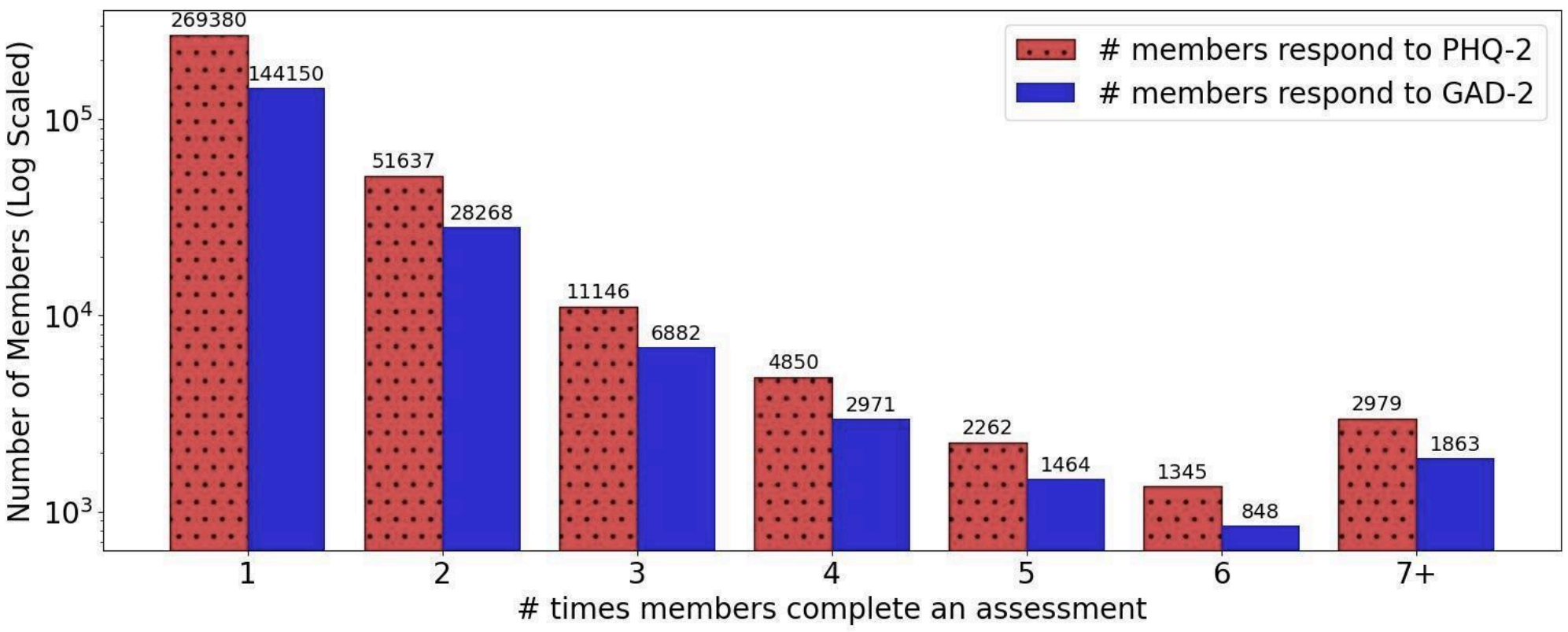

**Figure 1.** *Distribution of the number of times support-seekers completed the PHQ-2 and GAD-2 clinical assessments.*

To answer these questions, we conducted a study using natural language processing and machine learning techniques to analyze changes in support-seekers' depression and anxiety in a dataset spanning over two years containing over 8-million anonymous text-based support chats between volunteer support counselors and online support-seekers. We explore the effectiveness of support behaviors, like asking open-ended questions and providing reflections involved in motivational interviewing and expressing empathy, used by these OMHC volunteers and examine how these behaviors are associated with long-term improvement in support-seekers' depression and anxiety. Specifically, we answered the following foundational research questions:

***RQ*1.** *Does participating in support conversations in OMHCs improve people's depression and anxiety?*
***RQ*2.** *What volunteer characteristics and techniques used in these conversations affect mental health outcomes?*

### *Dataset*

To answer these questions, we used a dataset spanning January 2020 to May 2022 and contains over 8 million chats, involving over 1.5 million support-seekers, and 288 thousand support providers. A typical (median) chat lasts for 23 minutes and, over the span of our two-year dataset, support-seekers chatted with 12 distinct providers on average. All data was anonymized throughout the entire research process including analysis and no personally identifiable information was used in this study. None of the studies in this thesis deploy interventions or alter people's experiences on the platform in any way given the potential dangers and ethical concerns in doing so.

Our study's dataset includes 343,599 support-seekers' responses for PHQ-2 and 186,446 for GAD-2 between January 2020 to May 2022. Because we are interested in changes in mental health, we focus on the support-seekers who completed the questionnaires at least twice – 74,219 support-seekers for PHQ-2 and 42,296 support-seekers for GAD-2. In over 60% of questionnaires completed on the site, support-seekers show evidence of clinically significant symptoms of depression and anxiety (see Table 1). Descriptive statistics for the dataset are included in Table 1 while the detailed distribution of the number of responses per support seeker is shown in Figure 1.

**Table 1.** *Descriptive statistics of dataset. Note: A "chat session" refers to back-and-forth messages exchanged without a break of more than 24 hours. The length of a chat session is measured by the duration between the first and last messages exchanged.*

| | PHQ-2 | GAD-2 |
|---|---|---|
| **# of users, taken at least 2x** | 74,219 | 42,296 |
| **Mean score (out of 6)** | 3.55 | 3.58 |
| **% of users with significant symptoms (i.e., score ≥ 3)** | 63% | 61% |
| | **Obs. period, between PHQ assessments** | **Obs. period, between GAD assessments** |
| **Median # of chat sessions** | 2 | 3.20 |
| **Median # of messages, in total chat sessions** | 286 | 314 |
| **Median duration of total chat sessions (in days)** | 12.33 | 18 |

To answer RQ1, we did not pursue a randomized controlled trial or other interventional study (e.g., randomly withholding treatments) to investigate our research questions due to ethical concerns. Instead, we used a pre-post observational study design to infer the treatment effects of engaging in one or more chats on support-seekers' mental health. Specifically, we adopted an "observation-level" random-effects regression with propensity score matching to predict support-seekers' post-assessment mental health scores from their pre-assessment scores, depending on whether they chatted with a support-provider during the observation period and other independent variables.

In order to monitor mental health changes, we first selected an "observation period" as our unit of analysis. This period is defined by a support seeker completing two consecutive PHQ-2 or GAD-2 assessments. We refer to the first questionnaire in this sequence as the "pre-assessment", and the following one as the "post-assessment". These observation periods enable repeated observations for the same individuals at different time points; thus, this panel data allows us to use multilevel models like random effects to account for potential omitted-variable bias.

### *Mitigating Selection Bias via Propensity Score Matching*

To account for individual differences between people, such as their prior experience on the site, initial mental health status, or age, that may influence whether support-seekers engage in support chats, we used Propensity Score Matching (PSM). This quasi-experimental method helps emulate the conditions of Randomized Controlled Trials (RCTs) prior to regression analysis. PSM is recognized for its effectiveness in reducing the effects of confounding in observational studies and has been applied in online mental health contexts similar to our own.

In our study, each observation falls into the treatment group if the support-seeker engaged in at least one chat during the observation period. Although a possible alternative would involve using the total number of chats rather than splitting support-seekers into groups based on whether they had any chats, data limitations prevent us from creating multiple small groups. This alternative method could be explored in future work. The goal of PSM is to match each observation in the treatment group with a similar observation in the control group (support-seekers who did not participate in chats) based on characteristics like demographics and mental health status. To do this, PSM first estimates a propensity score for each observation – the likelihood of being in the treatment group given certain covariates – and then pairs observations from the treatment and control groups with similar scores. This pairing results in treatment and control groups that are comparable on these characteristics but differ in whether they received "treatment."

**Step 1. Propensity Score Estimation**

We used a logistic regression model to estimate propensity scores, predicting the likelihood of a support-seeker engaging in chats during a given observation period based on the covariates listed below.

Though demographic factors like age, gender, and ethnicity are often included in similar studies, our platform collects limited background information. One key demographic factor we could reliably measure is the age of support-seekers. The mean age of support-seekers is 24.52 years, with a standard deviation of 10.61. Age is the only demographic variable that seekers are required to provide upon signing up.

Another important factor is a help-seeker's prior experience on the platform. We operationalized this through three specific metrics.

We implemented a one-on-one matching strategy without replacement, based on the logit of the estimated propensity score. Each observation in the treatment group was paired with a counterpart in the control group with the closest logit within the specified caliper. The caliper was set at 0.2 of the standard deviation of the propensity score's logit, following recommended best practices – specifically, 0.205 for PHQ outcomes and 0.226 for GAD outcomes.

Following matching, we identified 43,249 matched pairs for the PHQ analysis and excluded 12,854 individuals in the treatment group who lacked an adequate match. The GAD analysis includes 26,224 matched pairs, with 9,747 people in the treatment group excluded due to insufficient matching. Most variables show an 80-90% reduction in bias after matching, and all covariates display less than five percent bias, a recommended threshold. No reduction in bias for pre-assessment mental health scores was observed, as the treated and control groups already had similar scores.

### *Regression Analysis*

After matching, we conducted a random-effects, multi-level regression with observations nested within support-seekers. Given that about half of the sample had only a single observation, fixed-effects regressions were deemed inappropriate. This regression predicts support-seekers' post-assessment mental health (i.e., PHQ and GAD scores) from their pre-assessment mental health status, based on whether they had at least one support chat during the observation period (treatment group) or not (control group). Including pre-assessment mental health scores in the regression models allows us to effectively examine how the independent variables predict changes in mental health. To test whether engaging in support chats has a stronger effect for individuals with worse initial mental health, we included an interaction term between participating in a chat and pre-assessment mental health. Additionally, we controlled for the number of support chats the seeker had before the observation period.

### *Findings*

Table 2 shows the results for RQ1, the effects of chats on changes in mental health. As a reminder, because the PHQ-2 and GAD-2 scales ask about the frequency of symptoms, positive coefficients in the regression analyses indicate that a variable was associated with an increase in symptoms (ie., worse mental health). In terms of control variables, as expected, both depression ($\beta = .67$, $p<0.001$) and anxiety ($\beta = .68$, $p<0.001$) were very consistent over the several weeks between pre- and post- assessments. In addition, support-seekers who had more on-platform conversations prior to the current observation period had lower post-assessment depression ($\beta = -.05$, $p<0.001$) and anxiety ($\beta = -.07$, $p<0.001$) scores.

**Table 2.** *Random-effects regression predicting support-seekers' post-assessment PHQ-2 or GAD-2 scores, comparing matched seekers who had or did not have a support chat during the observation period, controlling for their pre-assessment mental health and the number of chats they had prior to the observation period and the pre-assessment mental health X number of prior chat interactions.*

| | PHQ-2 Coeff. | Robust SE | GAD-2 Coeff. | Robust SE |
|---|---|---|---|---|
| Support-seeker experience in chats (log) | -0.052*** | 0.006 | -0.065*** | 0.008 |
| Pre-assessment mental health score (MHS) | 0.668*** | 0.004 | 0.683*** | 0.005 |
| Treatment: Participated in ≥ 1 chats during observation period | -0.126*** | 0.010 | -0.036** | 0.013 |
| Treatment × MHS | -0.155*** | 0.006 | -0.145*** | 0.007 |
| Constant | 3.620*** | 0.007 | 3.739*** | 0.010 |

| N (Respondents) | 53,915 | 32,222 |
|---|---|---|
| N (Observations) | 86,498 | 52,448 |

*p < 0.05; **p < 0.01; ***p < 0.001

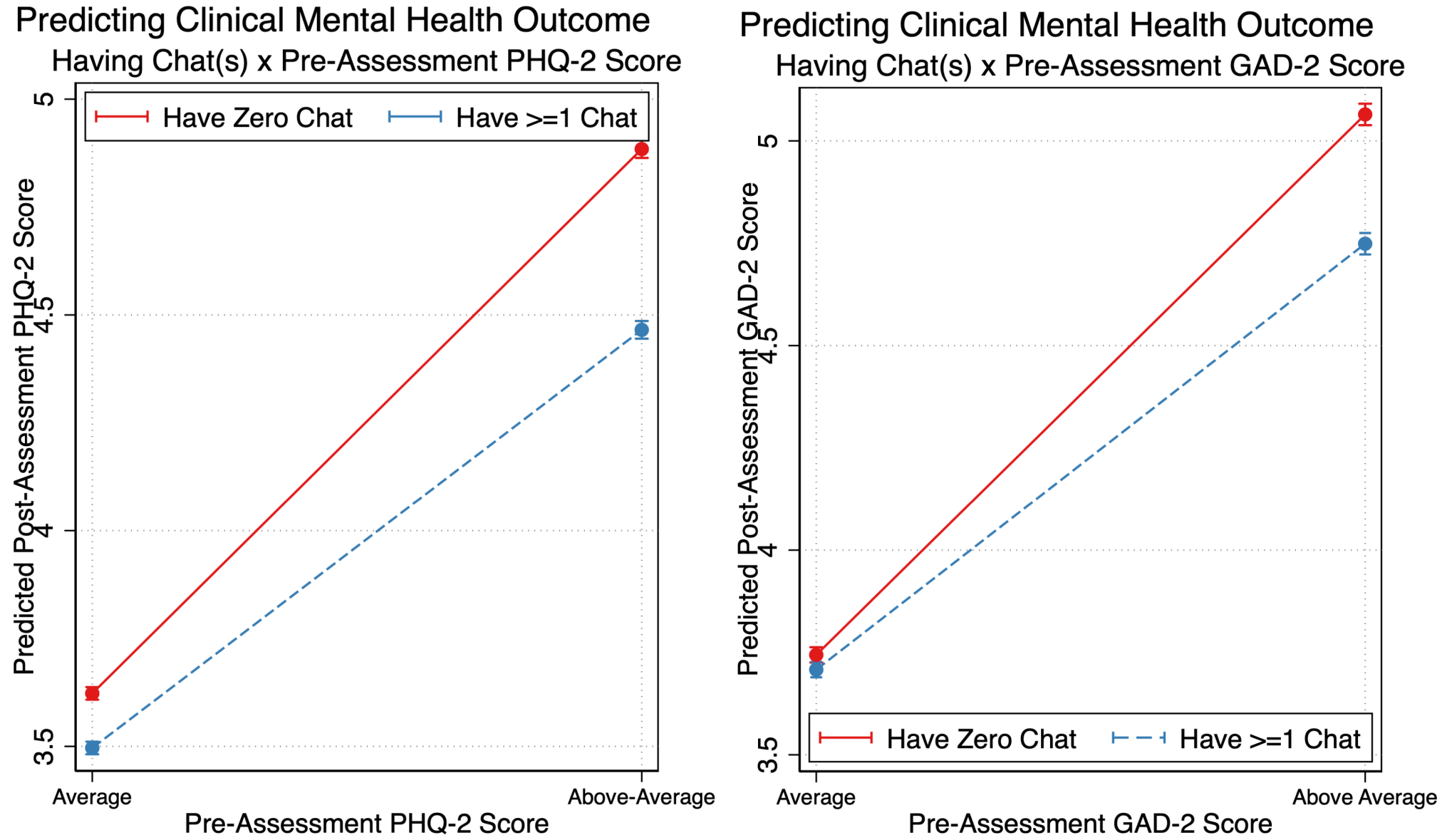

**Figure 2.** *Marginal plot that illustrates the interaction effect between treatment and pre-assessment score, for support-seekers who have average vs. above-average initial symptom severity. Note that a higher score indicates worse symptoms (i.e. greater symptom severity).*

Most importantly, the results show that the treatment – participation in at least one support-chat during the observation period – was associated with small but highly reliable improvements in mental health. Participation in support-chats was associated with improvements in symptoms of both depression ($\beta$ =-0.126, $p < 0.001$) anxiety, although to a lesser degree ($\beta$ = -0.036, $p < 0.01$). In addition, the Treatment X pre-assessment mental health interactions show the benefits of participating in the support chats were greater for those who initially had worse mental health for both depression ($\beta$ =-.155, $p<0.001$) and anxiety ($\beta$= -.145, $p<0.001$). As illustrated in Figure 2, the dotted blue lines consistently fall beneath the solid red ones. This indicates that support-seekers who participate in chats exhibit fewer symptoms of depression (PHQ-2) and anxiety (GAD-2) (indicative of improved mental health) than those who do not participate. In addition, the Treatment X MHS interactions were significant and negative for both the PHQ-2 scale ($\beta = -0.155$, $p < 0.001$) and the GAD-2 scale ($\beta = -0.155$, $p < 0.001$), indicating that participating in a chat was associated with improved mental health more for those who initially had more depression and anxiety symptoms respectively. Altogether, what we concluded is that participating in OMHCs through **chatting with volunteers improves mental health outcomes to a small but significant degree – and people who have more severe depression or anxiety derive more benefit.** This helps highlight how new mental health interventions not only increase access to mental health support, but also are actually effective (even to a small degree) currently.

These findings beg a second question: ***what*** is making online support communities work (or not work)? Since the findings for this question are not directly involved in our discussion on simulation in this dissertation, but may also be an open area for simulation and emulation to help improve OMHCs (*e.g.*, through informing training simulation for OMHC volunteers), we briefly summarize findings below and link to the full paper with detailed insights here.

Firstly, we collected consecutive PHQ-2 and GAD-2 assessments from support-seekers and arranged them in pairs, creating an observational period. Then, for support-seekers who had a support chat during the observational period, we measured factors that the prior literature suggests should be associated with changes in support-seekers' mental health. These include characteristics of the support providers and support-seekers before the observation period started and characteristics of the support conversations that occurred during the observations period. Finally, we used random-effects regression models to predict how the providers', seekers', and chat characteristics predict changes in mental health assessments (i.e., post-assessment controlling for pre-assessment).

For each observation period in which a support-seeker has at least one conversation, we used seven independent variables as listed below.

**Support-provider characteristics**

- ❖ **Chats with repeat support providers**: capturing the percentage of chats the support-seeker had with a support-provider they previously interacted with before the time of pre-assessment. This tracks the relationship between support-seekers and support providers, a crucial factor in counseling processes as suggested by previous studies (Ackerman & Hilsenroth, 2003; Wampold & Flückiger, 2023).

- **Support-providers' age**: a continuous variable representing the average age of the support providers interacted with by the support-seeker.
- **Support-providers' experience in chats**: a variable measuring the average total number of chats these support providers have engaged in from their registration until the time of pre-assessment.
- **Support-providers' tenure**: time since the support provider registered on the online support platform.
- **Support-providers' training modules**: the average number of awards for support providers for completing additional training modules by the platform.

**Support-seeker characteristics**: We measured two characteristics of the support-seeker before the start of an observation period.

- **Support-seekers' pre-assessment mental health scores (MHS)** (i.e., PHQ-2 or GAD-2): By including pre-assessment scores, this lagged dependent variable model effectively examines how the independent variables predict changes in mental health status (J. Cohen, 2013)).
- **Support-seekers' tenure**: As a measure of the support-seeker's experience with the platform, we included the time since the support-seeker registered on the online support platform. To account for skewed distribution, we used a log transformation.

We analyzed all communication the support-seeker had with one or more providers during an observation period and measured the following attributes of these conversations:

- **Total conversation turns**: The number of conversational turns during an observation period is a measure of the conversation length. This allows us to assess the "dosage effect" of peer-support chats on mental health.

- **Sentiment in support-seekers' language**: The positivity or negativity in the support-seekers' language during conversations
- **Support providers' therapeutic language**:
  - **Motivational Interviewing behaviors**: We measured seven Motivational Interviewing behaviors as defined in the MITI 4.2 manual (Moyers et al., 2016), including asking questions, offering affirmations, and reflecting the seeker's thoughts, using classifiers developed by (R. S. Shah et al., 2022).
  - **Empathy**: We assessed three components of empathy (emotional reaction, exploration, and interpretation) based on classifiers developed by (Sharma et al., 2020).

*Regression Analysis*

In order to understand how support-provider characteristics, support-seeker characteristics, and characteristics of the conversations predict changes in mental health, we used a similar random-effect model specification as in RQ1 with additional predictor and control variables. However, because the RQ2 analysis contains many more predictor variables and all predictors are continuous rather than binary, we cannot use propensity score matching as a pre-processing step. Results are shown in Table 3.

**Table 3.** *Random effects regression predicting support-seekers' ("seeker") post-assessment PHQ-2 and GAD-2 mental health score, from characteristics of support providers ("provider"), seekers, and chats.* *: $p<0.05$, **: $p<0.01$, ***: $p<0.001$

| Variable | PHQ-2 MITI | PHQ-2 Empathy | GAD-2 MITI | GAD-2 Empathy |
|---|---|---|---|---|
| **Independent Variables** | | | | |
| Constant | 3.526*** | 3.526*** | 3.737*** | 3.737*** |
| Provider's age | -0.036*** | -0.039*** | -0.019 | -0.016 |
| Provider's chats (log) | -0.001 | 0.003 | -0.012 | -0.011 |
| Provider's training modules (log) | 0.017 | 0.011 | 0.037*** | 0.030** |
| % chats with same supporter | 0.015 | 0.012 | 0.000 | -0.002 |
| **Seeker Characteristics** | | | | |
| Seeker's time on platform (log) | -0.005 | -0.007 | -0.010 | -0.010 |
| Pre-assessment MHS | 0.446*** | 0.446*** | 0.460*** | 0.460*** |
| **Chat Characteristics** | | | | |
| Total turns (log) | -0.023* | -0.021* | 0.012 | 0.023 |
| Total turns × MHS | -0.071*** | -0.071*** | -0.056*** | -0.056*** |
| Seeker's negative sentiment | 0.208*** | 0.208*** | 0.193*** | 0.193*** |
| Seeker's positive sentiment | -0.024** | -0.028** | -0.011 | -0.018 |
| **Provider Behaviors** | | | | |
| Provider gives information | 0.019* | | 0.017 | |
| Provider persuades w/ permission | 0.022** | | 0.033** | |
| Provider questions | 0.011 | | 0.005 | |
| Provider reflection | -0.022* | | -0.008 | |
| Provider affirmation | -0.007 | | 0.007 | |
| Provider seeks collaboration | 0.016* | | 0.001 | |

Table 3 shows that some characteristics of the support-providers were associated with statistically significant changes in seekers' mental health. Overall, we found that talking with older support-providers was associated with reductions in depression ($\beta$ = -.036, p<.001). Surprisingly, however, support-providers' experience on the site, including past chatting experience and having repeated interactions with the support-seeker during the observation period, was not associated with changes in either depression or anxiety (p>0.05). In fact, talking with support-providers who completed more training modules on the site was associated with increases in anxiety ($\beta$ = .037, p < .001). One possibility is that poorer support-providers are the ones who seek out more training, while another is that the training on the site is counterproductive. Given the correlational nature of the evidence, we cannot determine the causation behind this association.

We now turn to the associations of changes in mental health with support-seekers' characteristics. Support-seekers with a unit higher pre-assessment PHQ-2 score had higher post-assessment scores as well ($\beta$=0.446, $p$<0.001). Similarly, those with a unit higher pre-assessment GAD-2 anxiety score reported higher anxiety during post-assessment ($\beta$=0.46, $p$<0.001). The sentiment of support-seekers in their messages predicted changes in their mental health. In particular, support-seekers who expressed more negative sentiment in conversations during an observation period reported worse depression ($\beta$ =0.208, $p$<0.001) and anxiety ($\beta$ = 0.193, $p$<0.001) on their post-assessments. In addition, the associations with positive sentiment followed a similar but weaker pattern. Support-seekers who expressed more positive sentiment in conversations during an observation period reported slightly lower depression ($\beta$ = -0.024, $p$<0.01), but no change in their levels of anxiety ($\beta$ = -0.001, $p$>.05) on their post-assessments.

We now review the associations of changes in mental health with characteristics of the support conversations. Our analysis for RQ1 demonstrated reliable improvements in depression and anxiety associated with participating in online peer support conversations, with the improvements larger for those who initially had worse mental health. The RQ2 analysis shows analogous results among those who participated in the support-chats. We consider the length of the support chats as a "dosage" of a support conversation; people who participated in longer chats with more turns have slightly improved mental health for depression ($\beta$=-.023, $p$ <.05) but not anxiety ($\beta$=.012, $p$ >.05), with a larger effect size for those who initially showed worse symptoms. Having longer chats was more strongly associated with improvements in mental health for those who initially had worse depression ($\beta$=-.071, $p$ <.001) and for those who initially had worse anxiety ($\beta$=-.056, $p$ <.001). These interaction effects are illustrated in Figure 3.

In terms of the content of the support-conversations, our RQ2 results suggest that some MI and empathy behaviors predict improvements in mental health, but others are actually associated with declines in mental health. Of the seven MI behaviors, Reflections and Emphasizing Autonomy were significantly associated with improvements in depression symptoms. Specifically, when people received one standard deviation more Reflection and Emphasizing Autonomy during an observation period, their PH-2 depression scores improved by 0.022 (p<.05) and 0.02 (p<.05) units respectively. In contrast, when they received a standard deviation more Information, Persuasion with Permission, and Seeking Collaboration, their PH-2 depression outcomes got worse (p<0.05). However, we note that Seeking Collaboration and Giving Information both have only low to moderate accuracy in our MI classifiers (0.500 and 0.574, respectively), so both of these significant results may need to be interpreted with some caution. In terms of the components of empathy, only Emotional Reaction had a significant association with changes in mental health; Exploration and Interpretation did not.

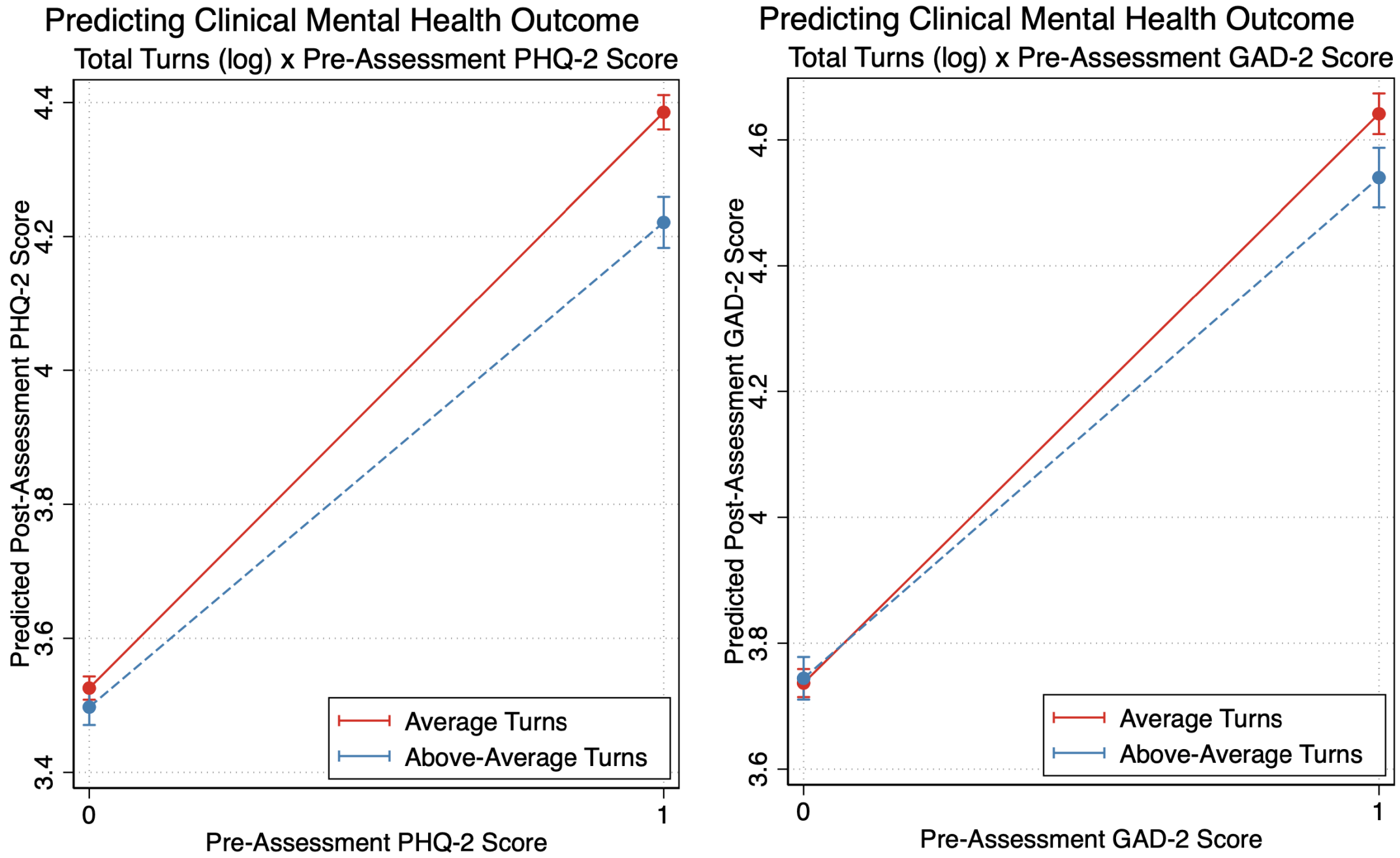

**Figure 3.** *Marginal plot that illustrates the interaction effect between total conversation turns and support-seekers' pre-assessment scores, comparing effects for those with average versus above-average initial symptom severity.*

Inconsistent with much research on the common factors that seem responsible for success in conventional psychotherapy (Norcross & Lambert, 2019), receiving Emotional Reactions during the observation period was associated with a small, but statistically reliable increase in depression (0.028 (p<0.001) points on the PHQ-2 score).

In terms of anxiety symptoms, none of the MITI codes was associated with statistically reliable improvements in anxiety. In contrast, like the results for depression, a standard deviation increase of Persuasion with Permission was associated with worsening of anxiety symptoms ($\beta=0.033$, $p<0.001$). Also like the depression analysis, support-seekers had worse anxiety when provided with more Emotional Reaction. One standard deviation increase in receiving Emotional Reaction during the observation period was associated with a small, but statistically reliable increase of 0.048 ($p<.001$) point increase in their GAD-2 score.

## 3.3 Takeaways

Our research strongly suggests that online peer support chats are beneficial for improving clinical depression and anxiety outcomes, albeit with a relatively small effect size. Additionally, we identified that there are some inconsistent results when it comes to techniques offline working online. Altogether, this provides us insight that online mental health communities are not only perceived to be useful (from prior literature), but they are also *effective* – even if only to a small degree.

Furthermore, our findings reiterate that the online mental health setting is also unique. Our study suggests that some things that work offline do work online, but not always. Rather than automatically translating mechanisms that work offline into the online care setting, HCI research requires a unique and careful understanding of the online setting specifically – studying online platforms, their users, and their needs carefully.

Our study generally contributes to the design of training for digital support platforms. In particular, relevant to this dissertation, we gained understanding that these communities have the potential to be effective for people's mental health (and thus may be important subjects to study and build upon). In our findings that the online setting is unique, we are also guided to then understand these communities' unique limitations.

# *Chapter 4*

# Discovering User Needs for Matching in Online Mental Health Communities

Clients of traditional therapy typically engage in an involved process for selecting a therapist given their unique preferences and individuals seek peer support from select individuals depending on the situation (Liddle, 1996; Manthei et al., 1982; Netzky, 1979; Procidano & Heller, 1983; Walen & Lachman, 2000). However, the selection process in OMHCs often relies on random or naive matching; users are helped on a first-come-first-serve basis or matched with one another based on simple criteria, such as solely the topic of discussion. For example, support forums such as Facebook's support groups are topic-based and rely on users to find these groups using search, trial-and-error, or word-of-mouth means. Other communities such as Wisdo and TherapyTribe also provide support groups based on topic. The most potential for more advanced matching methods exist on sites that rely on professional therapists online rather than free peer support, such as Betterhelp.com and Talkspace.com that offer intake forms for new users but do not provide details on how the intake form is actually used for matching. OMHCs that provide free and accessible peer support, such as 7cups.com, do not currently consider users' personal characteristics and preferences that have been shown to be vital to decision making in mental health resources (Coleman et al., 1995; Landes et al., 2013). As a result, algorithmic matching may have the potential to resolve these challenges by providing more optimal matches given users' characteristics, including for live chats, with minimal efforts on a user's end. However, it has not been explored in online mental health contexts.

We address this gap by studying how users of OMHCs currently feel about matching, and whether technologically-supported matching may improve community member experiences. As past work has demonstrated, technically sound systems may fail to resolve problems they were designed to tackle if they are inconsistent with — or even harm — important values and needs of the communities who use them (Lee & Baykal, 2017; Veale et al., 2018). Thus, it is crucial to first conceptualize and understand stakeholders' needs, values, and concerns to guide any algorithm design choices (Barrett et al., 2016; Matthews et al., 2014; Zhu et al., 2018). This may be particularly important for OMHCs due to the vulnerability of their members. As a result, we use a mixed-method approach to discover the experiences and values of users in OMHCs and, given this information, explore how we can improve the formation of positive online peer relationships through algorithmic matching. We use one of the world's largest online peer support platforms to conduct interviews with 12 users and analyze behavioral log data of nearly 200,000 chats among around 129,000 users. Our study found (1) current first-come-first-serve-based matching systems do not adequately support users' needs and capabilities; (2) users were overwhelmingly supportive of algorithmic matching for OMHCs, including being comfortable disclosing information to support algorithmic matching; (3) algorithmic matching may be especially valuable for vulnerable populations like LGBTQ+ and female support-seekers; (4) users suggested algorithmic matching systems be optional in order to alleviate potential concerns, such as connecting with the same group of people repeatedly. Our quantitative analysis also suggests that potentially striking improvement can be made through purposeful matching based on gender, age, and experience levels. This improvement could be especially salient for marginalized groups; for example, our interviews revealed that nonbinary support-seekers felt hesitancy towards speaking with male support-providers and our data analysis showed a remarkable chat rating improvement of 1.18 stars out of 5 when nonbinary support-seekers chatted with a nonbinary support-provider instead of a male support-provider.

## 4.1 Relevant Prior Work

Although past work has not examined the potential of algorithmic matching in OMHCs nor measured the effects of influential factors for matching in this context, some initial exploration has taken place on what users may find valuable in finding connections through online support systems. Andalibi and Flood studied a digital support platform that manually matched users together, and found that users felt compatible with those who had a shared identity and shared interests;

however, users felt some concerns with partners who had shared mental health issues in part due to potential comparisons to one another (Andalibi & Flood, 2021). People's unique goals and needs at the time of sharing health related information online may also affect who they choose to interact with (Newman et al., 2011; O'Leary et al., 2017). O'Leary et al.'s work touched on the value of some demographic factors in online peer support as well, such as their finding that women found groups consisting of all women yielded greater safety and comfort for sharing sensitive experiences (O'Leary et al., 2017).

There is also extensive research on matching people in traditional mental health services, consistently finding that people often have strong preferences for mental health providers who are similar to themselves although these similarities do not necessarily impact health outcomes. Social psychology research over decades has established the notion that people perceive those similar to themselves to be more trustworthy (Cabral & Smith, 2011; Newcomb, 1961; Simons et al., 1970) and likely to share their worldviews (Marks & Miller, 1987). Thus, research over many decades has explored how matching a client and therapists' race, language, gender, and other variables impact therapeutic outcomes (Felton, 1986; Jackson & Kirschner, 1973; Jones, 1978; Leong, 1986; McKinley, 1987). Although our paper addresses online peer support platforms, our work is grounded in the variables explored from past research in traditional therapy contexts.

Racial matching has been a main focus of past literature in matching for mental health. Coleman et al. conducted a meta-analysis of studies from 1970s to early 1990s, concluding that clients have strong preferences for choosing a therapist of the same race (Coleman et al., 1995); however, Maramba and Hall found that this preference did not yield significant effects on actual wellbeing outcome (Maramba & Hall, 2002). The effects of racial matching may differ between racial groups as well, most notably for African-American clients. Although studies do vary, some research has suggested that Black clients not only show significant preference for Black therapists but also may benefit from racial matching due to mistrust towards mental health services and its association with White or European American values (Maultsby, 1982). Flaskerud also found that Asian clients showed better therapeutic outcomes if their therapist shared other common languages with them (Flaskerud & Liu, 1990).

Gender has been presumed to be an important factor in choosing a therapist (Gornick, 1986). Felton et al. in the 1980s found that clients consistently prefer therapists of the same gender (Felton, 1986), and Jones et al. found that clients found female therapists more effective regardless of their own gender (Jones & Zoppel, 1982). A study measuring gender effects in treatment plans for adolescent substance abusers found that gender matching resulted in higher likelihood to complete treatment (Wintersteen et al., 2005). Wu and Windle found that specifically Asian male clients had higher retention with male therapists, but did not see the same effect among other racial groups or females (I. H. Wu & Windle, 1980). Flaskerud and Liu similarly found that the dropout rate significantly decreased with a male client and male therapist, but did not see this same gender effect with females (Flaskerud & Liu, 1991).

There has also been peripheral work in matching based on client needs. Yarborough et al. studied how to match mental health service type to clients' goals (Yarborough et al., 2016) while Vlahovic et al. found that matching users' expectations of emotional or informational support can have significant differences for their satisfaction (Vlahovic et al., 2014). Hartzler et al. studied what factors influence participants' choice on who to reach out to as a potential mentor in an online medical community, finding that sample posts on mentors' profile pages were the only significant predictor rather than their person-generated health data or demographic data (Hartzler et al., 2016).

Although algorithmic matching has not been introduced to OMHCs, it has been explored in other social matching contexts. Algorithmic matching has most notably been established successfully in The National Resident Matching Program for matching physicians to residency programs, as well as in the educational context such as matching students to public high schools in New York City and Boston (Abdulkadiroğlu, Pathak, & Roth, 2005; Abdulkadiroğlu, Pathak, Roth, et al., 2005; Roth & Peranson, 1999). Further research done in the education space has explored matching students together based on various factors to better academic outcomes. Campbell and Campbell studied how racially matching students to mentors led to better long-term performance like higher GPA, graduation rate, and rate of entering graduate education (Campbell & Campbell, 2007). Pairing students by gender, however, has had conflicting results; Fenwick and Neal finding that gender matching in group educational settings was predictive of outcomes while Stewart and D'Mello refuted that claim (Ding et al., 2011; Fenwick & Neal, 2001; Lockheed & Harris, 1984; Stewart & D'Mello, 2018). Researchers have also studied pairing students according to their performance levels (Sutherland & Topping, 1999). Day et al. found that pairing low performing trainees with high performing partners resulted in little benefit (Day et al., 2005), and similarly Fuchs et al. found that pairing people of similar performance ability together resulted in better quality work (Fuchs et al., 1998). Recent work by Robertson et al. studied the disconnect between algorithm designers' aims and the actual real-world impacts of matching algorithms for educational systems, stressing the importance of directly engaging with stakeholders to align values with real-world conditions (Robertson et al., 2021).

Matching based on people's personal characteristics has been explored in online dating and friendships as well. Hitsch et al. found that users often have strong preferences for dating people similar to themselves across numerous dimensions; most notably, racial matching is often highly influential despite often not being self-reported (Hitsch et al., 2010). Similar work has found that other dimensions of similarity, such as profile similarity and history of past relationships, lead to better recommendation success on online dating sites (Tu et al., 2014). Researchers have also explored platonic relationships, having developed algorithmic systems to optimize friend finding in online social networks based on matching user qualities like their lifestyles and personalities (Bian & Holtzman, 2011; Z. Wang et al., 2015).

Our research is at the intersection of this prior work. Although the effects of matching have been studied in traditional mental health resources, the potential for algorithmic matching in the online mental health context and its improvements on support-seeker experiences have not been studied. Thus, in this chapter we will address the key research question and its subparts:

**RQ: How can algorithmic matching form better online peer relationships in OMHCs?**

- ❖ **RQa:** *What are the challenges that users currently face in OMHCs?*
- ❖ **RQb:** *What are users' sentiments towards matching systems on OMHCs?*

❖ **RQc:** *What characteristics may be useful to optimize matching in OMHCs?*

## 4.2 Methods

We conducted our study through mixed-methods using qualitative and quantitative analyses. To understand users' challenges with current systems and how they perceive algorithmic matching, we conducted interviews followed by analysis of the behavioral log dataset to quantitatively evaluate the potential benefits of algorithmic matching and further explore the findings from our qualitative study.

### *Interviews*

To understand how users of OMHCs perceive algorithmic matching, we conducted semi-structured interviews with 12 users of the anonymized online platform. Participants were recruited by the research platform's staff, who emailed our interview invitation to interested users who were at least 18 years old and based in the U.S. Note that although interviewees were constrained to users in the U.S. due to our study's compensation restraints, our quantitative analysis includes all users regardless of country. All site privacy policies were obeyed during this process. Our participants are sampled across 4 dimensions: (1) account type, (2) gender, (3) age, (4) tenure on the site. As the research platform's user population skews female and over three-fourths of users are 25 years old or younger, our participants also reflect a higher proportion of females and young users. Details on the 12 interviewees are in Table 4,

**Table 4.** *Details for each of our interview participants regarding their account type, gender identity, age, and tenure.*

| | Account Type (support-) | Gender | Age | Tenure |
|---|---|---|---|---|
| P1 | -provider, -seeker | Male | 18 | <1 month |
| P2 | -provider, -seeker | Female | 19 | 4+ years |
| P3 | -provider | Male | 20 | 1-3 months |

| P4 | -provider | Female | 22 | 1-3 months |
|---|---|---|---|---|
| P5 | -provider, -seeker | Nonbinary | 23 | 1-2 years |
| P6 | -provider, -seeker | Nonbinary | 24 | 1-2 years |
| P7 | -provider, -seeker | Female | 25 | 1-3 months |
| P8 | -provider, -seeker | Female | 26 | 1-2 years |
| P9 | -provider | Female | 29 | 3-6 months |
| P10 | -provider, -seeker | Female | 36 | 1-3 months |
| P11 | -provider, -seeker | Female | 47 | 3-6 months |
| P12 | -seeker | Female | 53 | 4+ years |

Interviews were semi-structured, guided by a list of questions but allowed to deviate depending on topics participants may have introduced. All interviews took place remotely via Zoom or Google Hangouts in December 2020 and January 2021. Interviews lasted approximately 1 hour and participants were compensated with US$20. During the interview, participants were asked their challenges in using the current queue system (e.g., How do you normally connect with a support-seeker/support-provider and what are the challenges, if any, that you have experienced using those methods?). Participants then reflected on what qualities made certain conversations or chat partners a good match for them (e.g., What characteristics do you feel make certain conversations better or worse than others?). We also sought to understand participants' opinions on algorithmic matching, including their willingness to disclose personal information to OMHCs for matching purposes, likelihood to engage with an algorithmic matching system, and their needs for transparency in such a system. Lastly, we probed for our participants' reactions by showing them some preliminary quantitative analysis comparing how different pairings of support-seekers and support-providers perform. In some cases, this further elicited experiences from our participants that either countered or complemented our results.

All interviews were recorded and transcribed. Analysis was completed by a team of two academic researchers, who both identify as female and BIPOC. Based on Braun and Clarke's method of thematic analysis (Braun & Clarke, 2012), our analysis involved tagging topics in each transcript including both inductive codes and sensitizing codes from past research. Our iterative analytic cycle consisted of: (1) recording and transcribing interviews (2) coding transcripts (3) amalgamating codes (4) discussing codes as a team (5) highlighting themes (6) writing and revising memos. Our analysis continued until saturation, when new interviews gave no new insights. Our analysis ended with 225 codes, organized into 34 axial-codes, and summarized into 10 themes presented. We noted key quotes that illustrate our findings. A summary table of our themes and their axial-codes can be found in Appendix B.

### *Behavioral Logs*

For the study described in this chapter, we used a dataset from January 2020 to August 2020; note that this is a smaller dataset than we used in Chapter 3 (and we will use in Chapter 5). However, this dataset is still large-scale and includes roughly 200,000 chats, with around 93,000 support-seekers having given chat ratings.

Relevant to this chapter, the dataset includes user information such as all users' signup dates and birth years, as well as logs of users' actions such as every time a user started a new chat and blocked another user. The most vital part of the dataset for this study is the features used as proxies of chat outcome – chat ratings (see Chapter 3).

We utilize chat ratings to measure the outcome of a seeker-provider pairing. It is worth noting that there are three other possible metrics for measuring outcome: (1) "hearts" that support-seekers can send support-providers on specific messages in chat, (2) retention of a support-seeker, and (3) emotional wellness tests that the research platform sends periodically to support-seekers. The metrics of hearts, retention, and chat rating are not highly correlated with one another, indicating that they likely show different definitions of success in accordance with past literature (Cunha et al., 2019; T. Wang et al., 2023). Compared to chat rating, the number of hearts sent in a chat is a more ambiguous assessment of a support-seeker-support-provider pairing, considering it occurs at a message-level rather than chat session-level, can indicate simple acknowledgements rather than positive interactions, and is highly dependent on the number of messages in a chat (Shah et al., n.d.). As for retention, it is unfortunately an unclear metric in OMHC contexts as retention has been shown to be an indication of either failure or success of the community (Massimi et al., 2014) and poses issues of truncation (Breen & Fellow of the Royal Irish Academy Professor of Sociology Richard Breen, 1996). Lastly, the emotional wellness tests (PHQ-9 for depression and GAD-7 for anxiety (Kroenke Kurt & Spitzer Robert L, 2002; Williams, 2014)) are likely a useful indicator of long-term improvement and actual mental health outcome for support-seekers on the site. Unfortunately, these tests are scarcely filled out by users and evaluating change in well-being would require users to have filled out the assessments more than once; a total of 4.6k users filled out the PHQ-9 more than once and 1.3k filled it out more than twice over the 8-month period of our dataset, which is not a significant enough population of users to be adequate for our study's analysis.

### *Labeling missing demographic information*

For our behavioral log analysis, we focus on how both personal characteristics (gender, age) and experience (number of previous chats on the platform) impact support-seeker experiences, as these are characteristics we can identify through people's profile information and chats as opposed to other characteristics such as assessing support-providers' expertise in mental health or on specific mental health subjects.

Demographic information such as gender identity can be optionally added by users at a later time while using the site and thus is a sparsely populated field in our dataset. Because gender is a central element to past research done on matching in traditional mental health resources, we attempted to label users' gender identities using their chat data in order to evaluate gender effects for our study. Using 6 gender groups (female/male/nonbinary identities, transgender female, transgender male, transgender female/male unknown), we automatically labeled users' gender identities in the dataset if a user had stated their own gender identity in their chat messages. We utilized a variety of phrasing and gender terms to capture the variance in how people present their own gender. For example, a user who has stated their gender (e.g. "I'm", "I am") as a female-identifying term (e.g. "female", "woman", "girl") will be labeled as female-identifying. The full term list is included in Appendix C. This process can introduce conflicting labels, such as if users have said they are female and male at different times; in these cases, our labeling system takes a conservative approach by not labeling these users as any gender.

Our labeling process was highly accurate. Among all users, 34,423 users had both input their gender identity on their user profile and been gender labeled by our system. Only 0.8% of those users were given a mismatched label from our system. Additionally, we were able to significantly increase the number of users with a labeled gender identity. In our analysis dataset of 129,318 users, originally only 12,845 support-seekers (13.9%) and 160 support-providers (0.4%) had manually input their gender identity. Our method ended up increasing the number of users with a labeled gender identity to 17,316 support-seekers (18.7%) and 8,420 support-providers (23%).

Other demographic information like race is rarely given during chats thus, despite it being a central element to matching, we are unable to label users' race or ethnicity for analysis. Basic statistics of the dataset are in Table 5.

**Table 5.** *Descriptive statistics of dataset used in this Chapter 4 study.*

| | |
|---|---|
| # of chats with ratings | 195,832 |
| Average chat rating | 4.06/5.0 |
| # of users involved in ratings | 129,655 |
| # of distinct members in ratings | 92,909 |
| # of distinct listeners in ratings | 36,746 |
| % of users with gender labels | 20% |
| % of users with age | 100% |

## 4.3 Qualitative Results

We first provide an overview of our results from interviews.

Firstly, we found that users believed that the current matching mechanisms of the platform are insufficient and ineffective, which leads to both short-term and long-term negative impacts. In particular:

- ❖ Members already engage in a manual selection process in order to find relevant support-providers to meet their mental health needs; however, the manual selection process is both ineffective and inefficient.
- ❖ The lack of a matching system can lead to support-providers feeling incapable of adequately helping certain support-seekers.
- ❖ Ineffective matching can lead to fewer long-term relationships and reduced support-seeker commitment.

Secondly, participants shared their insights and opinions about algorithmic matching.

- ❖ Users were overwhelmingly supportive of algorithmic matching for OMHCs, including being comfortable disclosing information to support algorithmic matching.
- ❖ Users varied on the degree of transparency they would like in an algorithmic matching system.
- ❖ Users largely felt that algorithmic matching could be especially valuable for populations vulnerable to harassment, such as LGBTQ+ and female support-seekers.
- ❖ Members expressed that matching based on experience level would help alleviate frustrations arising from chatting with inexperienced support-providers.
- ❖ Users felt that the priority in matching on OMHCs should be to avoid the worst possible outcomes.
- ❖ Blocking inappropriate behavior in OMHCs was seen as necessary for the success of algorithmic matching.

Each of these findings is expanded on below.

### *Current challenges*

During our interviews, users generally expressed dissatisfaction with the current matching process. Many challenges that users mentioned regarding the process promote the need for algorithmic matching.

*Members already engage in a manual selection process in order to find relevant support-providers to meet their mental health needs; however, the manual selection process is both ineffective and inefficient.*

During our interviews, all participants regardless of their account type stated that support-seekers commonly come into the site with their own personal criteria for the type of support-provider they want to chat with. Most often, support-seekers reported seeking support-providers who had demographic qualities similar to themselves. This follows previous work showing that people often associate this similarity to shared worldview and greater feelings of comfort (Marks & Miller, 1987).

Because the queue is limited to just displaying support-seekers' username and topic tag, information can only be shared between the support-provider and support-seeker once they are already connected in the chat room. Although the platform's current policy states that users should not share personal information, all of our interviewees stated that they ask or have been asked general demographic information at the beginning of the chat in an effort by users to gauge their fit to one another. Participants reported that gender and age are commonly given information and can be barriers to whether a chat will continue.

> *"As a support-seeker, I like to talk to females more. I just feel more comfortable with females than I do with males...I do ask* [*the support-provider's gender*]." (*P*10)

> *"We're not supposed to normally reveal* [*our information*]*, but sometimes people feel more comfortable talking to a certain gender or age. It happens majority of the time,* [*members*] *will ask some personal information about you...I don't mind talking about personal experiences, letting them know I'm a female.*" (*P*7)

Although anonymity is important to creating safe spaces for vulnerability in OMHCs (Andalibi et al., 2016), participants expressed that knowing more about users during the matching process would help people better judge fit up front and gauge how to exchange support during their conversations. Additionally, the uncertainty of the queue disproportionately impacts support-seekers more due to the fact that the queue relies on support-providers choosing support-seekers to chat with and support-seekers have no control in the process.

> *"I feel like a lot of the times I queue up with someone, no idea anything about them...you kind of have to take cues rather than knowing directly the person's age or gender. That would probably be a little bit more helpful in selecting who to help, because maybe I'm not able to help females as much as males in certain situations, stuff like that."* (P3)

> *"As a support-seeker, it kind of feels awful with the current system, because you never know when you're going to get picked or how long exactly it's going to take for somebody to pick you up. Because you're being chosen at random, basically based on support-providers choosing you, you never ever know if you're going to get a good support-provider or not, you never know what kind of experience you're going to get."* (P6)

This system also results in many dropped conversations — if a support-provider ends up not matching the support-seeker's criteria, the support-seeker will leave the chat to go back onto the queue and wait for another support-provider. The trial-and-error process was described by our interview participants as tedious and time-wasting, and may cause lowered confidence in participating for both support-providers and support-seekers. As P3 states, reflecting on his experiences as a support-provider:

> *"When I message someone, they're like 'if you don't mind me asking, are you a boy or a girl?'. I'm like 'I'm a boy'. And then they say 'Ok, well I want to talk to a girl'...that's fine, but again that kind of wastes their time...it delays their time and my time and then sometimes when they do see I'm a male, they don't respond afterwards. I'm assuming they didn't want me then."*

P1 has had similar experiences as a male support-provider, mentioning, *"Usually when I tell them I'm male, they don't want to talk to me, they just disappear."*

Furthermore, ineffectiveness is further exacerbated if the topic tag is not utilized by a support-seeker. According to our interviews, support-seekers sometimes do not list a topic tag if they *"just want to have a chat with someone to get to know each other"* (P2) or their issue is *"not covered by the categories"* (P5). However, all of our participants who were support-providers stated they primarily choose support-seekers to chat with based on the topic tag listed, if available, and therefore actually avoid taking chats without a topic tag because *"you don't know what you're going to get"* (P6). P4 avoids tagless chats, reflecting, *"I'm definitely a little bit more wary. It's so much more comfortable, especially in an anonymous chatting room, to know what's going to happen."* P2 echoed a similar opinion, *"If I'm familiar with the topic, then I will click on that support-seeker. Sometimes when people don't have a topic...I don't feel like I'm capable."* Thus, support-seekers who do not choose a topic tag wait longer on the queue. Again, these asymmetries stem from only support-providers getting a choice on who to chat with, and may be alleviated by automated matching that gives equal power to both support-seekers and support-providers, as P12 mentions,

> *"I think really for [this platform], [algorithmic matching] is a good idea because there's a lot of support-providers that pick and choose, you've got this group of support-seekers that nobody will take their chats."*

Given the challenges that support-seekers shared during the interviews, it is clear that the community shows a core need for more purposeful matching. As we have seen, this need is so widespread that users actually violate community rules about sharing personal information to work around current limitations in order to create a type of selection process amongst themselves. Migrating the matching process to be algorithmic or automatic can help support the community's goal in more systematic and time-efficient ways.

### *The lack of a matching system can lead to support-providers feeling incapable of adequately helping certain support-seekers.*

Although interviewees stated that normally the support-seeker is the one to leave a chat from unmet criteria, support-providers also have their own criteria for what kinds of support-seekers they feel they can support. Given that support-providers have sole ability to start chats with support-seekers by selecting them off of the queue, most support-providers reported choosing a support-seeker to chat with primarily based on the support-provider's experience with the support-seeker's topic, while continuing a chat was dependent on demographic factors.Although some participants mentioned that demographic factors such as gender and sexuality were more important to support-seekers rather than support-providers — as P11 summarizes: *"For the support-provider, it doesn't matter much...everyone needs some help sometimes"* — the common consensus among interview participants was that factors such as the age difference between a support-provider and support-seeker was a vital factor to whether a support-provider was capable of proceeding with a chat. In particular, younger support-providers (P1, P2, P3, P4) reported that they found it difficult to help support-seekers who were significantly older due to either inexperience with the topics that older people struggle with or from natural generational differences.

> *"It's tough to help with someone who's much older...and especially the cultural barrier now with just colloquium and the way people engage with each other. I would understand why it would be better to have someone talk with you who is your age...I've come across marriage problems and managing their life with their spouse and their job. It's like, I'm still in college and I'm not married. That's already two things I don't have in common with this person. It ends up being difficult to even understand what they're going through because I don't have my own experience with it. "* (*P*3)

> *"They always ask me my age and sex to see if they are comfortable talking with me.* [*If not then*] *they will just find someone else that is close to their age range...there are some older people on there, and I'm only* 19. *If they want to talk about something more serious, I cannot talk about that with them. So I will just tell them to find someone else with the correct age group because I just don't feel I am capable."* (*P*2)

P4 reflects on her experiences observing conversations among older support-seekers and younger support-providers in forums and group chats:

> "[*Young support-providers*] *are all trying to figure out what life looks like. And we're hearing all these older people being like 'we still have problems'...and even we can't figure out what life is all about. It's harder for younger support-providers to listen to older support-seekers. I know that I don't have a* *weight of experience or wisdom that I think is going to be the most useful to an older support-seeker."*

However, when P4 starts a chat only to find out later that she is speaking with an older support-seeker, she finds herself in a difficult position of not wanting to abandon the support-seeker:

*"I don't have really any experience with* [*certain topics*]*...it doesn't put me in a position where I can be the most active support-provider...but I'm not backing out of it. They do say you can, you know, tell them to contact another support-provider, which is really nice. But I do feel a sense of personal responsibility that once you click on that person, you shouldn't just bail."*

On the contrary, older support-providers (P5, P6, P8, P9, P11) stated that they could adequately handle the issues of both younger and older support-seekers. One exception was P7 who expressed that she has faced challenges relating to support-seekers much younger than herself:

*"I would say that I can relate to both* [*older and younger support-seekers*]. *But with teenagers...it's a little bit harder. You have to remember how you were when you were in high school and how you felt because it feels like their problems may be a little bit smaller than other people's problems...you don't want to become negative towards them. You don't want to do that."*

*Ineffective matching can lead to fewer long-term relationships and reduced support-seeker commitment.*

We have discussed how the lack of purposeful matching leaves users unable to effectively find suitable chat partners. Importantly, it can also cause longer term consequences on the community as a whole by reducing users' belief of whether they can obtain adequate help.

As we have mentioned, retention to the community is not necessarily a clear indicator of success in the OMHC context. However, previous work has studied how specifically returning to the same therapist or health provider shows indication of satisfaction with help received, and that these long-term relationships can lead to more self-disclosure and better outcomes for clients (Hill et al., 1996; Wellman & University of Toronto. Centre for Urban and Community Studies, 2001). As a result, we asked interviewees whether they were able to form lasting relationships with other users, returning to the same support-provider or support-seeker to chat repeatedly. Overwhelmingly, interviewees stated that they did have up to a few people that they repeatedly chatted with, but these were rare occurrences. Interviewees spoke highly of these relationships, and expressed that they led to more periodic check-ins with one another either during times of need or just to casually chat.

> *"I have some long term support-seekers that I talk to. We talk about ongoing problems, not necessarily very deep issues. There's this one girl I talked to a lot and we just check in every week. I just kind of help her stay on track...I feel like that's kind of the best, like a light but serious sort of relationship."* (*P*8)

> *"I get people messaging me back after we've talked days before, and asking like 'do you mind chatting with me, I'm really struggling with this again'. I'll reach out to them sometimes too if I've had a long chat with them and if I feel like, they still weren't at a resolution or anything, just to check up on them."* (*P*9)

> *"There are some that go especially well...and if I see that they're on, I'll go back and message them and see how they're doing. It's awesome."* (*P*10)

Interviewees agreed that there were many benefits of repeated interactions with the same user(s), including greater willingness to open up or engage in self-disclosure:

> *"After I've gotten to know someone and had repeated conversations with them, then I would be more comfortable and more open with what I said. And I would say a lot more to someone I've been talking with for 8 or 9 months than someone where this is the first or second time we've chatted."* (*P*12)

However, these relationships were widely reported to be infrequent occurrences and difficult to find. Participants, at most, stated they had only a few of these longterm chat partners even for those who had been on the platform for years. Some have never returned to the same chat partner. P12's experience from over 6 years on the platform reflects this:

> *"There were a couple* [*long lasting relationships*]. *There was one that we talked for well over a year. I've been through, I don't know, probably* 50 *different* [*support-providers*]."

> *"I think part of it is, I'm older than a lot of the people on* [*the site*], *so I have a different perspective. So they get on, you know, and they want to talk about how*

> *their parents are so demanding. And I'm like, I'm that parent...most of the time, I think it'd be more helpful if it was someone my age...that's usually the way it goes, you know, if I get someone young, it flip flops and I almost become the support-provider"* (*P*12)

Ineffective matching through the queue may cause not only a lack of lasting relationships, but also long-term consequences on support-seeker commitment. P12 stated that she had stopped using the live support-provider system completely due to not receiving adequate support from repeatedly getting support-providers who were too young or irrelevant for her needs:

> *"I've gotten to that point over the years that I just gave up using support-providers. The one last time I did try to use a support-provider was on Election Day...this [support-provider] happened to be a pro Trump-er from Canada. You can imagine I did not cope well....they just lectured me on what my country was doing to the world."*

These effects of unmet needs on long term use of OMHCs is consistent with previous work studying why dropout occurs in therapy. Studies have found that people often stop utilizing mental health resources due to lack of improvement, perceiving that the resource could not provide any better help, and mismatching demographic qualities particularly for clients belonging to minority groups (Roos & Werbart, 2013).

### *Users' opinions about algorithmic matching*

Given our interviews, we summarize below how users felt regarding algorithmic matching as well as common themes that users expressed were important in such a system.

*Users were overwhelmingly supportive of algorithmic matching for OMHCs, including being comfortable disclosing information to support algorithmic matching.*

All but one interviewee were supportive of algorithmic matching and felt it would improve the limited process in OMHCs currently. Interviewees largely felt that algorithmic matching would most help the inefficiencies of finding helpful partnerships in OMHCs. For example, P1 stated: *"If the system knows their gender,* *age, location, it'd be a lot easier to match them with someone best suited towards them...rather than putting them with someone that they don't necessarily need.*" P3 expressed that purposeful matching is "*what's best for the support-seekers, just to decrease the variability and experience with other things.*"

P11, who was the sole interviewee doubtful about algorithmic matching, felt it may still suffer from similar challenges to the current system:

> "*I think it would have the same problems. It always depends on what people put in about themselves which can be misleading or random. The improvement should be the support-provider having higher transparency with the support-seeker.*"

However, it is worth noting that after further discussion, P11 echoed many mechanisms that would be in place with algorithmic matching:

> "*It would be helpful for there to just be filters. That would make the support-provider's job a lot easier. The support-provider getting more information from the* [*support-seeker*]. *One problem is that members look for specific support-providers — certain gender, age. It would be better to have that displayed so* [*support-seekers*] *and* [*support-providers*] *don't waste their time.*"

In addition to supporting algorithmic matching conceptually, all participants stated they would disclose any reasonable information needed for such a system such as demographics and mental health experiences. In fact, while all participants stated that they already disclose information to their chat partners in the current system, everyone was surprisingly equally or more comfortable disclosing information to the platform itself for algorithmic matching. No participant had any concerns, including data privacy, with giving this information to the platform.

*Users varied on the degree of transparency they would like in an algorithmic matching system.*

Although participants were very supportive of contributing information to be matched algorithmically, their opinions varied on whether a platform should display users' information and why they were matched with one another.

As we have discussed previously, users actually already engage in self-disclosure to varying degrees so that users may gauge who they are talking to. Only 2 participants expressed during the interviews that they currently feel hesitation towards sharing detailed information to their chat partner. Surprisingly, this caution did not result from privacy or safety concerns — instead, both users stated that they were hesitant to share too much information in case it changed their chat dynamic.

> *"Usually if I do get a question* [*about my age*]*, I will say 'I'm in my* 20*s'. But I don't say a specific age because I want to try to keep it as vague as possible...It's more for the sake of dynamics, I'm not too concerned about safety...if the member kind of creates their vision of me in their head for them to talk to I feel like that's more effective."* (*P*8)

> *"I don't disclose much...I'm not worried anybody's gonna come to my house, none of that information is there. I think people talk to you differently when they can picture more of who you are...I like being just kind of this formless entity that's allowing people to express their emotions."* (*P*4)

We then asked our participants whether they would want an algorithmic matching system to be transparent about their information, displaying why two people were matched in the chatroom and the information used during the matching process. Of our 12 interview participants, 7 people expressed that they would definitely want that information to be available, 3 people were not opposed but expressed they were not particularly interested in knowing why they were matched, and 2 people stated they would be supportive only if users could choose which information could be revealed. No participants were opposed entirely to the algorithmic matching system being transparent about its process and the users' information to their chat partner. P7 is one participant who was supportive of the system being fully transparent, but also mentioned a potential drawback of the chat dynamic being altered if her age is always revealed:

> *"Age can make it so that somebody treats you in a different way. So if there's someone who's* 50 *years old they might not feel comfortable talking to someone who's in their* 20*s because they don't feel like they can relate on the same level of life...I have a lot of things that I've gone through in my life that most people my age haven't gone through. So my life experience would be more of someone in their* 30*s or* 40*s...it's not really going to be an accurate picture just by age."*

Notably, both participants who would be conditionally supportive of showing information if they could choose the information shown were nonbinary. Both expressed that having choice over revealing the information individually was important for situations that they did not feel comfortable disclosing their gender identity.

> *"I identify as nonbinary and a lot of times, if I'm talking about my stress during a pandemic, for instance...my gender doesn't matter at all to the topic. The way I'm envisioning it, the site would be like 'Do you care about* [*seeing*] *the gender of your support-provider?'...I think* [*having choice*] *would be nice. I think a lot of* [*support-seeker*] *would probably find that to be acceptable as well. If it was just buttons of like, "Are you okay with us sharing this information? Yes or no?" I feel like a lot of* [*support-seeker*] *would think that's a good system."* (*P*5)

> *"I don't know that as a* [*support-seeker*] *if I would feel comfortable telling a support-provider* [*my gender identity*] *unless I know that their gender was also listed as being nonbinary...if I'm in a bad mental health state, I want to make sure the support-provider is going to be okay with that sort of thing...I would want to know* [*why we were matched*] *if it's information that I personally choose to reveal. I would be okay if somebody is shown my age if I can see theirs, I don't care about that."* (*P*6)

Given these findings, users do not seem to fully agree on the degree of transparency that an algorithmic matching system should have. Thus, these systems may need options for people including selection for which information they are comfortable being shown to their chat partner or used for matching purposes.

*Users largely felt that algorithmic matching could be especially valuable for populations who are vulnerable to harassment, such as LGBTQ+ and women support-seekers.*

Many users brought up that algorithmic matching could be particularly useful for populations that have been historically vulnerable to stigmatization that prevents them from seeking support (Administration et al., 2001; Augustaitis et al., 2021; Bockting et al., 2013; H. S. Kim et al., 2008). These populations, including female and gender-variant members, may especially benefit from gender matching in order to avoid harassment or abusive chats.

Participants widely reported that female support-seekers often prefer female support-providers due to greater comfort. P4, a female support-provider, commented, *"I usually get a little scared when somebody asks [if I'm a girl] but it's usually because the person on the other line is also female and has certain issues that she wants to talk to another female about"* while P9 mentioned that since she is a female support-provider, female support-seekers tend to be *"more open about whatever the issue might be."* Both of our male participants also mentioned that female support-seekers typically *"aren't looking for a male"* (P3) and often *"just want to talk to another female because they're more comfortable"* (P1).

Our participants reflected on the importance of matching for the LGBTQ+ communities, given that better help can be given if the support-provider has similar life experiences to the support-seeker.

> *"Openness to LGBTQ issues is often important. If I'm talking about my boyfriend, [support-providers] assume my gender is a girl...if someone is trying to talk about questioning their gender, I would think that most people like that would want someone who either has real life experience with that, or has family members or friends or people around them who have struggled with that, and who are open about it, and who were accepting. And I've known a couple people to be like, hey, I want someone who's in the LGBTQ community who is trans or nonbinary, that just makes them more comfortable"* (*P*5)

> *"Most of my friends and my partner as well are transgender. I'm on the cusp of gender topics. With situations like LGBTQ issues, it is very helpful to know if a person is in the LGBTQ community."* (*P*8)

P5 expressed that generational differences can sometimes contribute to negative experiences of the gender-variant community:

> *"There was a chat I had with someone who was in his 40s. I had mentioned struggling with my gender and wanting to be nonbinary...I could tell the reaction he had was because of his age, like he just grew up in a generation where nonbinary people weren't open and he wasn't educated. So his reaction to me being nonbinary was basically like, 'Oh, it's fake, you're just being stupid.' Basically he just told me that I'm a stupid kid, I don't know what I'm talking about...I know people who are nonbinary who are in their 40s and 50s, they're fine."*

P5, who also struggles with lifelong disability, additionally added that matching can be important for those seeking support on disabilities:

> *"I'll mention that I have a lifelong debilitating disability.* [*support-providers*] *say, 'Oh yeah, I wear glasses so I can understand.'...they're trying to understand from their perspective, instead of putting themselves in my shoes and having empathy...I have support-providers, though, who do struggle with similar things to me, and they tend to relate to me a lot better...I like support-providers who actually have some sort of real life experience with issues that are similar to mine"*

Interestingly, participants also mentioned that LGBTQ+ support-seekers may be wary of male support-providers in particular. P5 commented that they are "*definitely uncomfortable*" with male support-providers and "*a lot of the bad support-providers that I've had are cisgender males*". P8 expressed that her experiences with male support-providers have been positive, but understood the hesitance for LGBTQ+ people such as her partner and friends:

> *"From what I've heard, non binary-identifying people feel apprehensive about talking to cis males off the bat. There's often a barrier there to begin with."*

*Members expressed that matching based on experience level would help appease their frustrations arising from chatting with inexperienced support-providers.*
Our interviewees who were experienced support-seekers on the platform reported difficulty chatting with less experienced support-providers. In particular, most participants who had been on the platform as a support-seeker for more than a year (P5, P6, P8, P12) expressed that they had become increasingly dissatisfied with the current queue as they became more cognizant of what kind of support-providers they wanted. The more experience a support-seeker gained on the platform, the pickier they became about their support-provider matches.

> *"As a* [*support-seeker*]*, I have definitely gotten more critical. I know what talking to a good support-provider is like versus a bad support-provider."* (*P*8)

> *"My evaluation of chats has changed as time goes on. At the beginning, it was great to just have somebody to listen. And then I got a little more picky as time went on. I got to the point where I was comfortable saying, you know, this is not right for me. This is not the right support-provider."* (*P*12)

> "[*Over time*], *I'm very picky. Getting a new support-provider, I'll give them a chance. But it's kind of like, new support-providers have like one strike. Then I'm done with you*." (*P*5)

Several participants expressed that incorporating experience level into algorithmic matching would improve experiences significantly. P1 mentioned that "*pairing with a support-provider that has as much experience as* [*the support-seeker does*], *that would be better than just putting them with a support-provider that just started off*". P6 felt that shortcomings of the current queue largely were a result of the system not considering tenure time,

> "*That's basically what the problem is. The support-providers who tend to do the general queue requests tend to be newer support-providers and they don't have as much experience...If* [*members*] *are just getting a random support-provider every time they're going to have more negative experiences*."

P12, who has been a support-seeker for several years, suggested that experienced support-providers may be important to pair with support-seekers losing faith in the system similar to herself.

> "*Put* [*experienced support-providers*] *with someone like me, that had been around for a long time that you can look and see has, you know,* 30 *support-providers and none of them are repeats. Why is this* [*support-seeker*] *not coming back to these support-providers? Just to restore their faith in the site, even if it's just one or two chats, but to let them know that there are some decent support-providers out there*."

Similarly, P7 also expressed that factoring in experience levels alleviates a concern of hers regarding constantly matching with bad or new support-providers through algorithmic matching, stating "*The only thing is with auto-matching, it feels like a lot of the time you're always getting people who have just started. That's kind of discouraging because a lot of the time it's their first or second chat…*[*matching new support-providers with new support-seekers*] *would be one solution, I like that idea*."

*Users felt the priority in matching on OMHCs should be to avoid the worst possible outcomes.*

When asked about the primary objectives of algorithmic matching systems, the overwhelming majority of our participants felt that avoiding the worst outcomes was the most vital in OMHCs. Participants expressed that having extremely bad experiences in OMHCs may deter people from seeking help, cause even more distress for support-seekers, or change users' behavior on the site from then on.

> *"If you're gonna match people that have extreme negative experiences, then they both can walk away [from the platform]...if you can at least get just a normal conversation, it would be a lot better."* (*P*10)

> *"I think we should just try to avoid the worst situation...realistically, we cannot guarantee people's experience on the platform, but we can try to do the least harm for everyone."* (*P*2)

> *"I would say that it's definitely important to keep people away from the extreme negatives. Because I know, as a support-provider, if somebody is in distress and they really want to talk to someone...that could put them more in distress. So at least have a level where they wouldn't get more in distress."* (*P*7)

> *"It isn't that big of a deal if it isn't, like, the best support-provider out there."* (*P*1)

Avoiding harmful situations may be especially important in a support-seeker's early stages on OMHCs, especially for vulnerable populations. P6, a nonbinary user, discussed how their first chats on the platform were marked by harassment towards their gender identity, which ended their willingness to reveal or discuss their gender from then on:

> *"Avoiding the extreme negative would be better...I did experience some transphobic and abusive behavior from [support-seekers]. And that was very upsetting to me as a new support-provider...that's basically the main reason that I only take requests that are tagged with certain topics, because I don't trust the general queue as it is right now…[the transphobic experiences] as a new support-provider put me off basically entirely of talking about my gender identity at all on the website. I can't trust. The one time as a member I did try to talk about my identity, they tried to give me very, very bad and dangerous advice."* (*P*6)

*Blocking inappropriate behavior in OMHCs was seen as necessary for the success of algorithmic matching.*

As Saha et al. (Saha et al., 2020) describe, OMHCs are susceptible to antisocial behavior like trolling and harassment similarly to other online communities but moderating against this negative behavior can be complex due to the especially vulnerable population of OMHCs. Unfortunately, all of our participants reported experiencing trolling or abusive behavior while chatting, occurring more regularly for our female interviewees. According to participants, the lack of adequate prevention of these inappropriate behaviors is a key challenge to having productive conversations from the queue.

Our participants recall their negative experiences:

> "*It goes both ways with* [*support-providers*] *and* [*support-sekers*]*, there are some that sexualize the chats. I've experienced that quite a lot. They always ask your gender, then they will decide to continue the conversation or not, depending on gender. I told them that I'm not going to engage in those chats, if they don't listen to me I just report them.*" (*P*2)

> "*There were a lot of guys, they got support-provider accounts just to try to hook up with female* [*users*]*. They were very sexualized conversations.*" (*P*12)

Some users mentioned workarounds they do to avoid these behaviors from the queue, such as avoiding certain topic tags or judging by username.

> "*Certain categories are more reliable to have actual people that want to talk about problems. 'Depression', they really want to talk about whatever it is that's going on. But I've noticed that 'Relationship Stress', 'Loneliness', a lot of the time end up having people who want to have conversations that aren't appropriate. I've also noticed there are people that continuously use the general request queue. There's one* [*user*] *specifically that changes their username...the first part is always "noMales"...it was some guy and as soon as you say, this isn't appropriate, then he calls you all sorts of names. For the most part, I try to stick towards certain topics that are less likely to bring out something like that.*" (*P*9)

> "*There are still people who use* [*the site*] *as like, a sex chat line. You can usually tell from the username...I've never picked up a chat that I thought was going to be inappropriate, just by telling by the username. So definitely I see the majority of* [*inappropriate*] *chats I've seen are put out onto the general queue. They're kind of just sitting there waiting to be picked up.*" (*P*4)

As a result, a primary concern of users was how algorithmic matching could impact the prevalence of inappropriate behavior in OMHC. Participants largely felt that the current reporting system does not adequately deter bad behavior due to users "just logging out and making a new user and doing the exact same thing over again" (P9). As a result of the distrust in the current reporting system, support-providers felt worried that they could automatically match with the same abusive users repeatedly and not have sufficient means to avoid these chats.

> "*It's kind of concerning like auto-matching that is going to keep* [*inappropriate*] *people coming up in my inbox. If you're going to do auto-matching, there has to be some way to improve how people are able to just regenerate new accounts and not have to deal with the consequences from their behavior.*" (*P*9)\\

> "*I think* [*algorithmic*] *matching takes some free will for support-providers out of it...like what if I get matched to a person who I know is using it as a sex line? Do I have to pick it up? That'd probably be the biggest problem.*" (*P*4)

Thus, an improved system for preventing and blocking inappropriate behaviors may be necessary for migrating to algorithmic matching.

*Users preferred if algorithmic matching was an optional system.*

While we have discussed the concern of users about being automatically matched with users exhibiting inappropriate behavior, another potential issue is how connecting with the same group of people can potentially stifle opportunities to learn new skills. 4 of our participants brought up this concern, including P2 who expressed "*I would like to talk about other things as well besides just talking about the same topic all the time*" and P4 who shared,

> "*I think auto matching takes the capacity for growth for support-providers out of it...To its credit, you want to have people have the most meaningful conversations possible and you don't want anyone to walk away feeling like they haven't been listened to. I just think it can create a pool where you're never going to have different perspectives that could possibly help you.*"

However, all four users who expressed this sentiment felt that making algorithmic matching be optional in addition to the queue system alleviated this concern. P2, who previously felt concerned about talking about the same topics all the time, felt that the option to still pick off of the queue would help her vary her topic choices adequately, stating, "*Yes, it does* [*alleviate my concern*]. *I like having two options instead of one, if this was it I actually think I will use both* [*the queue and algorithmic* *matching*]". Similarly, P4 stated: "*I think* [*making auto-matching optional*] *helps that concern, that's a really great option*."

During our interviews, we also asked all participants how they felt about the current manual queue, a mandatory algorithmic matching feature, and a hybrid approach where users could choose between the two. All 12 interviewees preferred the hybrid approach where users have an option between algorithmic matching and the manual queue. Participants reported that algorithmic matching would be useful for getting good quality matches and seeking long-term partnerships while using the queue system could help give variety to their conversations or find a chat partner quickly. Most commonly, participants cited that giving options in general for users would be positive.

## 4.4 Quantitative Results

Throughout our interviews, it was clear that users perceived matching gender, age, and experience level as impactful for their chats. Using behavioral log data, we measured whether these qualities actually do significantly impact chat outcomes.

In this section, we utilized OLS regression with the dataset described in Chapter 3 to show how various factors directly impacted chat rating outcomes, and then presented these results to the research site's stakeholders given the ease of interpretability of OLS. We used the statistical software package Stata to run regressions on support-seeker and support-provider gender, age, and experience level individually as independent variables while utilizing chat rating as our dependent variable; all qualities were organized into categorical variables (i.e. gender groups, age groups, experience level groups).

**Table 6.** *Distribution of gender and age among all support-seekers and support-providers.*

| | Support-seekers N = 92,713 | Support-providers N = 36,605 |
|---|---|---|
| **Gender** | | |
| Female | 12.6% | 14.1% |
| Male | 5.0% | 7.9% |
| Nonbinary | 0.2% | 0.3% |
| Transgender (Unknown) | 0.5% | 0.4% |
| Transgender Female | 0.2% | 0.1% |
| Transgender Male | 0.3% | 0.1% |
| Unknown | 81.0% | 77.0% |
| **Age (in Years)** | | |
| 15–18 | 42.6% | 25.2% |
| 19–25 | 34.8% | 48.1% |
| 26–30 | 11.0% | 14.2% |
| 31–35 | 5.1% | 5.6% |
| 36–40 | 2.6% | 3.0% |
| 41+ | 4.0% | 4.0% |
| Unknown | 0.0% | 0.0% |

For reference, Table 6 shows the general distribution of gender and age groups among support-seekers and support-providers and Table 7 shows the distribution of experience level for support-seekers and support-providers among chats. We provide descriptive statistics of the average chat rating and total number of chats by each of these groups in Figure 4. We then present our regression results, in which all tables have labels for the control group (baseline) and statistically significant ($p<0.05$) values. Descriptive statistics of the counts for each support-seeker-support-provider group pairing is presented in Appendix D as well.

**Table 7.** *Distribution of experience level among all 195k chats. Experience level is measured by a user's number of previous chats. The 0.2% of members with unknown experience level is due to missing log data and has a negligible effect on our analysis.*

| Experience Level | Support-seekers | Support-providers |
|---|---|---|
| 0–1 chats | 33.4% | 6.5% |
| 2–10 chats | 26.1% | 14.4% |

| 11–100 chats | 23.7% | 36.0% |
|---|---|---|
| 101–1,000 chats | 14.0% | 32.8% |
| 1,000+ chats | 2.6% | 10.3% |
| Unknown | 0.2% | 0.0% |

| Gender | Avg. chat rating | # of chats |
|---|---|---|
| Female | | |
| Member | 4.16 | 39100 |
| Listener | 4.20 | 46832 |
| Male | | |
| Member | 4.20 | 18658 |
| Listener | 3.89 | 34859 |
| Nonbinary | | |
| Member | 4.14 | 734 |
| Listener | 4.46 | 1002 |
| Trans (F/M Unkn.) | | |
| Member | 4.20 | 1172 |
| Listener | 4.24 | 1644 |
| Trans Female | | |
| Member | 4.28 | 1119 |
| Listener | 4.02 | 1485 |
| Trans Male | | |
| Member | 3.06 | 1479 |
| Listener | 4.30 | 1353 |

| Age | Avg. chat rating | # of chats |
|---|---|---|
| ≤ 18 yrs. | | |
| Member | 4.20 | 73936 |
| Listener | 4.19 | 52917 |
| 19-25 yrs. | | |
| Member | 4.02 | 66948 |
| Listener | 4.04 | 81015 |
| 26-30 yrs. | | |
| Member | 3.94 | 24896 |
| Listener | 3.93 | 31687 |
| 31-35 yrs. | | |
| Member | 4.02 | 12252 |
| Listener | 4.0 | 12437 |
| 36-40 yrs. | | |
| Member | 3.87 | 6608 |
| Listener | 4.0 | 6519 |
| 41+ yrs. | | |
| Member | 3.83 | 11165 |
| Listener | 4.07 | 11230 |

| Experience | Avg. chat rating | # of chats |
|---|---|---|
| 0-1 Chats | | |
| Member | 3.99 | 65243 |
| Listener | 3.97 | 12601 |
| 2-10 Chats | | |
| Member | 4.19 | 50997 |
| Listener | 4.08 | 28135 |
| 11-100 Chats | | |
| Member | 4.1 | 46298 |
| Listener | 4.04 | 70285 |
| 101-1000 Chats | | |
| Member | 4.05 | 27394 |
| Listener | 4.07 | 63910 |
| 1001+ Chats | | |
| Member | 3.43 | 5129 |
| Listener | 4.12 | 20130 |

**Figure 4.** *Descriptive statistics broken down by group for each of our analysis features. Note that for gender, the chat rating and number of chats includes those with users of unknown gender*

## *Matching based on gender*

**Table 8.** *Results of OLS regression measuring the effects of gender on chat rating. Each model in the table shows the regression analysis for the given group. The coefficients reported in the table show the difference compared to the baseline, which is with female support-providers (the highest population of listeners on 7 Cups). Note that this table only shows chats where we know both seeker and providers' genders. All gender groups are mutually exclusive and 20% of all users have a labeled gender.*

| | Member Group<br>*Coeff. (Std. Error)* | | | | | |
|---|---|---|---|---|---|---|
| | **Model 1 Women**<br>*N=17,719* | **Model 2 Man**<br>*N=8,391* | **Model 3 Non Binary**<br>*N=320* | **Model 4 Trans (W/M Unknown)**<br>*N=489* | **Model 5 Trans Woman**<br>*N=473* | **Model 6 Trans Man**<br>*N=546* |
| *Baseline (Woman Listener)* | Coeff = 4.33 | Coeff = 4.35 | Coeff = 4.21 | Coeff = 4.29 | Coeff. = 4.39 | Coeff. = 3.13 |
| *Man Listener vs. Baseline* | **-0.357*** (0.022) | **-0.395*** (0.034) | **-0.515*** (0.178) | -0.673 (0.144) | **-0.366*** (0.137) | 0.105 (0.174) |
| *Nonbinary Listener vs. Baseline* | 0.142 (0.098) | 0.122 (0.159) | **0.673*** (0.241) | **0.658*** (0.312) | 0.393 (0.374) | **1.874*** (0.852) |
| *Trans (W/M Unkn.) vs. Baseline* | 0.031 (0.076) | 0.051 (0.110) | 0.253 (0.398) | -0.031 (0.293) | -0.464 (0.374) | 0.096 (0.639) |
| *Trans Woman vs. Baseline* | -0.081 (0.086) | **-0.253*** (0.118) | 0.192 (0.449) | -0.019 (0.415) | -0.017 (0.489) | 1.208 (0.639) |
| *Trans Man vs. Baseline* | -0.137 (0.082) | 0.036 (0.134) | 0.692 (0.449) | **0.708*** (0.328) | 0.608 (0.969) | 0.446 (0.722) |

*$p < 0.05$

As mentioned, users felt algorithmic matching could be especially important for vulnerable populations. Our results in Table 8, which show the effect of different genders of support-providers with each support-seeker group, support this notion. Some groups of support-providers, such as nonbinary support-providers, are overall better than others, such as male support-providers. However, we see that even good support-providers who receive high ratings from most or all gender groups do perform particularly well given certain matches, such as nonbinary support-providers with other nonbinary and transgender support-seekers.

Compared to the control group of female support-seekers with female support-providers, female support-seekers on average rate male support-providers a moderate 0.36 stars out of 5 lower. This may reflect our female interviewees' experiences of feeling more safe and comfortable speaking with a female support-provider. Additionally, we see evidence of the reflections of our participants that support-seekers of the gender-variant community may be particularly cautious when speaking with male support-providers, as well as preferring support-providers who are also gender-variant. Nonbinary support-seekers rate nonbinary support-providers noticeably well, 0.673 stars higher than the baseline group (female support-providers). This is a stark contrast compared to when nonbinary support-seekers chat with male support-providers, which results in ratings of 0.515 stars lower than the baseline. Thus, these results show a potentially striking improvement that can be made — on average, nonbinary support-seekers rate chats a remarkable 1.19 stars out of 5 higher if chatting with a nonbinary support-provider instead of a male support-provider. Transgender support-seekers had an even more extreme effect — compared to speaking with male support-providers, on average transgender (female/male unknown) support-seekers rated chats 1.33 stars higher with a nonbinary support-provider and 1.38 stars higher when matched with a transgender male support-provider. Based on these gender effects, there is evidence that purposeful matching can create especially positive change for LGBTQ+ populations and can take into account the preference of support-seekers belonging to vulnerable populations to have support-providers similar to themselves.

We also found through our interview study in Section 4.1.1 that users will disclose their gender in chat in order to manually assess whether the match fits their needs. As an additional analysis, we fit an OLS to identify the effects on chat rating when users' genders are known or unknown in our dataset (i.e. has gender labeled through the process described in Section 3.2.3). The results of the OLS predicting chat rating given the two binary independent variables of (1) whether the support-seeker's gender is known and (2) whether the support-provider's gender is known follow our findings from the interviews. Chat rating is significantly affected when users' genders are labeled, showing a coefficient of 0.129 ($p < 0.001$) when a support-seeker's gender is known and a coefficient of 0.028 (p < 0.001) when a support-provider's gender is known. Given that our gender labeling process relies on users having revealed their gender in their chat messages and users reported disclosing their gender in chats in order to find better matches on the platform, this result follows our expectations as well as our interview study's findings.

### *Matching based on age*

**Table 9.** *Results of OLS regression measuring the effects of age on chat rating. Each model in the table shows the regression analysis for the given group. The coefficients reported in the table show the difference compared to the baseline, which is with the youngest listener group 18 years or younger. For example, 41+ support-seekers with 41+ support-providers performed 0.216 points better (on the 5-point rating scale) compared to the baseline of 41+ support-seekers with the youngest support-providers (* $p < 0.05$).*

| | **Member Group** *Coeff. (Std. Error)* | | | | | |
|---|---|---|---|---|---|---|
| | **Model 1** **≤ 18 yrs.** *N=73,936* | **Model 2** **19-25 yrs.** *N=66,948* | **Model 3** **26-30 yrs.** *N=24,896* | **Model 4** **31-35 yrs.** *N=12,252* | **Model 5** **36-40 yrs.** *N=6,608* | **Model 6** **41+ yrs.** *N=11,165* |
| *Baseline (≤ 18 yrs.)* | Coeff = 4.20 | Coeff = 4.08 | Coeff = 4.11 | Coeff = 4.22 | Coeff. = 3.88 | Coeff. = 3.89 |
| *19-25 yrs. vs. Baseline* | 0.004 (0.014) | -0.062 (0.040) | **-0.163*** (0.061) | **-0.235*** (0.087) | -0.066 (0.121) | -0.138 (0.096) |
| *26-30 yrs. vs. Baseline* | **-0.086*** (0.019) | **-0.199*** (0.042) | **-0.216*** (0.063) | **-0.276*** (0.090) | -0.013 (0.124) | -0.091 (0.099) |
| *31-35 yrs. vs. Baseline* | -0.035 (0.031) | **-0.130*** (0.046) | **-0.214*** (0.069) | -0.093 (0.097) | 0.007 (0.135) | 0.005 (0.106) |
| *36-40 yrs. vs. Baseline* | 0.002 (0.043) | **-0.123*** (0.051) | **-0.283*** (0.076) | -0.164 (0.106) | 0.120 (0.143) | 0.111 (0.116) |
| *41+ yrs. vs. Baseline* | -0.009 (0.032) | -0.084 (0.047) | -0.080 (0.069) | -0.071 (0.098) | 0.183 (0.133) | **0.216*** (0.105) |

*$p < 0.05$

**Figure 5**. *Comparison of average chat rating given the age difference between member and listener*

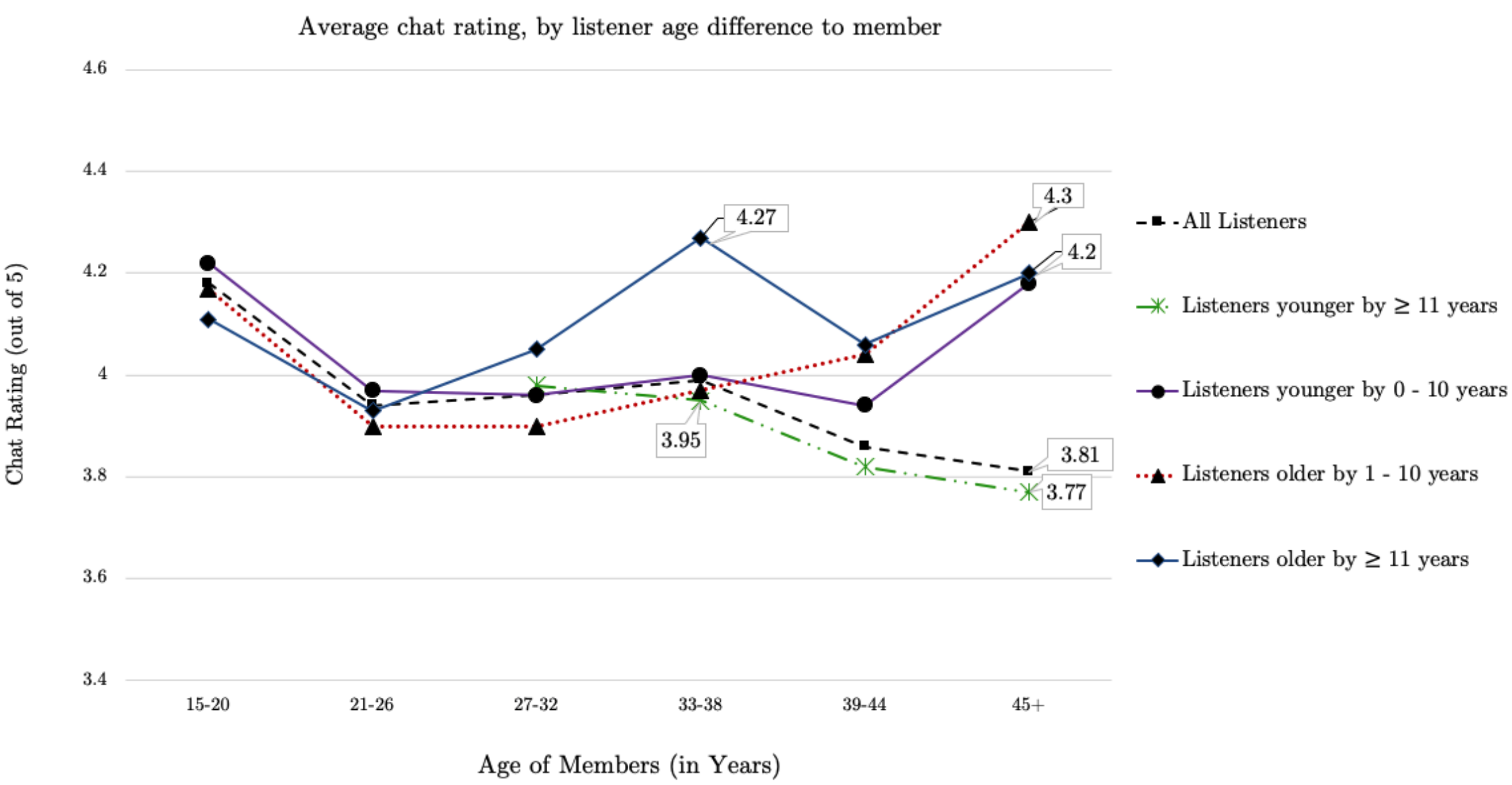

We conducted a similar analysis on age shown in Table 9. We grouped users into age groups following the World Health Organization's population standardization, although adjusted slightly so the youngest age group consists of youth users (18 years or younger); on the platform, youth users are restricted to only speak with other youth or with "verified" adults who have completed extra training modules, which also contributes to a noticeable drop in the number of chats between youth support-providers and adult support-providers.

Our findings reflect the reported struggles of young support-providers and older support-seekers. In particular, we see evidence of support-seekers in their mid-30s and older showing more dissatisfaction with younger support-providers; for example, the average ratings from support-seekers who are mid-30s and older to support-providers who are 18 years or younger is around 3.88 stars out of 5. By contrast, these ratings increase when chatting with support-providers who are of similar age.

We also show in Figure 5 how the average chat rating changes with the support-provider’s age difference compared to the support-seeker. In general, support-seekers in the youngest age groups rate support-providers similarly regardless of the age difference. However, chat ratings by older support-seekers such as those in the 33-38 and 45+ age groups show significantly more sensitivity depending on support-provider age. For example, the average chat rating by support-seekers who are 33-38 years old increases by a drastic 0.32 when chatting with a support-provider who is older by 11 years or more as opposed to a support-provider who is younger by the same amount. Similarly for support-seekers who are 45+, there is a detrimental impact when chatting with a support-provider who is significantly younger; paired with a support-provider who is older by either 1-10 years or 11 years or more results in high chat ratings of 4.3 and 4.2, respectively, while having a support-provider who is instead younger by 11 years or more results in a low average rating of only 3.77. We also check on the main effect of support-provider age on chat rating to analyze whether older support-providers are just better overall. The results shown in Table 9 show that the average ratings for older support-providers are around average or below average, making us further confident that the age difference between support-seeker and support-provider is indeed the driving force behind these disparities in chat ratings.

Given these results and our interviews, we conclude that age difference is particularly important when support-seekers are in their 30s or older. Generally, younger support-seekers have a similar experience with support-providers of similar age or older, while older support-seekers have a significantly better experience when matched with a support-provider who is older as opposed to a support-provider who is much younger than themselves.

### *Matching based on experience level*

We labeled users' experience levels to investigate the reported challenges of experienced support-seekers chatting with inexperienced support-providers. Since many users have had an account for a long time but have not used it often, we defined experience level by how many chats a user had at the time of a chat rating rather than their sign-up date for this analysis. Given that the skill and experience gained for a support-provider is likely most salient near the start of their tenure (i.e. the difference between support-providers with 0 versus 100 chats is larger than the difference between a support-provider with 500 versus 600 chats), we constructed the experience level buckets following the power law distribution often seen in online community behavior (Johnson et al., 2015). Our results are shown in Table 10.

**Table 10.** *Results of OLS regression measuring the effects of experience level (number of previous chats on 7Cups) on chat rating. Each model in the table shows the regression analysis for the given member group. The coefficients reported in the table show the difference compared to the baseline, which is with the least experienced listeners who have only 0 or 1 chats on 7 Cups (* $p < 0.05$).*

| | **Member Group** *Coeff. (Std. Error)* | | | | |
|---|---|---|---|---|---|
| | **Model 1 0-1 chats** *N=65,243* | **Model 2 2-10 chats** *N=50,997* | **Model 3 11-100 chats** *N=46,298* | **Model 4 101-1000 chats** *N=27,394* | **Model 5 1001+ chats** *N=5,129* |
| *Baseline (0-1 chats listener)* | Coeff = 4.07 | Coeff = 4.0 | Coeff = 3.87 | Coeff = 3.77 | Coeff. = 2.57 |
| *2-10 chats listener vs. Baseline* | **-0.051*** (0.024) | **0.242*** (0.032) | **0.238*** (0.037) | **0.284*** (0.052) | **0.699*** (0.144) |
| *11-100 chats listener vs. Baseline* | **-0.110*** (0.021) | **0.176*** (0.029) | **0.196*** (0.034) | **0.261*** (0.048) | **0.774*** (0.133) |
| *101-1000 chats listener vs. Baseline* | **-0.111*** (0.021) | **0.192*** (0.029) | **0.263*** (0.034) | **0.329*** (0.048) | **0.972*** (0.133) |
| *1001+ chats listener vs. Baseline* | **-0.075*** (0.144) | **0.248*** (0.033) | **0.288*** (0.038) | **0.347*** (0.053) | **1.289*** (0.144) |

*$p < 0.05$

We found that the largest effects on chat rating occur when very experienced support-seekers chat with very experienced support-providers versus very inexperienced support-providers. Experienced support-seekers have extremely negative experiences with inexperienced support-providers, as seen in the baseline rating of 2.57 stars out of 5 between support-seekers with 1001+ chats and new support-providers with 0 or 1 chats. These experienced support-seekers give a significantly higher rating with support-providers who have more experience chatting, such as 0.774 points higher with support-providers between 11 and 100 chats on the platform and a striking 1.289 points higher with support-providers having 1001+ chats. Furthermore, support-seekers who were the most experienced in general rated chats poorly compared to newer support-seekers. These results echo our participants' reflections in the interviews that they became generally more picky over time. On the other hand, less experienced support-seekers (2-10 and 11-100 chats) are overall consistent with support-providers of varied experience level, although with generally more positive effects with more experienced support-providers; this may be a natural result as experienced support-providers may be better at chatting in general. Interestingly, this effect did not appear for the least experienced support-seekers (0-1 chats). However, our participants provided us insight that this is likely due to blocked or reported users who often create new accounts constantly to rejoin the platform. These users may rate all chats negatively to troll or may rate a support-provider badly if they do not engage in the support-seeker's harassing behavior.

## 4.5 Design Implications

Matching to produce better support-seeker experiences in OMHCs has not been explored previously, despite the rising popularity of online peer support. OMHCs are particularly important for helping vulnerable or stigmatized populations, which are also populations that may most require purposeful matching to avoid unsafe, uncomfortable, or abusive environments online. Our study is holistic, deriving its findings from a combination of user interviews and behavioral logs. This work provides grounding for the future of algorithmic matching in OMHCs, including investigation into the unmet needs of users, core values of community members that are vital to the deployment of automated systems, and evaluation of how incorporating user characteristics can benefit support-seeker experiences.

Our work has demonstrated that users not only have strong preferences for those that provide them aid through OMHCs, but also the quantitatively measurable benefits on support-seeker experiences through matching users on these dimensions. Our study also explored users' possibilities of implementing algorithmic matching, concluding that users are overwhelmingly supportive and willing to engage with an algorithmic approach to matching. As a result, we urge for better matching mechanisms in OMHCs, which are quickly growing in size but have currently limited means to optimize support. We detail below various suggestions and design implications of our work for the creators and designers of OMHCs to improve their communities' matching systems. Building matching systems for OMHCs involves many complex considerations and trade-offs. As a result, we organize our recommendations for algorithmic matching into three levels: algorithm-level, interface/process-level, and deployment-level.

### *Algorithm-level Considerations*

We begin by discussing considerations for designers of algorithmic matching in the OMHC context.

#### *Features*

Firstly, we suggest that algorithmic matching for OMHCs consider all three elements of our quantitative study — gender, age, experience level — as possible elements for matching users. We also believe there is valuable future work in modeling the effects of these elements jointly, in addition to our exploratory work on their individual effects.

In addition to all three of these features showing significant effects on chat outcome in our log analysis, all three features emerged as important to users in our interview study to some degree. For example, gender was particularly important to vulnerable populations, including the LGBTQ+ community, who must navigate unique challenges of stigmatization and harassment in these online communities. We have also seen that support-seekers generally have a better experience on the platform with support-providers of similar age, and that older support-seekers have more negative experiences with significantly younger support-providers. As a result, algorithmic matching may often match people who are similar to one another together. In the case of experience level, we note that although the most experienced support-seekers felt growing frustration over time with chatting with inexperienced support-providers, matching inexperienced support-seekers with inexperienced support-providers could potentially create negative experiences for those just entering the community. As a result, accompanying training modules, chatbot interactions, or greater requirements for live chatting with users are likely good features to add to live OMHCs such as our research site's platform. Requirements such as a set number of forum posts or training modules where beginner support-providers chat with more advanced support-providers for practice would be good steps in this direction. However, it is vital that training modules are not so extensive such that they deter support-providers from engaging with the community at all and quitting the platform. Additionally, although our analysis did not assess the impact of topic on matching, we recommend that topic continues to be a central feature in OMHCs for identifying good chat partners due to our interviewees using this feature often to find suitable matches in the queue and support-provider viewpoints that support-seekers who do not have a topic tag can be difficult to assess fit for.

*Avoiding Worst Outcomes*

It is worth noting that, despite many factors being important to matching, platforms are still largely dependent on the available population of users providing aid; there will not always be ideal matches for the support-seekers seeking help. As a result, we recommend that algorithm designers prioritize avoiding the most negative pairings of users. Nearly all interviewees in our study stated that mediocre conversations are not problematic nor is always having the best experiences necessary; instead, it is most important to avoid situations that could cause extremely negative consequences on support-seekers to avoid worsening their emotional states, deterring them from using the platform again, or impacting their feelings of comfort using the site from then on. As a result, actions such as blocking and reporting on OMHCs should be utilized as indicators of extremely negative experiences for users, and can be potential outcome metrics for assessing algorithmic matching protocols.

*Balancing Waiting Times and Accurate Matching*

A potential tension with implementing algorithmic matching is the balance between waiting times for support-seekers seeking help and finding an optimized match for these support-seekers. Several participants expressed that the queue normally moves quickly and one participant mentioned that algorithmic matching could result in longer wait times while the system waits for a relevant support-provider to come online. As a result, an algorithmic matching system should display the expected waiting time for support-seekers (as the platform system does currently).

During our study's interviews, support-providers mentioned their appreciation for how the current system gives support-providers autonomy in how many support-seekers to chat with and who to chat with. We suggest that OMHC systems continue to allow users, whether seeking or providing help, to chat with multiple people at once. Users providing help should be able to set their own capacity for handling multiple chats, which would allow the platform to efficiently intake users seeking help.

*Allowing for personalization*

Past research has found that users greatly value being able to customize online community tools for their own needs and desires, and personalization of these tools help give users a sense of control as well as showcase an individual's sense of identity (Marathe & Sundar, 2011). Although our findings indicate that algorithmically matching users has the potential to increase efficiency and support-seeker experiences, it is vital that online platforms remain conscious of users' unique needs regardless of their demographic factors or background. Moreover, users' needs may change day-to-day or even chat-to-chat. Thus, any algorithmic matching system should rely first and foremost on users' preferences that they may input.

For example, one of our oldest interview participants who is a support-seeker mentioned that an algorithm would likely match her with someone of her same age, but at times she may want a younger perspective regarding issues with her son. As a result, we urge designers of OMHC matching systems to ensure that users can input their own preferences for the support-provider (for example, support-seekers specifying the support-provider's gender and age) as well as be able to change these preferences easily. However, it is likely that many users will not input all preference options (i.e. a user may prefer a certain age of support-provider but no preference for gender) or may opt for the system to fully algorithmically match them — in this case, we believe algorithmic matching that is built on historical data analysis is useful for forming a suitable matching to supplement any user input preferences.

### *Interface/Process-level Considerations*

#### *Making Algorithmic Matching Optional*

A sizable portion of support-providers stated that they would also be worried about lacking variety in chat partners or chat topics if using an algorithmic system. As a result, we suggest making algorithmic matching optional in addition to the manual current system. All our participants who feared a lack of variety stated that having algorithmic matching be optional would alleviate this concern. Since we envision the system to allow support-providers to input which topics they are willing to be matched on, it may be important to allow these settings to be easily reconfigurable such that support-providers could choose to talk about different topics at any given time.

#### *Transparency of Algorithmic Matching Systems*

At the forefront of building algorithmic matching for OMHCs should be considerations for transparency of users' information. Past work has found that users' participation in online communities is heavily influenced by their trust in the community; platforms with greater transparency over how they are using users' information may benefit from mitigating users' hesitations in engaging with new tools and systems of the platform (Benlian & Hess, 2011).

When asked about whether the platform should display users' information to one another once they start a chat through an algorithmic matching system, most participants in our study were supportive. This transparency was important to users for gauging who they were speaking to off the bat and avoiding assumptions about their chat partner, as well as simply fulfilling curiosity.

However, some concerns were brought up that we feel should be incorporated into a matching system. A couple of participants had concerns regarding revealing their demographic information, either due to the potential of changing the chat dynamic or opening up opportunities for harassment, but stated that they would be willing to reveal this information if they felt their chat partner would be accepting; for example, our nonbinary participants stated they would want their information to be shown if they knew they were speaking with someone also in the LGBTQ+ community. Our previous suggestion for a system that considers user preferences as a priority for their match may be able to provide this comfort, as users would have the ability to essentially restrict their matches to certain communities for increased feelings of safety and comfort. Systems should also allow for users to engage with a manual selection process, such as the site's current queue or a profile search system, so users have the option to not disclose any information about themselves to the platform or other users.

We note that our interviewees were open to disclosing their demographic information to the platform but varied more on whether to reveal this information to their chat partner. Thus, we recommend that matching systems also provide a granular level of transparency such that users are able to decide whether to show their information to their chat partner at all as well as which pieces of information is acceptable to show. It is essential that OMHC platforms deploying matching systems also ensure that users are properly informed on safe boundaries of information disclosure so that their privacy is not at risk of violation, given that users regularly disclose personal information in chat and none of our interviewees expressed any concerns with disclosing demographic information to the platform. There exists some conflict between users who want to remain anonymous and those who would provide their personal information conditional on their chat partner providing their information as well, and a key benefit of algorithmic matching is its ability to utilize users' information for the purposes of finding a good match without necessarily showing this information to their partner.

Interface-level changes may also incorporate further transparency but in less expensive ways than algorithmic and system-level changes. For example, changes to the interface like displaying users' demographic information at the time of users picking their chat partners may alleviate many pain points brought up in our study. However, we note that interface-level changes may exacerbate challenges around unwanted behaviors by requiring user information to be openly disclosed to the wider population, which algorithmic matching may instead alleviate. Our interview participants also indicated that search methods to find user profiles are inefficient despite greater autonomy over who to chat with, and users need to scour huge numbers of profiles that often lack important information like gender, age, race, or sexuality needed for an effective search (Felton, 1986; Landes et al., 2013; Liddle, 1996; Prescott et al., 2020). As a result, we suggest that communities using user profiles and search features allow for their users to voluntarily show demographic information (i.e. gender, age, race, sexuality) along with implementing search filters, which is likely to help users more effectively find chat partners if they choose to do so manually.

### *Deployment-level Considerations*

Given that algorithmic matching is not incorporated into the online mental health context, designers of these communities must experiment with different matching mechanisms and protocols depending on their community's unique goals and users. However, continuous implementation and iteration on these mechanisms in real communities may disrupt the existing community and even pose potential dangers to users' health. As a result, we recommend that builders of matching systems approach experimentation with simulation of different algorithmic matching protocols.

Agent-based modeling, which can yield understanding of agents and their actions under different conditions, has been used in past work to reveal trade-offs in different design decisions for online communities, including to study the effects of social influence and information propagation (Alvarez-Galvez, 2016a; Du et al., 2017; Plikynas et al., 2015b; Z. Tan et al., 2011; van Maanen & van der Vecht, 2013). Agent-based modeling can be used to apply social science theories and understand trade-offs in designing new tools for online communities (Ren &

Kraut, 2010). Thus, simulation relying on agent-based modeling can navigate the complex effects that different algorithms have on a community's outcome metrics and how these effects work towards the community's unique sets of goals, and allows community stakeholders to iterate on these algorithms before deployment without continuously changing the real-time experiences of their users. For example, communities may wish to simulate filter-based methods given our findings that vulnerable groups are subject to harassment on the platform and that these groups often wish to chat with those similar to themselves, as well as the trade-offs of these hard filters with impacts on the broader community. Another possibility for OMHCs that allow users to block one another is to simulate matching for optimizing lowering blocking numbers between users, which would align with our finding that avoiding the worst outcomes is most important to OMHC users. We also believe one strong possibility as the basis for algorithmic matching on OMHCs is to rely on the applicant-proposing deferred-acceptance algorithm, which we have previously discussed as having been established in contexts such as school matching and The National Resident Matching Program, which produces a stable matching assignment; the OMHC matching problem similarly consists of two types of agents (in this case, support-seekers and support-providers) who each have preferences and personal features for matching, akin to the stable marriage problem.

# *Chapter 5*

# Building Agent-Based Simulation for Designing Online Mental Health Communities

People are increasingly turning to online mental health communities (OMHCs) for mental and emotional support (Stratford et al., 2019; Turner, 1999; Whittaker et al., 1997). OMHCs are a practical and accessible way for users to receive both informational and emotional support on a variety of mental and emotional concerns (Naslund et al., 2016), with communities offering general support for any need or support for specified health issues. For example, communities such as 7 Cups provide general 1-on-1 peer counseling chats, whereas platforms such as BabyCenter provide targeted support resources for pregnant women (Baumel et al., 2016; Gui et al., 2017). OMHCs have been found to be vital in maintaining and improving people's well-being, such as reducing depression, fostering meaningful relationships, and increasing trust in mental health treatment.

However, despite the ability of OMHCs to yield meaningful and positive relationships between users, they currently rely on naive methods (ie, first come, first served and solely based on topic of discussion) for support-seekers to find these relationships without consideration of users' unique characteristics and preferences. Prior work suggests that current matching systems do not adequately support users' needs and capabilities; ineffective matching can also lead to fewer long-term relationships and reduced support-seeker commitment (Fang & Zhu, 2022). Moreover, this lack of purposeful matching methods may be particularly harmful for marginalized communities, who have both strong preferences for mental health care providers with a similar background and particular reliance on online communities for support (Administration et al. 2001; Augustaitis et al., 2021; Bockting et al., 2013; H. S. Kim et al., 2008; Yadavaia & Hayes, 2012). Given these challenges, intelligent forms of matching can provide more optimal matches with minimal effort on a user's end (Fang & Zhu, 2022).

However, matching is a complicated mechanism design problem (Abdulkadiroğlu, Pathak, & Roth, 2005). It is challenging to meet all the possible matching goals between support seekers and providers, and prioritizing one goal might lead to worse outcomes in other goals. Given these challenges, tools such as agent-based simulation that have long been used to apply social science theories to the design of human-computer interaction systems are useful for revealing the complexities and various trade-offs in matching protocols for online community designers (Ren & Kraut, 2014). Importantly, running these low-cost, virtual experiments can predict community members' likely reaction to alternative design choices without disrupting existing community dynamics. Thus, agent-based modeling enables researchers and community designers to pin down factors leading to desirable outcomes and understand how design choices affect behavioral outcomes by modeling the intervening processes.

In this study, we answered the following research question: how can we experiment with new matching algorithms for OMHCs? Specifically, we sought to experiment with these new algorithms without harming or disrupting the existing community. To do this, we created a simulated “sandbox” based on agent-based modeling that allows community stakeholders to play with different matching algorithms and helps designers consider complex trade-offs in building new mechanisms for their community. To build this simulation based on a real OMHC, we collaborated with one of the world’s largest peer support platforms. We used the platform’s data set to accurately replicate the platform in our simulation and then experiment with alternative protocols in the simulation to analyze how these new matching policies affect users’ experiences. We created and tested seven new algorithmic matching policies based on prior work studying OMHC users’ needs: (1) First Come First Serve; (2) Last Come First Serve; (3) similarity-based matching, which uses cosine similarity to prioritize support seekers and support providers of similar features; (4) gender-based matching; (5) age-based matching; (6) topic-based matching; and, finally, (7) filter-based matching, which focuses on protecting teenagers and gender minority groups.

Our work contributes the application and example use of agent-based simulation to uncover the effects of alternative policies in the design of online community matching. Exploring different matching protocols has the potential to disrupt the particularly sensitive population of OMHC users through implementation and iteration processes; given this, our work showcases the benefits of instead using agent-based simulation to reveal the impacts of various algorithms. In addition to applying simulation to the OMHC context, we contribute practical findings for matching design in OMHCs, such as how optimizing based on topic can improve chat experiences for vulnerable communities, as well as trade-offs, such as how using algorithmic matching can increase the quality of conversations but also the waiting time for support seekers.

## 5.1 Methods

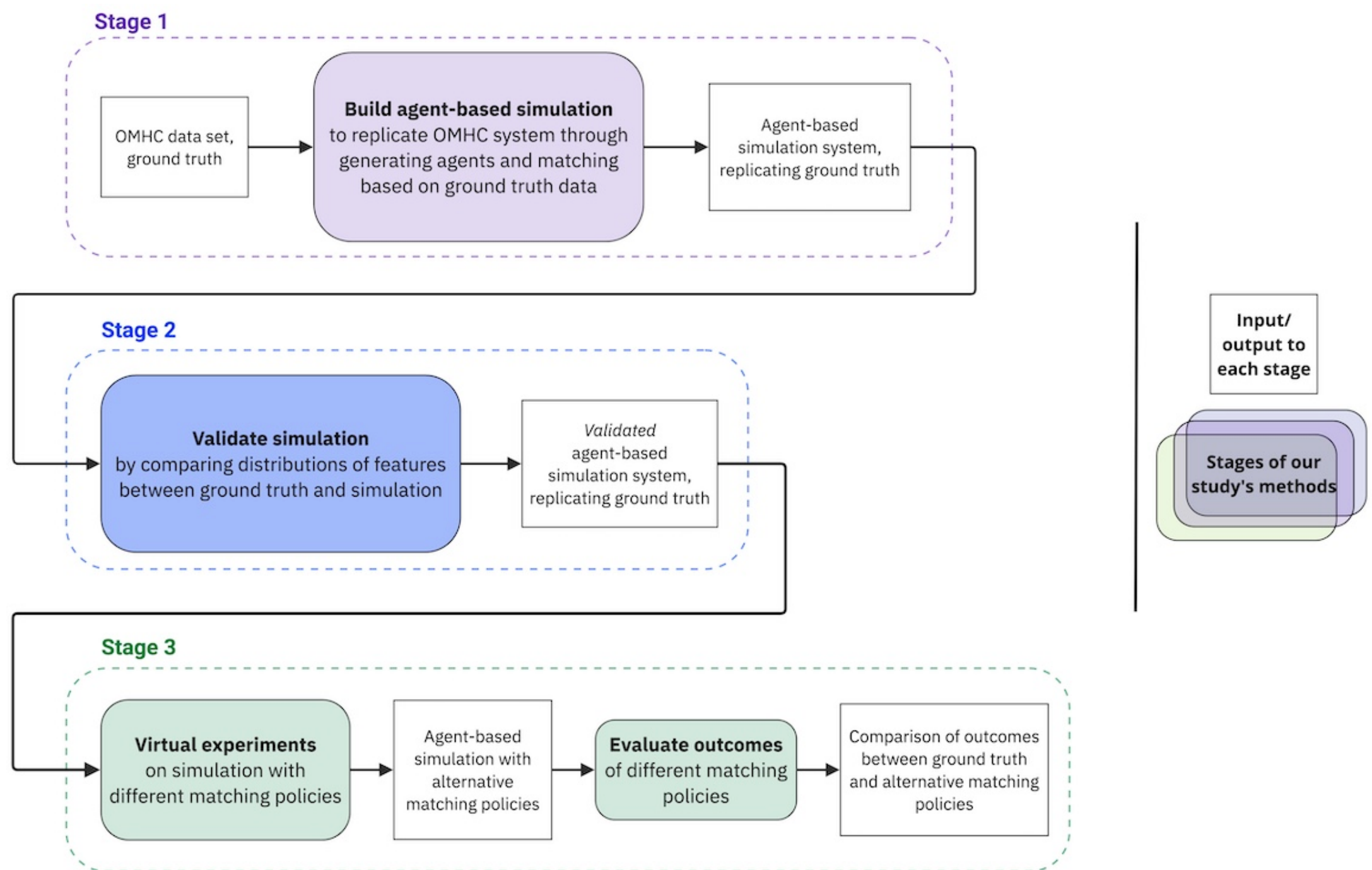

**Figure 6.** Workflow diagram showing inputs and outputs for the three stages of building and experimenting with agent-based simulation: (1) building an agent-based simulation, (2) validating the agent-based simulation, and (3) conducting virtual experiments using

the validated simulation. OMHC: online mental health community.

To showcase how agent-based modeling can be used to experiment with matching policies in the online mental health context, we followed 3 stages, as outlined in Figure 6.

### *Building an Agent-Based Simulation: Assumptions*

We first built a simulation based on agent-based modeling to replicate the current matching mechanisms of our study's research site and used outcome prediction models to validate this replication. In the following sections, we outline the assumptions made to build our simulation, all of which were implemented based on the real OMHC data set. Overall, we found that there was an extremely high correlation (ie, Pearson coefficient) between our simulation and the real research site. As all features had a high correlation, in the following sections, we show figures for only a few features as visuals of how our simulation performed compared to the research site.

#### *Simulation Period*

In our agent-based simulation, both support seekers and volunteer counselors have the possibility of being matched during each "round" or simulation period. We determined that a simulation period of 1 minute was fit for our model as 1 minute is temporally granular enough to yield quick matching of users and derivable from our empirical data.

#### *Generation of Agents*

Our simulation consisted of 2 types of agents: support seekers and volunteer counselors. Each simulation period (ie, each minute) generates new support seeker and volunteer counselor agents that are eligible for matching. Agents are considered "online" (ie, available to chat) immediately when they are generated. To determine the number of agents generated, we analyzed our study's data set from January 2020 to April 2022 to find the average number of online support seekers and volunteer counselors for each simulation period over a week (ie, 10,080 min).

When generated, each agent is also given several personal characteristics that may be significant to matching (gender, birth year, and topic of interest) [18]. We analyzed the OMHC data set to find the distribution of these characteristics at each simulation period and assign agents' characteristics so that the simulation's distribution was identical to the distribution found in the real OMHC data. Birth year was part of the raw data set, giving us complete and accurate real data to draw from. However, we had to conduct a labeling process for users' gender identity as a minority of users input their gender on their profile manually. Following prior work, we labeled gender according to whether a user had self-identified their gender in chat logs (ie, "I am a female" or "I am non-binary"), which has been found to be highly accurate and only mislabel gender for 0.8% of OMHC users [18]. Using this process, we were able to label the gender of 35% of support seekers and 50% of volunteer counselors in our data set. We then applied the distribution of gender among those whose gender was known to our generated agents so that all agents had a gender characteristic. In terms of topic, we used a previously built and validated topic classifier [48] on all chats in our data set to find the distribution of topics. The topic classifier we used from the work by Wang et al [48] was built using the same data set as our study. Using Empath [46], a tool that uses neural word embeddings for generating and validating lexical categories in large-scale text data, Wang et al [48] tuned Empath's model by feeding in "seed words" of 18 popular topics that support seekers discuss. The top 18 topics are romantic relationships, dating, pandemic, self-improvement, suicide, depression, parents, anxiety, family, stress, lonely, overwhelming, sexuality, LGBTQ, intimacy, home, dissociative identity, and health. Similar to assignment of the previous personal characteristics, we applied the real OMHC data set's distribution of topics to assign support seeker agents a topic of interest and counselor agents a list of up to 3 topics of interest that they have more experience or interest in talking about. The most frequent topic discussed on the site in our data set was "self-improvement," followed by "dating," "parents," and "depression." Our simulation's frequency distribution among all topics was the same, with a Pearson correlation coefficient of 1.

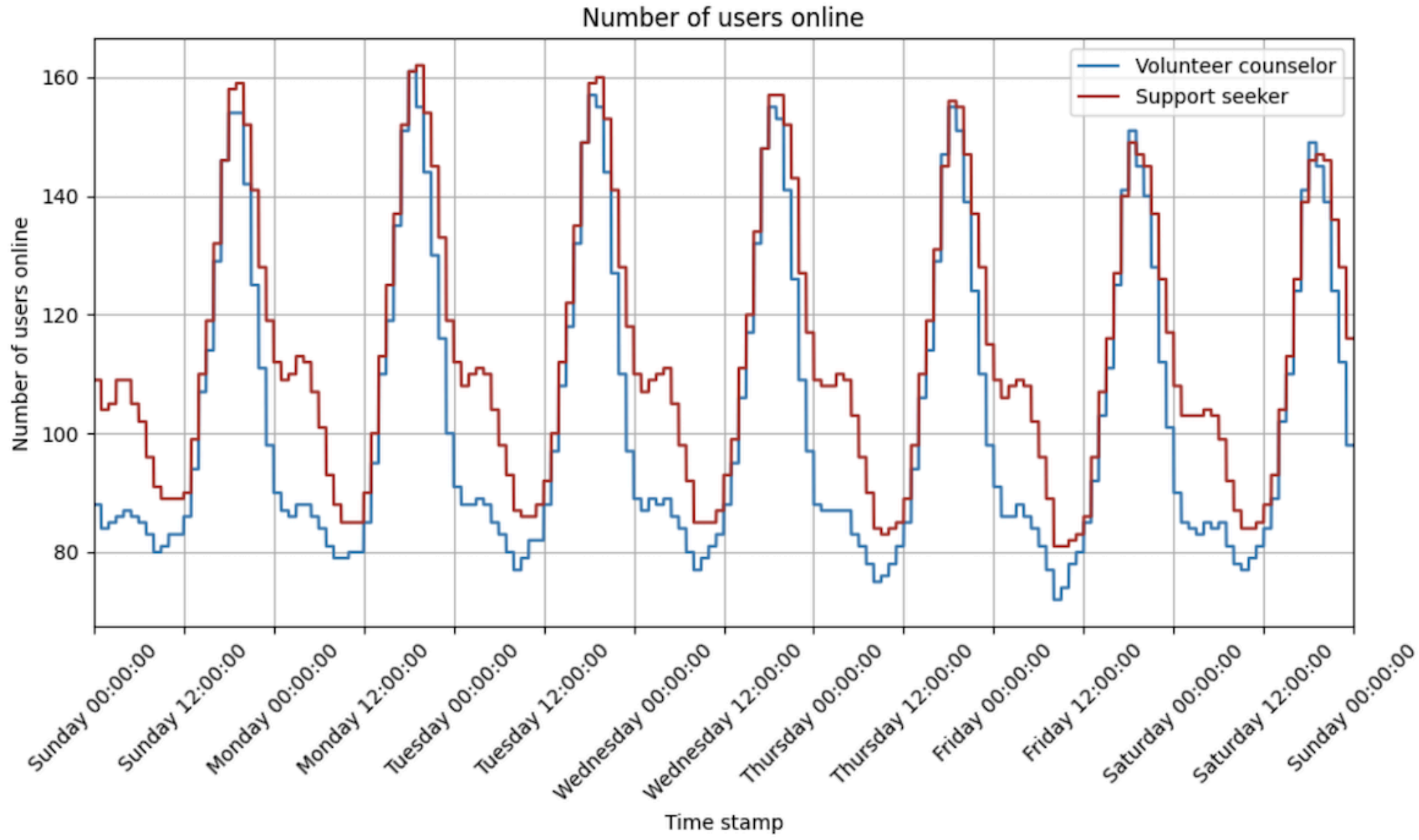

**Figure 7.** *A line graph showing the number of support seekers and volunteer counselors who are online in each simulation period. The number of support seekers always exceeds the number of volunteer counselors, with the number of support seekers and volunteer counselors who are online at any given minute ranging between 81 and 162 (mean 113.26, SD 22.56) and between 72 and 161 (mean 102.49, SD 25.07), respectively.*

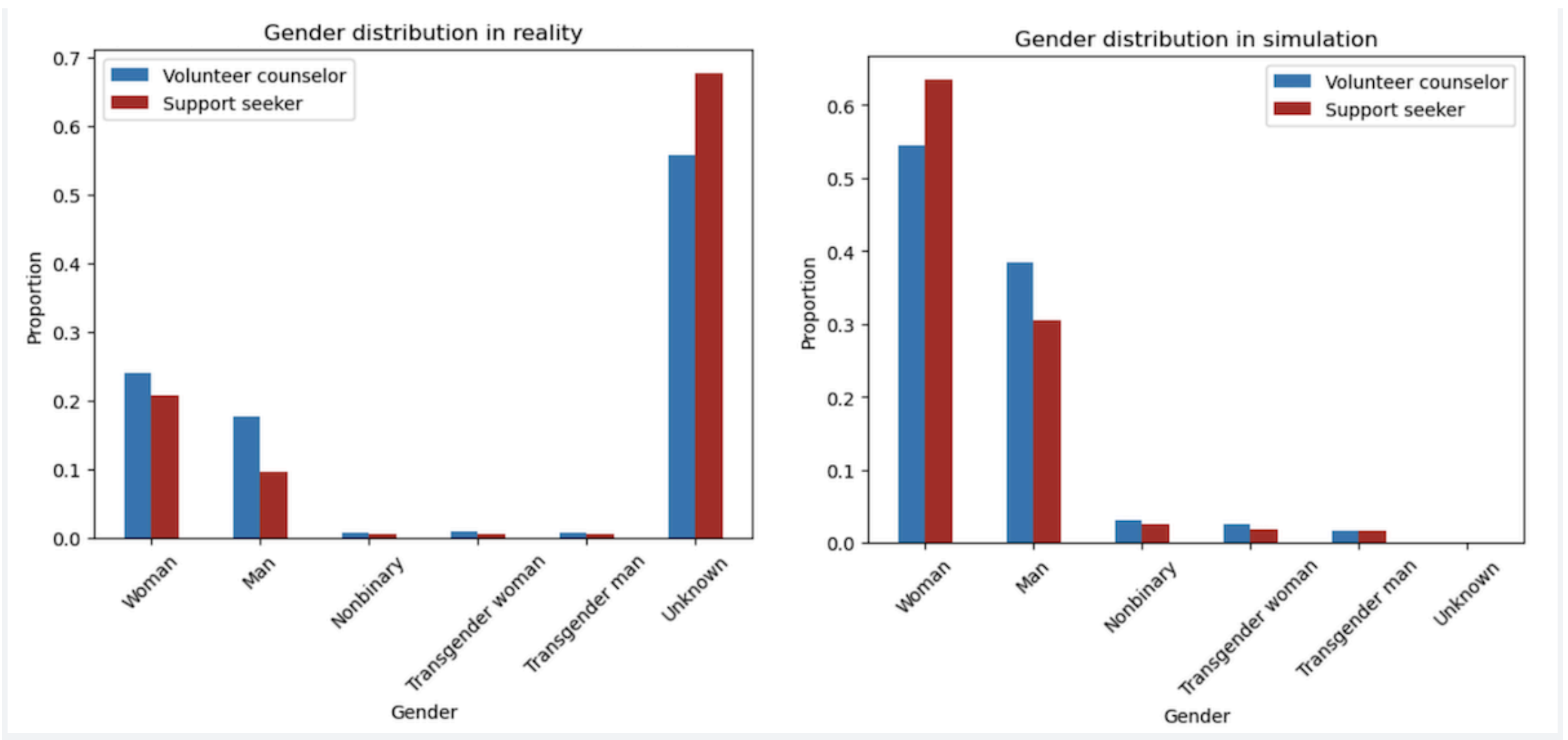

**Figure 8.** *Sex distribution in reality (left) versus our simulation (right). In the ground truth data set, many support seekers and counselors had unknown*

*gender. In contrast, all agents were assigned a gender in our simulation. We assigned gender according to the gender distribution of known genders (female, male, nonbinary, transgender female, and transgender male) in the ground truth data set.*

*Patience Level*

As support seekers may leave the site while waiting for a chat, we also replicated the "patience level" of support seekers. All support seeker agents were given a number of minutes that they are willing to wait to be matched before they cancel their request (ie, leave the platform). Support seeker agents go offline when the number of simulation steps in which the support seeker remains unmatched exceeds their patience level. Similar to previous characteristics, we assigned patience levels to support seeker agents according to the distribution we found from analyzing the data set for how long support seekers wait until canceling their chat requests in the queue. We assigned support seeker agents a patience level according to the distribution of time for support seekers to cancel their chat requests in the data set. The mean patience level in reality is 4.15 (SD 3.26) minutes, whereas our simulation's patience level had a mean of 4.16 (SD 3.27) minutes. The Pearson correlation coefficient was 0.991.

*Chat Length*

Once a support seeker and volunteer counselor are matched, their chat length is set in minutes, and volunteer counselor agents go offline after a chat ends. We set the chat length in our simulation to follow the distribution of conversation length found in the log data. The distribution of chat length in reality found through log data versus our simulation's distribution had a Pearson correlation coefficient of 1. The mean chat length in reality is 17.67 (SD 15.44) minutes, whereas our simulation's chat length had a mean of 17.67 (SD 15.42) minutes.

*Matching Support Seekers and Volunteer Counselors*

Finally, we describe the decision of matching a support seeker and volunteer counselor together to chat.

All agents who are online and not chatting in the current simulation period are considered available to be matched in the current round. In each simulation period, all volunteer counselor agents are presented with a list of all available support seeker agents and pick a support seeker to chat with depending on the matching policy. In the replication of the research site's system, volunteer counselors pick a support seeker to chat with randomly following an exponential distribution model for the time it takes to make their choice (Figure 9).

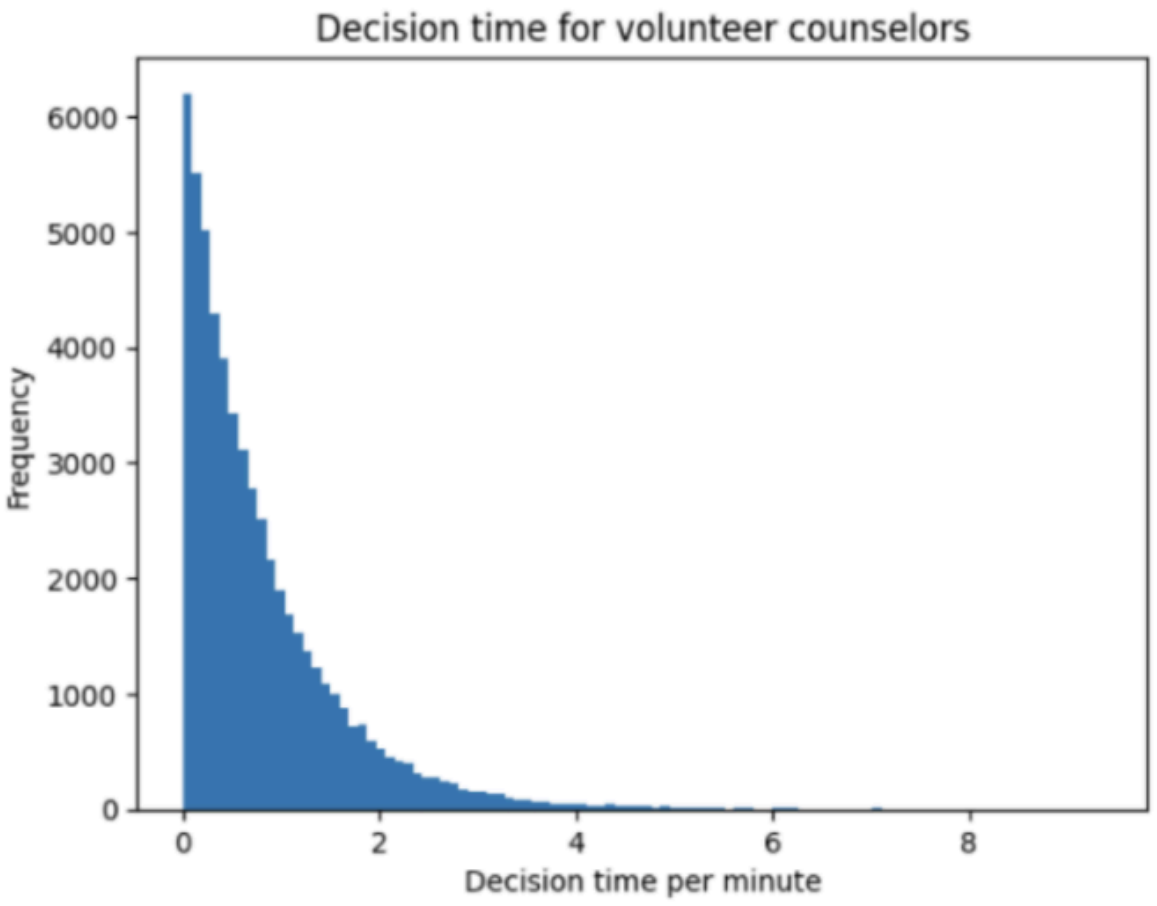

**Figure 9.** *Distribution of decision time of volunteer counselors, which is modeled using an exponential distribution with* $\lambda$ *of* 1.25

### *Building an Agent-Based Simulation: Prediction Models*

Although there are several outcome metrics that could be used to assess a support seeker's experience in a chat, the ability of a support seeker to give the chat a rating of 1 to 5 stars is the most direct indicator of the match. In addition, past work has found that users place more value on avoiding the worst chat experiences rather than aiming for the best given that extremely negative chats may include bullying and harassment and also may deter users from returning to the site [18]. Blocking another user is the main action that users take when they have a bad chat on the research site and is the most obvious reflection of a negative support seeker and volunteer counselor relationship. Both support seekers and volunteer counselors on our study's OMHC can block each other through the chat interface and provide a reason for blocking.

Note that, although our data set includes good indicators of matching quality, there is no existing data that can be used to calculate matching quality for simulated results. Therefore, we created 2 outcome prediction models—a chat rating prediction model and a blocking prediction model. In the validation process, we used those 2 models to test whether our replicated simulation was an accurate representation of the current system. In our virtual experiments, we used these prediction models to evaluate the effectiveness of our designed matching algorithms as well.

### *Chat Rating Prediction*

We gathered all chat ratings in the data set, with 80% of samples from the data set used as our training set whereas the remaining 20% were used as our testing set. We balanced the training set using the synthetic minority oversampling technique, which generates artificial samples for minority classes based on existing samples using the k-nearest neighbor algorithm [12]. After using the synthetic minority oversampling technique, we ended up with 16,730 samples for each of the 5 chat rating classes (1 to 5 stars).

As independent variables in our chat rating prediction, we used inputs of both volunteer counselors' and support seekers' gender, birth year, and topic. Our experiments included random forest, logistic regression, support vector machine, and decision tree models to find the best model to predict chat rating. Note that accuracy in this context means that the output must equal the true chat rating and is considered incorrect if it outputs any of the other 4 chat ratings. Given that random forest outperformed all other models in accuracy and F1-score at a 0.72 accuracy and 0.72 F1-score, we used the random forest classifier for our chat rating prediction model.

### *Blocking Prediction*

In creating our training data set for building the blocking prediction model, we labeled a support seeker and volunteer counselor pair as 1 if at least one person blocked the other and as 0 otherwise. Similar to our chat rating prediction model, we used input fields of gender, birth year, and topic. For all agents, we used an 80%/20% split of samples from the data set for training and testing, respectively.

We proceeded again with random forest classification as it resulted in the best performance, with the highest precision and recall scores of 0.91 accuracy and 0.91 F1-score.

### *Validation of the Agent-Based Simulation*

During the validation phase, we calculated and compared distributions of the features in the following sections between the real OMHC system data and our simulation's data. We then reported the Pearson correlation coefficients [43] between the real and simulated data distributions.

#### *Number of Users Online*

To validate the replication of the OMHC system, we split the data set into a 6-month training set and a 2-month test set. Figure 10 compares the number of online support seekers and volunteer counselors in each minute between the training set and test set in a period of 1 week. We found similar distributions between the training and test set, with Pearson correlation coefficients of 0.974 and 0.982, respectively.

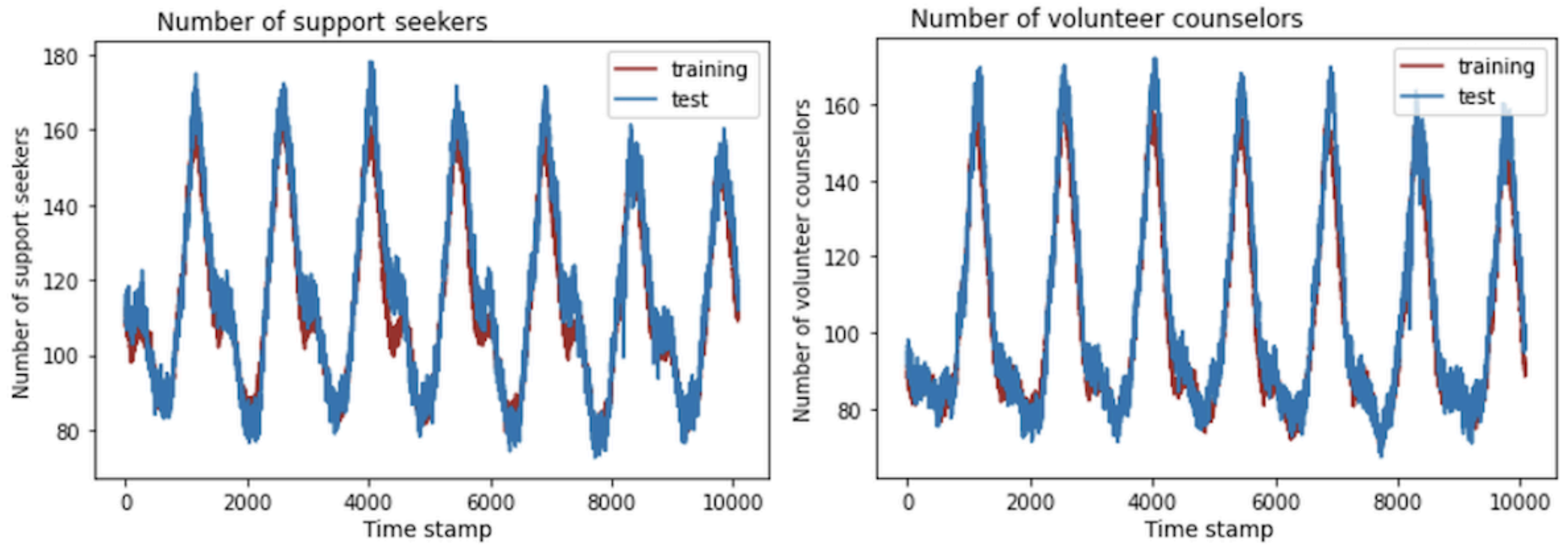

**Figure 10.** *Distribution of number of support seekers (left) and volunteer counselors (right) in the training and test sets. The Pearson correlation coefficient of the number of support seekers between the training and test sets was* 0.974*, whereas the Pearson correlation coefficient of the number of volunteer counselors between the training and test sets was* 0.982.

*Chat Ratings*

We used our chat rating prediction model to compare the distribution of chat ratings between our simulation and the ground truth. We found that we accurately simulated rating distributions on the research site, with a Pearson correlation coefficient of 0.99 between our replication and the ground truth (Figure 11). Similarly, we validated our replicated system using the blocking prediction model. Using the blocking prediction model, we found similar distributions of pairs that engage in blocking to those in the ground truth, with a Pearson correlation coefficient of 0.949, as shown in Figure 11, leading to further confidence that our simulation is an accurate representation of the current platform's system.

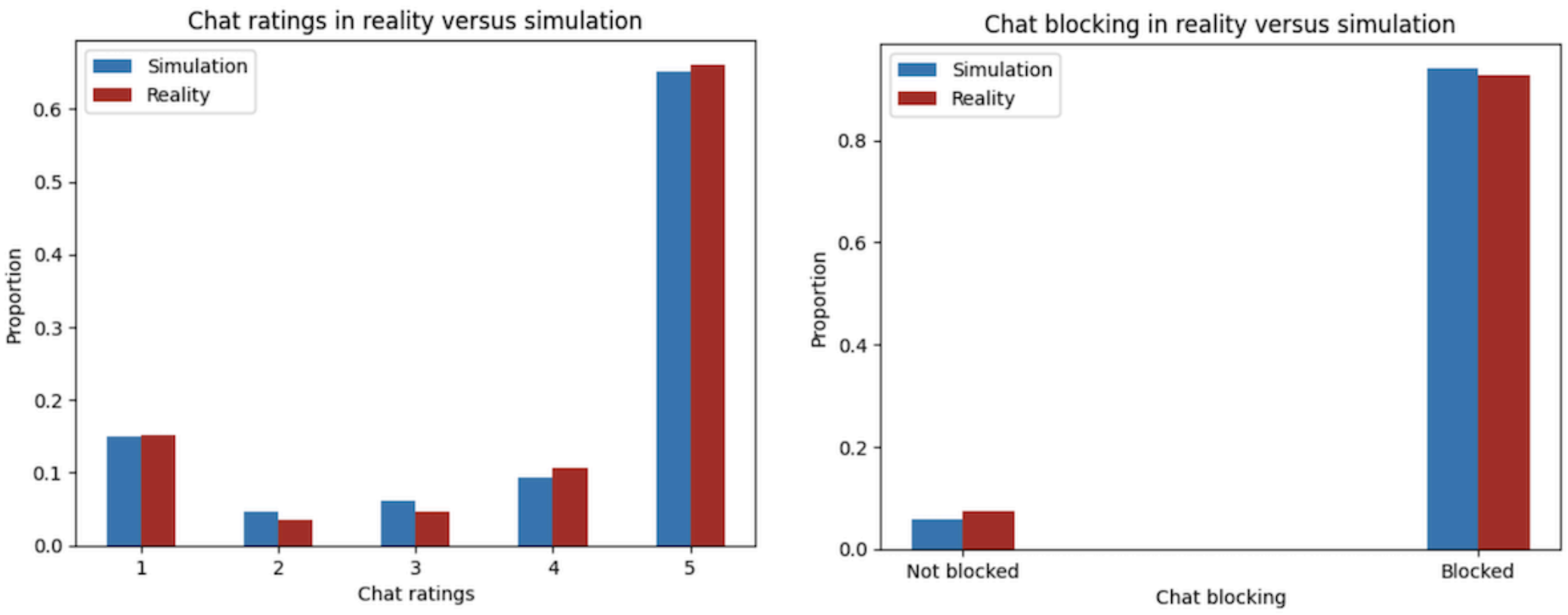

**Figure 11.** *Comparison of chat ratings (left) and blocking (right) between reality and our simulation. Our simulation's proportions of* 1*-star to* 5*-star ratings (shown in blue) were, respectively,* 14.96%, 4.64%, 6.08%, 9.26%, *and* 65.06%, *whereas the ground truth's proportions (shown in red) were* 15.18%, 3.51%, 4.56%, 10.63%, *and* 66.12%, *respectively. Regarding pairs that resulted in blocking, the ground truth proportion of blocked support seeker and volunteer counselor pairs was* 5.3% *compared to our simulation's proportion of* 5.86%.

*Waiting Time*

In terms of waiting time for support-seekers in our simulation, we found that the distribution of waiting time for support seekers who are matched with a volunteer counselor was similar to the distribution in the real online community system, although with a lower SD. The mean waiting time in reality is 3.2 (SD 2.9) minutes, whereas our simulation's waiting time had a mean of 3.2 (SD 2.02) minutes. The Pearson correlation coefficient was 0.995.

*Matching Rate*

We also compared the proportion of support seekers who were matched with a volunteer counselor to the total number of support seekers available to be matched. Note that support seekers may cancel their chat request if they are unmatched for longer than their patience level. To evaluate the ground truth, we analyzed the proportion of chats taken by volunteer counselors on the research site to the total number of requests by support seekers in the research site queue. We found that our simulation resulted in an overall 78.35% matching rate in a week, whereas the ground truth showed an average of 83.27% matching rate across all weeks in our data set. Although our simulation's matching rate was slightly lower than that of the ground truth, the likely explanation is that our simulation framework only allows support seekers and volunteer counselors to engage in one chat at a time, whereas the research site allows volunteer counselors to take multiple chats at once if they desire. Given this, we found that our simulation's matching rate was acceptable and still closely resembled the actual state of the research site.

## ***Virtual Experiments***

Next, we applied our simulation as a "sandbox" to test different matching algorithms. In the following sections, we review the matching algorithms in our experiment and their outcomes on the metrics of chat ratings, blocking, waiting time, and matching rates.

*Applicant-Proposing Deferred Acceptance Algorithm*

Our research problem consisted of 2 types of agents (support seekers and volunteers) with preferences and personal characteristics for matching. As a result, we considered it akin to the stable marriage problem, which seeks to find a stable matching between 2 classes of elements where both sides have an ordering of preferences. Thus, we used the applicant-proposing deferred acceptance algorithm as our matching algorithm, adapted from the established matching method for New York public schools (Abdulkadiroğlu, Pathak, & Roth, 2005).

In each simulation period, there are 2 sets of agents: support seekers (`M={m1, m2,..., mn}`) and volunteer counselors (`L={l1, l2,..., ln}`). Each support seeker has an ordered preference list—`P(m)={l1, l2,..., lm}`—where a support seeker's first choice is volunteer counselor l1, their second choice is volunteer counselor l2, and so on.

1. Each support seeker "applies" to their highest-ranked volunteer counselor according to their preferences, and each volunteer counselor "holds" their highest-ranked application and rejects the rest.
2. At any stage at which a support seeker has been rejected, they "apply" to their next most preferred volunteer counselor whom the support seeker agent prefers (if one remains). Each volunteer counselor holds their most preferred set of applications and "rejects" the rest.
3. The algorithm stops when no rejections are issued and each volunteer counselor is matched to the applicants they are holding.
4. Any agent who is not matched in this simulation period is marked as "waiting" and continues to be available for matching in the next matching period (along with any newly generated agents). Multimedia Appendix 1 shows the pseudocode for the aforementioned algorithm.

*Matching Algorithms*

The key design choice in algorithmic matching is to construct the preference lists of support seekers and volunteer counselors. Unlike school or physician matching with limited options, hundreds of volunteer counselors are available for any support seeker at any given time. As it is impractical to ask each support seeker to rank each counselor (and vice versa), preferences must be generated using rules or prediction models. The 7 algorithms are described in this section.

We chose methods based on our previous understanding of the research site support seekers' needs and consultations with the research site's leadership, including the chief executive officer and lead engineer. In general, the research site team expressed that their priorities were to optimize satisfaction on the platform (ie, chat rating) without reducing the general matching rate. It was important to the research platform that matching protocols allow for better experiences for support seekers but also that the research site could continue to serve most of its large support seeker population. Platform leadership was also particularly interested in gender-based protocols given that gender was a key factor in how people chose volunteer counselors. In addition, a "hard filter"–based protocol was suggested to try to mitigate harassment. Community leaders and our research team decided against directly optimizing for outcome metrics, such as rating-optimized or blocking-minimized protocols, for validity reasons; our study's prediction models (refer to the Building Agent-Based Simulation: Prediction Models section) use these same metrics (eg, rating and blocking), which would create validity issues if we were to make evaluations on protocols that directly optimize for these metrics.

Each algorithm varies in how it "recommends" a support seeker to each volunteer counselor in each simulation period—volunteer counselors have a 90% chance of taking the recommendation and a 10% chance of random selection.

- **First Come First Serve**: Support seekers are ranked in volunteer counselors' preference lists by decreasing waiting time. The goal of First Come First Serve is to improve the number of successful matches and prioritize serving support seekers who arrive to the queue first [9].
- **Last Come First Serve**: Support seekers are ranked in volunteer counselors' preference lists by increasing waiting time. The goal of Last Come First Serve is to minimize queue size when support seekers are more likely to leave the queue the longer they wait.
- **Similarity-Based Matching**: We used 3 dimensions of access to us that have been shown in previous literature to be important for matching purposes: gender, age, and topic of discussion. Using gender, age, and topic, we defined a vector for each agent with their gender identity, birth year, and topic of interest to calculate 2 agents' similarity using the cosine similarity between their vectors. Agents with higher similarity are ranked higher in other agents' preference lists.

  As reviewed previously, past work has found that clients often seek therapists similar to themselves among multiple dimensions. In terms of gender, prior work has found that it is one of the most important factors in choosing a support provider in online platforms in both client preferences and outcomes, and especially so for gender minority groups (Felton, 1986). In addition, people who are closer in age (especially for older populations) are better suited for one another in OMHC support (Fang & Zhu, 2022) given their similarity in experiences and communication. Finally, having a therapist that is knowledgeable and willing to discuss the client's needs and issues is vital to the therapeutic relationship. Topic relevance is a widespread idea in the space of online communities; for example, social media sites regularly infer a user's interests to recommend them relevant content (eg, purchase suggestions and personalized advertising) using someone's previous behaviors on the site (Bian & Holtzman, 2011; Z. Wang et al., 2015). Similarly, in our case, we judge a counselor's expertise on topics based on their past chats. Each support seeker is assigned a topic of interest that they wish to chat about according to distribution based on the real OMHC data set. On the basis of the top 3 topics by frequency for each volunteer counselor in our OMHC data set, we assign each simulated volunteer counselor a top-3 topic list. Volunteer counselors with a relevant topic in their list are ranked higher in the respective support seekers' preference list, and vice versa.

In addition, we include matching protocols exploring each of these 3 dimensions (gender, age, and topic) individually.

- ➢ **Age-Based Matching**: Support seekers and volunteer counselors who are closer in age to each other are ranked higher in each other's preference lists.
- ➢ **Gender-Based Matching**: Support seekers and volunteer counselors with the same gender identity are ranked higher in each other's preference lists.
- ➢ **Topic-Based Matching**: Volunteer counselors are prioritized in a support seeker's preference list if the counselor's topic list contains the support seeker's topic of interest; similarly, support seekers are prioritized in volunteer counselors' preference lists if the topic of interest is in the counselor's expertise. In other words, if a support seeker's topic is in a counselor's topic list, then we label a feature of "expertise matching" as 1 (and 0 otherwise) and use expertise matching as the only matching criteria.

- ❖ **Filter-Based Matching**: Suggested by stakeholders in our study's research site, we implemented filter-based methods to prioritize the protection of 2 vulnerable groups—namely, in this context, teenagers and gender minority groups. The filter-based method includes 3 different pools for agents: underaged (aged ≤18 years) pool, gender minority (noncisgender women or noncisgender men) pool, and all others. Volunteer counselors can only select support seekers who are in the same pool as themselves. However, apart from this limitation on what support seekers are available to volunteers, volunteer counselors can pick anyone (ie, random selection) following the existing protocol on the site's simulation (refer to the Matching Support Seekers and Volunteer Counselors section).

## 5.2 Findings

Our study's primary contribution is the application of agent-based simulation to the online mental health context. However, we also review in the following sections the outcomes of experimenting with 7 specific algorithms using our simulation sandbox. Although these are just 7 possible protocols for matching in this context, in the following sections, we review our comparison of their outcomes to show the kinds of findings and trade-offs revealed through an agent-based simulation.

The outcome metrics used to evaluate the algorithms are defined as follows:

- **Average rating**: average rating for all pairs in the simulation predicted by the rating prediction model (the higher the better)
- **Percentage of blocked pairs**: percentage of pairs in the simulation that were predicted as "blocked" (the lower the better)
- **Matching success rate**: the ratio of support seekers who were successfully matched (ie, chatted with a volunteer counselor) to all support seekers in the simulation (the higher the better)
- **Average waiting time (matched)**: average waiting time of support seekers before they were matched in the simulation (the lower the better)

- **Average waiting time (unmatched)**: average waiting time of support seekers whose waiting time exceeded their patience level and quit (the higher the better)

We ran 1-tailed t tests to see whether different algorithms' outcome metrics were statistically significant when compared to the replication protocol. We indicate both the standardized P value threshold of .05 as well as a more conservative threshold of .001 in the tables, focusing our discussion on the findings with P<.001.

**Table 11.** *Outcome metric results for different algorithms' performances. Significant outcomes when compared to the replication of the research site (first row) are indicated.*

| | Rating, mean | Blocked pairs (%) | Matching success rate (%) | Waiting time for matched clients (min), mean | Waiting time for unmatched clients (min), mean |
|---|---|---|---|---|---|
| **Replication of the research site** | 4.05 | 5.86 | 78.91 | 3.19 | 3.69 |
| **First Come First Serve** | 4.06 | 5.82 | 81.93[a] | 3.68[a] | 2.73[a] |
| **Last Come First Serve** | 4.04 | 6.11 | 74.81[a] | 2.61[a] | 4.69[a] |
| **Similarity based** | 4.04 | 7.37[a] | 79.36[b] | 3.34[a] | 3.53[a] |
| **Age based** | 4.02[a] | 7.22[a] | 79.97[a] | 3.40[a] | 3.41[a] |
| **Gender based** | 4.07[b] | 6.04 | 81.81[a] | 3.61[a] | 2.86[a] |
| **Topic based** | 4.31[a] | 4.26[a] | 80.70[a] | 3.45[a] | 3.29[a] |
| **Filter based** | 4.03[b] | 6.21[b] | 61.30[a] | 3.24[a] | 3.86[a] |

[a]*p*<.001.
[b]*p*<.05.

**Table 12.** *Average chat rating results for different algorithms' performances among different groups. Listed are the mean ratings for each group. Significant outcomes when compared to the replication of the research site (first row) are indicated. Our simulations showed that the similarity-based protocol (and its subprotocols) and the filter-based protocol had significant effects on the average chat rating among demographic groups. For example, both filter- and similarity-based protocols raised the average chat rating for minority groups but lowered it for nonminority groups.*

| | Adults | Underaged group | Non–gender minority group | Gender minority group |
|---|---|---|---|---|
| **Replication of the research site** | 4.00 | 4.28 | 4.06 | 3.80 |
| **First Come First Serve** | 4.01 | 4.31 | 4.07 | 3.84 |
| **Last Come First Serve** | 3.98 | 4.31 | 4.05 | 3.83 |
| **Similarity based** | 3.95[a] | 4.46[a] | 4.04[a] | 4.20[a] |
| **Age based** | 3.94[a] | 4.39[a] | 4.03[a] | 3.76 |
| **Gender based** | 3.94 | 4.32 | 4.06 | 4.14[a] |
| **Topic based** | 4.26[a] | 4.53[a] | 4.33[a] | 4.06[a] |
| **Filter based** | 3.92[a] | 4.39[a] | 3.98[a] | 4.69[a] |

[a]$p<.001$.

**Table 13.** *Blocking results for different algorithms' performances among different groups. Shown are the percentage of pairs that resulted in blocking behavior from either party. We indicate statistically significant findings. Our simulations showed that the similarity-based protocol helped reduce blocking for underage and gender minority individuals but raised blocking for adults and the non–gender minority group. The topic-based protocol showed a statistically significant reduction in blocking for adults, the underaged group, and the non–gender minority group. The filter-based protocol yielded better results only for the gender minority group, with a worse performance for all other groups.*

| | Adults (%) | Underaged group (%) | Non–gender minority group (%) | Gender minority group (%) |
|---|---|---|---|---|
| **Replication of the research site** | 6.73 | 1.85 | 5.45 | 12.31 |
| **First Come First Serve** | 6.66 | 1.91 | 1.86 | 12.5 |
| **Last Come First Serve** | 7.03 | 1.86 | 5.66 | 13.13 |
| **Similarity based** | 8.69[a] | 1.25[a] | 7.34[a] | 8[a] |
| **Age based** | 8.29[a] | 1.88[a] | 6.77[a] | 14.32[b] |
| **Gender based** | 7.02 | 1.52 | 5.89[a] | 8.64[a] |
| **Topic based** | 4.97[a] | 0.99[a] | 3.76[a] | 11.84 |
| **Filter based** | 7.38[a] | 2.35[a] | 6.66[a] | 0.44[a] |

[a]$p<.001$.

[b]$p<.05$.

### *Average Rating*

Overall, the results of our virtual experiments followed intuition in that different matching policies served different goals. Average chat rating was one of the metrics of the most interest to our research team in evaluating algorithm outcomes given its interest to our community stakeholders as a good proxy for community satisfaction on the site. We found 2 reliable (P<.001) results in which the topic-based protocol yielded significantly higher ratings than the replication protocol and the age-based protocol had significantly lower ratings.

The topic-based protocol performed significantly better than the replication of the research site (P<.001) but also all other algorithms. While all other algorithms output average ratings between 4.02 and 4.07 out of 5 stars, the topic-based protocol had a 4.31 average rating out of 5 stars—yielding a statistically significant increase of 6% over an already high baseline replication. Regarding the age-based protocol, it had a statistically significant but marginal decrease in average chat rating of 0.7% compared to the replication. Generally, we found that the age-based protocol performed well for minors (Table 12) but did not have any improvement for adults (who make up most of the population). As a result, age similarity seems to matter for young people connecting but not for the general adult population. We discuss more about the breakdown of protocols across demographic groups in the Results by Demographic section.

### *Blocking*

In general, blocking rates remained low for all protocols, including in the replication. However, we did observe one statistically significant improvement among protocols, particularly when using the topic-based protocol to match pairs; the proportion of support seeker and volunteer counselor pairs who engaged in blocking (from either party) was reduced from 5.86% to 4.26%. On the other hand, similarity-based matching (combining age, gender, and topic) and the age-based protocol yielded a marginal increase in the blocking rate.

### *Matching Success Rate*

Regarding overall matching success rate, we found that First Come First Serve, as expected, led to the highest matching success rate, whereas Last Come First Serve had one of the lowest matching success rates. Given that Last Come First Serve as a matching protocol is aimed at keeping queue size small but not necessarily serving the most people, it follows intuition that we observed a low waiting time for matched pairs but overall relatively fewer successfully matched pairs (we explore this further in the Waiting Times section).

Worth noting is that all protocols other than similarity-based matching showed statistically significant results ($P<.001$) compared to the replication when it came to the matching success rate. First Come First Serve and age-based, gender-based, and topic-based matching were all protocols that resulted in higher matching success rates. Last Come First Serve and filter-based mechanisms resulted in statistically significantly lower matching success rates for the community. Filter-based matching resulted in the most striking difference—a 22.3% reduction in overall matching success rate; although we find in Tables 11-13 and discuss later in this chapter that there are many benefits to a hard filter–based mechanism for certain groups, our simulation showed a substantial trade-off when it came to the number of people in the community able to find support chats.

### *Waiting Times*

Waiting times were generally within 1 to 2 seconds of each other when comparing across protocols. However, we note an intriguing finding in that First Come First Serve initially showed counterintuitively one of the longest waiting times for people who were successfully matched. However, upon further reflection, this result actually was to be expected. First Come First Serve matched the greatest number of people; therefore, more people with higher patience levels and who, thus, had higher waiting times were also able to be matched. This increased the average waiting time of successful matches. Thus, the average waiting time for unmatched agents only included the agents who were impatient (short patience levels); as a result, the average waiting time for matched clients seemed higher, and the average waiting time for unmatched clients seemed lower. The opposite case occurred for Last Come First Serve; there were short waiting times for successfully matched support seekers, but few support seekers were successfully matched. Thus, we note for community stakeholders that evaluating the average waiting time requires consideration of the matching success rate in conjunction rather than being evaluated in isolation.

### *Results by Demographic*

We show in Tables 2 and 3 a breakdown of different algorithms' performances for average chat rating and blocking percentage by demographic groups of adults, underaged people, and non–gender minority groups. Note that we do not have racial information for this analysis, so we are limited to commenting on the effects on only gender and age matching for this study. We also provide the results for matching success rate and waiting times in Multimedia Appendix 1. In general, the largest differences were observed for underaged people and gender minority groups, showing that algorithmic matching has the greatest potential effect for these often marginalized, excluded, and vulnerable groups.

When breaking down protocols by their performances among different groups, we observed several trade-offs on how algorithms perform on majority versus minority groups. It also allowed us to see that the current state of the research site has drastically different effects on various demographic groups. For blocking rate, we noted that the replication of the research site had a significantly higher blocking rate of 12.31% among gender minority groups—an alarming rate that is more than double that of non–gender minority groups. This follows prior work finding that LGBTQ+ communities on OMHCs have a higher risk of harassment (Craig & McInroy, 2014; Fish et al., 2020; McConnell et al., 2017). This finding also allowed us to see the major degree of improvement allowed for by the algorithmic matching as we found that the filter-based algorithm (which restricts gender minority support seekers to only being paired with also gender minority counselors) led to a striking improvement of 96% for blocking (from 12.31% to 0.44%) and 23% for average chat rating. Thus, our simulation results give quantitative support to previous qualitative findings that have suggested that LGBTQ+ support seekers prefer a counselor of a similar identity. In particular, the reduction in blocking behavior is promising regarding protecting minority or vulnerable groups from experiencing unwanted behaviors or relationships.

There are significant trade-offs with algorithm choice that emerged from our breakdown of results by demographic group. As reviewed previously, the filter-based algorithm performed exceptionally when it came to gender minority groups. It resulted in an impressive average chat rating of 4.69 out of 5 stars for gender minority groups, a 23.4% improvement over the replication protocol. It also predicted a very low proportion of pairs who blocked one another for both minors and gender minority groups (2.35% and 0.44%, respectively, compared to the replication's results of 6.73% and 12.31%, respectively). However, it notably performed poorly overall when it came to chat rating and blocking among adults, underaged groups, and gender majority groups. In fact, the filter-based algorithm resulted in a worse performance compared to the replication of the research site for all other groups for both chat rating and blocking; this indicates that the hard filter approach results in worse user experiences for most of the site's population compared to if there was no algorithmic consideration in matching at all. This result may complement previous work that found that LGBTQ+ users in OMHCs feel safer speaking with other LGBTQ+ users [18] but also reveals a likely trade-off for lower overall satisfaction among majority groups on a platform such as the research site. We observed a similar trade-off when it came to the overall similarity-based model. Compared to the replication of the research site, the similarity-based protocol showed a significant decrease in average chat rating for adults and non–gender minority groups (−1.3% and −0.4%, respectively) but a significant increase for underaged people and gender minority groups (+4% and +10.5%, respectively). Regarding blocking, there was an almost equal rise in blocking rates for adults and non–gender minority groups (+29% and +35%, respectively) as there was a decrease in blocking rates (a positive result) for underaged people and gender minority groups (−32% and −35%, respectively). However, note that, except for gender minority groups, blocking rates were relatively low among all other groups to begin with.

However, despite this, one of the most notable takeaways from our simulation results is that this trade-off between the experiences of minority and majority groups was not necessarily the case with all protocols. We found that the topic-based protocol performed well overall even among the marginalized communities we studied despite no consideration of gender or age in its matching criteria and improved the experiences of the general community. When compared to the replication protocol, we found that all results of the topic-based protocol, with the exception the insignificant finding for the blocking rate of gender minority groups, were improvements on both chat rating and blocking among all groups; other outcomes similarly showed improvement or a similar performance to that of the replication of the research site, as shown in our full results in Multimedia Appendix 1. Regarding underaged people, the topic-based protocol (average chat rating of 4.53 and blocking rate of 0.99%) actually outperformed the hard filter–based protocol (average chat rating of 4.39 and blocking rate of 2.35%), which is the most restrictive protocol for prioritizing the matching of demographic groups. As a result, a key takeaway from these results is that directly using demographics of gender and age as matching criteria is not necessary to improving the experiences of the targeted groups. Although topic-based matching does not include any demographic consideration, it still led to significant improvement for groups that are marginalized or vulnerable due to their demographics (age and gender) and shows that improving the experiences of minority groups is not necessarily exclusive to improvement for the overall population as well. However, we note that our findings of demographics' effects on matching are limited to gender and age; racial matching is an important factor for traditional services, but our analysis lacked enough racial data, and thus, we cannot draw conclusions on the role of any racially or culturally based matching.

## 5.3 Discussion and Takeaways

### *Agent-Based Modeling for Designing OMHCs*

In this work, we showcased our important contribution of applying agent-based simulation to the OMHC context and provided a framework for how it can help the creators and designers of online communities weigh the various trade-offs when building mechanisms to best help their support seekers find meaningful relationships online. Our study suggests that different goals can be achieved through different algorithmic choices for these communities, from optimizing the quality of conversations to the protection of gender and age minority groups on these platforms.

Our research was guided by previous literature that revealed how algorithmic matching is particularly beneficial in the online mental health context and the numerous considerations that are necessary for effective partnerships to aid users' well-being (Fang & Zhu, 2022). Through simulating the algorithms and mechanisms of the research site, we found that there are numerous trade-offs to be made in deciding how to match users together in OMHCs. For example, communities aiming to optimize the number of users that engage in chats may wish to experiment with First Come First Serve methods given simulation results that have the highest overall matching rate. Alternatively, algorithmic matching may lead to better conversations between users but also increase the time that users must wait for a match. Our simulation revealed that there is indeed substantial improvement that can be made in the quality of conversations using algorithmic matching, such as a >6% increase in the average chat rating just using topic-based matching. However, users may become impatient while waiting for a "best fit" chat partner and log off the site, and thus, fewer people overall are helped by the platform. We see this reflected in our experiments—for example, First Come First Serve is able to match the most people, with a matching rate of nearly 82% despite being the simplest algorithm without optimization for users' characteristics.

We also considered and offered the benefits of agent-based modeling in conversations with the stakeholders of the research site. Although these conversations introduced other important considerations in algorithmic matching implementation, such as the computational resources and refactoring of the codebase to implement matching, the stakeholders of our study's research site found our results insightful and beneficial to the community. In particular, the site's leadership expressed that simulating algorithmic matching helped them weigh the effects and trade-offs of different design choices for their site without disrupting the existing community through continuous algorithmic testing. Other conversations were also had between our research team and the site's leadership, such as which outcome measures were most crucial in evaluating the platform's success for support seekers. The research site has begun developing an algorithmic matching system for their platform based on this work.

It is also worth noting that the research site's stakeholders wanted to know more about the reasoning behind the numerical results for each matching algorithm. As a result, we believe that it is important that the presentation of these experiments is accompanied by model explanation and algorithmic transparency as well to best help community stakeholders understand results and make decisions based on these types of simulations. Indeed, this suggestion is supported by a plethora of past research that has found that model transparency is necessary for trust and understanding when deploying new algorithms in online communities for both community decision makers and community members (Eslami et al., 2019; Juneja et al., 2020; Kemper & Kolkman, 2019). We have released our simulation as open source at https://github.com/liuyuhan1997/Agent-based-Simulation-for-OMHC-Matching.

### *Matching Criteria for Online Support Chats*

It is important to note that there does not exist a universally best design for all communities; instead, the choice of algorithms and mechanisms for mental health communities is specific to the community's context, its goals, and its support seekers [30,48]. However, although our experiments are not necessarily generalizable to every OMHC, they do provide some initial insights into making algorithmic choices for these communities.

Past work in traditional mental health services (eg, therapy) has found that using demographics for matching clients and providers is important for outcomes, particularly so for minority and vulnerable groups (Felton, 1986; McConnell et al., 2017). Our findings in the online community context support this notion in that matching based on gender and age indeed improved outcomes for underage and gender minority groups. However, a more surprising result from these experiments in our agent-based simulation model is that matching approaches that do not use demographics as matching criteria can also serve to protect vulnerable groups. As discussed, we found that matching solely based on topic resulted in high average ratings and low blocking percentages among all groups and was an improvement on the baseline. Indeed, topic-based matching outperformed for average rating, proportion of blocking, and the matching success rate when compared to the similarity-based protocol that also included age and gender as matching criteria. However, it is possible that using race or cultural background as a matching criterion could lead to improvements given their importance in research on matching in traditional mental health contexts. However, regarding our findings on gender and age, topic as an effective metric for matching people in vulnerable groups makes some intuitive sense; people who are teenagers, for example, are likely to share similar issues compared to older adults (eg, parenting or divorce), and people who are from gender minority groups may also use an OMHC to seek support for gender issues, transitioning, or other general LGBTQ+ questions [18]. As a result, topic may inherently connect people who not only are of similar demographics but also have similar experiences to one another—this includes those who are not necessarily in what our simulation considered a vulnerable group (minors or gender minority groups), and thus, we observed the topic-based protocol performing well on the overall population largely due to its improvement for majority groups in addition to minority groups. It is important to note the privacy risks when it comes to people's personal information given to online platforms, and there is some evidence that people have reservations when it comes to revealing this information [18]; thus, our work importantly reveals that bettering people's experiences is not necessarily reliant on more demographic or other personal information about them.

We note another surprising result: similarity between people among all features—age, gender, and topic—in our similarity-based protocol generally did not result in significant improvements compared to other protocols. Supporting prior work on the importance of demographic features [3,18,19,24], our evaluation of outcomes by demographic features (Tables 2 and 3) showed improved experiences for the group that a protocol used as a matching feature (ie, the gender-based protocol for gender minority groups or the age-based protocol for minors). However, the general similarity-based protocol that used all features (age, gender, and topic) only showed marginal improvement among underaged and gender minority groups when it came to blocking proportions (Table 13), albeit slightly more substantial improvements for the same groups when it came to average chat rating (Table 12). When evaluating the findings in Tables 12 and 13, we hypothesize that simply matching on all characteristics is not necessarily conducive to an overall improvement for vulnerable groups; instead, intentional choice must be made for selecting certain features (such as gender for gender minority groups) to aid their experience. Overall, we found that just as much or even more improvement was observed for vulnerable groups when solely matching on one characteristic, such as gender for gender minority groups or age for underaged people, and even more so that topic was as good or even better as a sole matching criterion compared to cosine similarity among all features. However, we also acknowledge a potential issue in our simulation's use of cosine similarity as it may not be the best way to measure similarity between people in this context; further work may be necessary to test other similarity measurements. In addition, as mentioned previously, the lack of racial and cultural background data for our study may have limited our findings.

# Part 2

*Simulation for Self-Care Technology*

---

*The following chapters are based upon the following works*:

> ***Fang, A.***, *Chhabria, H., Maram, A., & Zhu, H.* (2025). *Social Simulation for Everyday Self-Care*: *Design Insights from Leveraging VR, AR, and LLMs for Practicing Stress Relief. In Proceedings of the* 2025 *CHI Conference on Human Factors in Computing Systems* (*CHI* '25). *Association for Computing Machinery, New York, NY, USA, Article* 716, 1–23. *Honorable Mention.*
>
> ***Fang, A.***, *Chhabria, H., Maram, A., & Zhu, H.* (2026). *Social Simulation for Integrating Self-Care*: *Measuring the Effects of Contextual Environments in Augmented Reality for Mental Health Practice. Under Review for the* 2026 *CHI Conference on Human Factors in Computing Systems* (*CHI* '26).

---

*While Part One focused on improving virtual communities through simulation-based experimentation, Part Two extends these insights to individual self-care practices. Just as online support platforms benefit from structured, intentional design, individuals managing stress and anxiety can also benefit from structured and controlled ways for improving their mental well-being. However, similarly to Part One, experimentation in real-world high-stress situations can be intimidating and ineffective — people may struggle to apply techniques in the moment, and unsuccessful attempts can reinforce feelings of helplessness rather than improvement.*

*Building on the findings from Part One, which highlight the importance of tailoring interventions to user needs, this section explores how immersive simulation can provide individuals with a low-risk environment for experimenting with self-regulation techniques. By leveraging VR, AR, and LLM-powered interactions, I examine how people engage with simulated stress scenarios and how different levels of realism, interactivity, and guidance shape their ability to develop coping skills. Through this work, I aim to establish simulation as not just a tool for system designers, but also as a method for individuals to gain confidence and skill over managing their mental health in everyday life.*

# *Chapter 6*

# Understanding Users' Needs for Technology-Supported Self-Care

When faced with stress and anxiety, taking care of ourselves is challenging. At least 70% of Americans experience physical and mental symptoms of stress and 58% of people aged 18 to 34 feel most days contain an overwhelming amount of stress. However, only 37% of people believe they manage this stress effectively (Stress in America 2023, 2023). Although mental health resources have become more accessible through advancements like virtual therapy and online communities (Chester & Glass, 2006; Kauer et al., 2014), most people do not have constant access to these resources when going about their personal, work, or social lives and inevitably encounter stress or even panic. Stress is not only difficult to feel, but also affects our ability to fulfill other aspects of our lives as family members, friends, professionals, and community members (Maslow, 1943).

Human-Computer Interaction (HCI) research has extensively explored mental health care needs and systems over the past few decades (Balcombe & De Leo, 2022). Despite the many people who are not able to manage their own stress effectively, though, there continues to be a notable gap in applying HCI technology to practicing everyday self-care. One potential avenue for offering people a safe environment to practice their stress relief skills is with *simulation technology* – technological systems that aim to replicate real-world behaviors, agents, or processes in a virtual environment (Hollan et al., 1984; Park et al., 2022; Tambe et al., 1995). Virtual simulation can aid in stress-relief practices by providing immersive, controlled scenarios that allow individuals to safely engage with and manage stressors. By recreating realistic social and environmental dynamics, users can potentially practice coping mechanisms in a low-risk environment, a preventative self-care measure that can improve emotional regulation and preparedness for real-life situations (Carson & Kuipers, 1998).

In our work, we leverage virtual reality (VR), augmented reality (AR), and large language models (LLMs) to develop social simulation in order to simulate common everyday environments for users to practice self-soothing – specifically, box breathing, which is a technique well-documented in literature for its effectiveness in treating stress, anxiety, and panic (Clark et al., 1985). Using eight low fidelity prototypes, we conducted *prototype-driven interviews* with 19 participants to explore their reactions to using simulation for practicing stress relief for the everyday. Specifically, our work explores the research question:

**RQ: What are the needs, risks, and design considerations in using simulation technology to help practice management of everyday stress?**

Our findings indicate that many people lack effective methods for self-care and found our simulation prototypes useful for building skills to deal with everyday stressors; most expressed they would incorporate this use of simulation into their real-world practices as a self-care practice tool. We also studied different dimensions for designing social simulation for self-care, including modality, interactivity, and mental health guidance features. Participants particularly preferred AR due to its integration with real surroundings, which made it easier to transfer skills to real-world contexts and engage physically with the environment. However, participants also saw accessibility as a valuable asset of LLM-powered, purely text-based simulation despite lower immersion, given its likely greater accessibility on-the-go. Our interviews also uncovered trade-offs in designing social simulation for self-care, such as strong preferences for more realism alongside concerns about potential trauma from hyper-realistic designs. Finally, our work highlights key design considerations for using LLMs in social simulation, such as users' concerns about the effects of LLM-generated recommendations on their sense of agency and real-world transferability of skills.

## 6.1 Methods

In this section, we describe the three key design dimensions — modality, interactivity, and guidance needs — that guide our prototype development. Using these prototypes, we conducted prototype-driven interviews and analyzed these interviews using thematic analysis.

Our prototype development uses an approach akin to parallel prototyping in order to generate richer insights into people's design preferences and lead to higher quality design recommendations. Given our exploratory goals to uncover users' needs and constraints, we designed eight different prototypes rather than testing one "optimal" solution so we could thoroughly explore the design space. While we did probe participants after each prototype (*e.g.*, "*What are your initial impressions?*", "*Would you use this?*"), we also focused on asking overarching thoughts and design feedback after participants had finished seeing all eight prototypes (*e.g.*, "*Which of these designs do you prefer, if any, and why?*", "*How realistic did these simulations feel?*").

### *Key Design Dimensions*

We designed prototypes following three main dimensions guided by our research question and past studies in immersive technology:

**Modality**: We aimed to investigate how the form of the simulation — whether VR, AR, or text-based — and thus its level of immersion affects users' perceptions of the technology in managing stress.

**Interactivity**: We also ask what level and types of interaction users desire within these simulations – whether through dialogue, body language, or environmental feedback. Our prototypes range from non-interactive, static environments to fully interactive settings where users engage in real-time, responsive conversations with LLM-powered avatars.

**Guidance Needs**: Our study explores how to help people practice active soothing mechanisms, such as breathing, during moments of high stress. However, some past research in VR/AR has found success in just facilitating exposure (without any incorporated self-care techniques) to help build resilience. We explore how users felt practicing stress relief techniques superimposed on the simulation of real-world environments; in particular, we selected box breathing as our guidance intervention given its positive effects for stress and relative simplicity to learn for new users. As a result, for each modality we include prototypes without any breathing intervention and those with intervention (**`VR-Guided`**, **`AR-Guided`**, **`Text-Guided`**).

Through these three dimensions, we designed our prototypes with the goal of providing insights into how different aspects of simulation can contribute to social simulation for self-care.

### *Design Prototypes*

Although everyone experiences everyday stress, what triggers that stress varies from person-to-person. To better understand participants' feelings about our prototypes in the context of their own personal experiences, we designed prototypes around three distinct scenarios that have been simulated in prior research, with the aim for at least one to resonate with participants as a stressor. Due to interview time constraints given our many prototypes, we did not show participants interactive controls for AR and instead asked participants to imagine given the system they had just previously interacted with in VR. Participants were also shown screenshots of prototypes to better visualize the various scenarios and interfaces.

1. **Public Speaking Q&A.** Public speaking is the most common fear. Given the practical challenge in asking participants to give a full speech, we opted to ask participants to imagine themselves giving a speech and interact in a post-talk Q&A setting instead. In this scenario, the user is asked to imagine they have just finished giving a speech on any topic of their choosing and answer questions from a 12-person audience.
2. **Social Party.** We selected to simulate a crowded social situation given that social anxiety is the most common anxiety disorder. In this scenario, the user is asked to imagine they just moved to a new area and have been invited to a party. They are met with a crowded room of people and are greeted in conversation with 4 other party guests.
3. **Interpersonal Conflict.** We simulated interpersonal conflict (in this case, with a roommate) due to most people having encountered it but lacking the opportunity to practice the complex dynamics of conflict resolution in an emotionally safe setting. Stress during conflict has also been shown to be a barrier to resolution. In this scenario, the user is asked to engage in conflict resolution with their roommate about a topic of their choosing.

Based on our design dimensions of modality, interactivity, and guidance needs, we created the following eight prototypes for each of the three scenarios. We go into more detail on the implementation in the following sections. Example screenshots are shown in Figure 12.

- **VR-Static**: a non-interactive scene in virtual reality. Avatar(s) and the user do not engage in any dialogue interaction.
- **VR-Interactive**: same as **VR-Static**, except avatar(s) and the user can engage in dialogue. Users can pull up a keyboard to type or dictate their replies.
- **VR-Guided**: same as **VR-Interactive**, with the addition that the user can press a button on their controller at any point to pull up a breathing guidance orb. The user can use the breathing guidance at any point for as long as they wish.
- **AR-Static**: a non-interactive scene.
- **AR-Interactive**: same as **AR-Static**, except avatar(s) and the user can engage in dialogue similar to the VR version.
- **AR-Guided**: same as **AR-Interactive**, with the addition that the user can press a button on their controller at any point to pull up a breathing guidance orb. The user can use the breathing guidance at any point for as long as they wish.
- **Text-Interactive**: users interact with a text-based bot that role plays as an audience/party guests/a roommate, depending on the users' scenario.
- **Text-Guided**: same as **Text-Interactive**, with the addition of text output of breathing guidance (i.e. "*Breathing Guidance*: *Inhale for* 4 *seconds*. *Hold for* 4 *seconds*. *Exhale...*").

### *Prototype Development*

Our method involves developing usable, interactive prototypes, as opposed to using storyboards, in order to generate more detailed insights as participants could tangibly experience the immersive capabilities of social simulation. We go into more detail below on building the prototypes, including the visual and audio components of VR/AR and the LLM-powered dialogue.

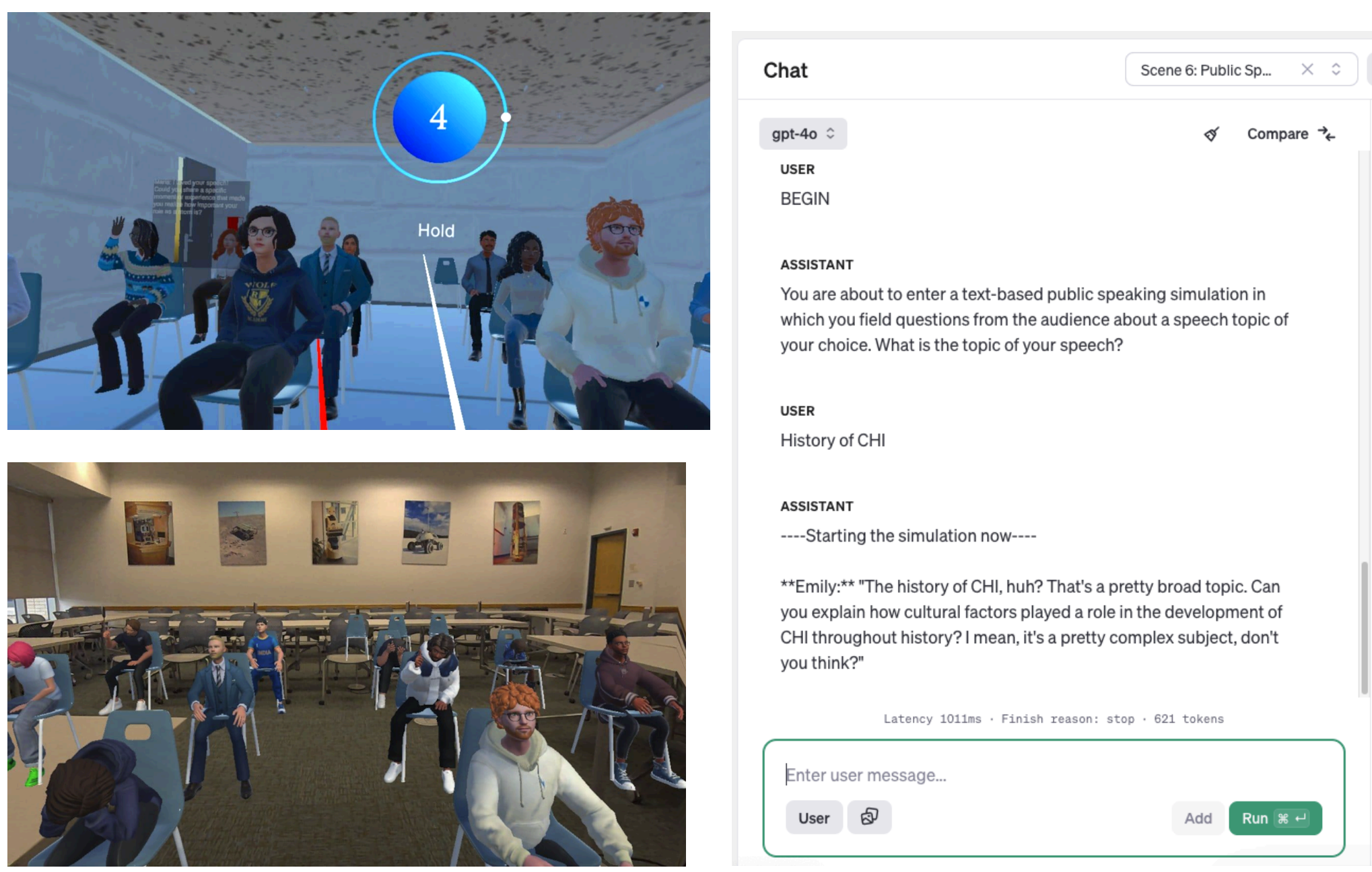

**Figure 12.** *Examples of different prototypes, shown here for the public speaking scenario, for VR-Guided (left top), AR (left bottom), and text (right).*

**Environment.** For all 18 VR/AR builds (three VR and three AR, across three different scenarios) we built the prototypes using the Unity game engine and deployed them on a Meta Quest 3 headset. All avatars were created using Ready Player Me where the first three authors selected diverse avatars and appropriate clothing for the scenario (e.g. casual outfits for the social party). Additionally, we looped sound clips that suited each scenario using freely available audio from online. Specifically, we included soft classroom and hallway noises for the public speaking scenario, incoherent talking and dance music for the social party scenario, and layered air conditioning and street noises for the interpersonal conflict scenario.

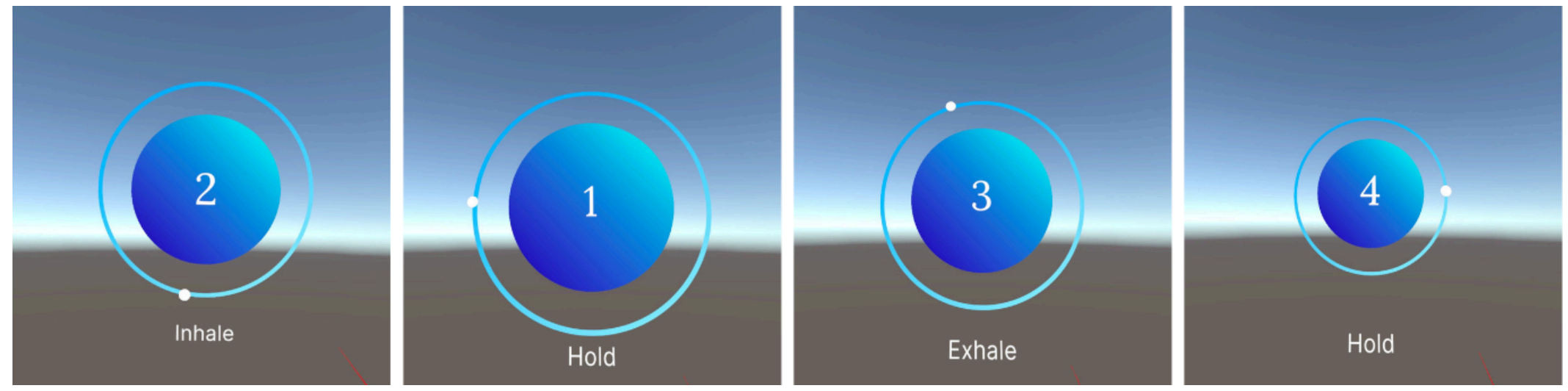

**Figure 13.** *Onboarding system of box breathing guidance. The user is asked to inhale, hold, exhale, and hold for 4 seconds each. The bubble expands and contracts following the breath, with a countdown timer for 4 seconds for each stage.*

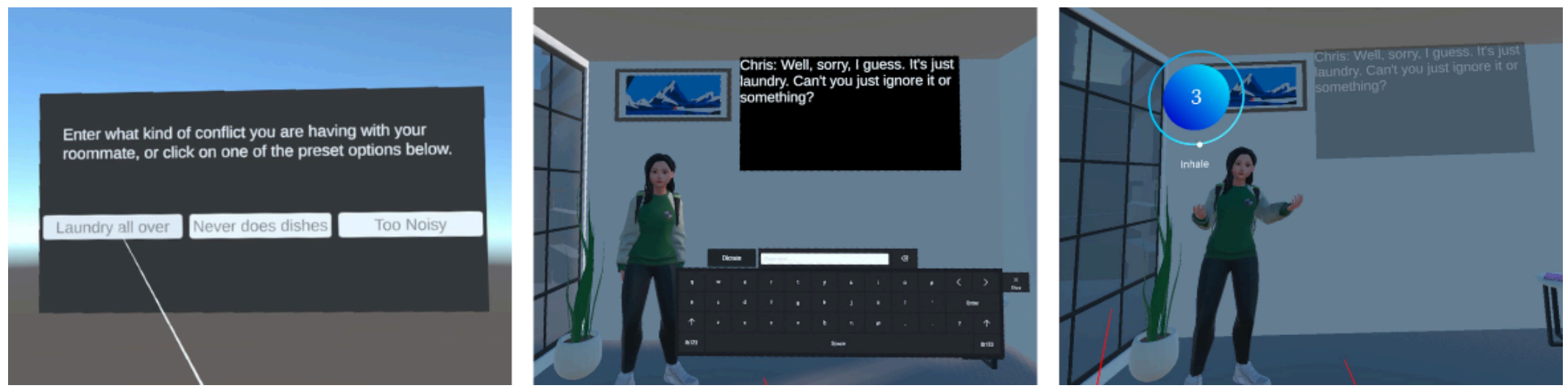

**Figure 14.** *Example of interaction flow in the VR prototype for roommate conflict. The user can select a topic (left) then begin a dialogue with the generated scene (middle). At any point, the user may trigger the breathwork guidance and conduct box breathing while the simulation is paused (right).*

**Dialogue.** Dialogue for all prototypes is generated using GPT-4o, the most recent GPT version at the time of this study. Our work contributes to the existing but little work on using LLMs to generate realistic avatar dialogue for interactive extended reality. We used consistent prompting patterns across text-based, VR, and AR prototypes. To ensure natural interactions, the first three authors iterated on prompting patterns to create realistic exchanges. For instance, we asked GPT to generate character personas with seed examples of conversational styles to create realistic and relevant interactions. We also ensured that GPT would output dialogue from individual characters so they spoke one-at-a-time in order to not overwhelm the user. For VR/AR prototypes, we used the OpenAI-Unity package to integrate GPT-4o responses to avatars in VR/AR. For the text-based prototypes, participants interacted directly with GPT-4 via the OpenAI Playground. We conducted pilot testing and gathered feedback from a psychiatry practitioner before proceeding with participant interviews.

**Interaction and Controls.** The text-based prototypes were created and shown to participants in OpenAI Playground, so participants interacted with GPT directly using a laptop keyboard. For VR/AR interaction, participants interacted with avatars by using the Quest 3 controllers to type on a VR keyboard, or could use speech-to-text dictation that we implemented using WhisperAI. Participants are stationary in the simulation, although they can turn to look anywhere 360-degrees.

**Guidance System Development.** For VR/AR prototypes, the box breathing guidance was informed by prior research to use cool colors, abstract imagery, and expansion and contraction to guide the breathing pattern. Given past work's recommendations, we also showed users a dedicated onboarding system (without any scenario simulation) for users to familiarize themselves with box breathing before entering the simulation; screenshots of the box breathing guidance system are shown in Figures 13 and 14. Users could access breathing guidance at any point by pressing a button on the Quest 3 controller. For **`Text-Guided`**, its guidance implementation is output just as text instructions in OpenAI Playground; we prompt GPT to output breathwork at recommended moments of high stress. We ask users their opinions on this LLM-powered guidance instruction versus manually-selected guidance (like in VR/AR) in our interviews.

### *Semi-Structured Interview*

We conducted prototype-driven interviews with 19 participants. Interviews were semi-structured, guided by a list of questions but allowed to deviate depending on topics participants may have introduced. The interviews took place in-person, lasted approximately 1 hour, and were audio-recorded.

#### *Participants and Recruitment*

We aimed to recruit participants of diverse backgrounds, so eligibility was only restricted to those 18 years and above and required in-person presence for the study. We recruited participants through putting up physical flyers in the local community, posting on social media, and snowball sampling. Interested respondents were asked to complete an intake form that asked for their basic demographic information of gender identity, age, and racial identity. Interview participants' demographic information is listed in Table 14. In total, we interviewed 19 participants, who we met in-person. Participants were compensated with a $20 Amazon gift card. This study was approved by the appropriate Institutional Review Board (IRB).

***Table* 14.** *Demographic and exposure conditions for participants* (*N*=19).

| | Gender | Race / Ethnicity | Age | Exposure |
|---|---|---|---|---|
| P1 | Cisgender Man | Asian | 21 | Public Speaking |
| P2 | Cisgender Woman | Asian | 24 | Public Speaking |
| P3 | Cisgender Woman | White | 21 | Social Party |
| P4 | Cisgender Man | Hispanic, Latino, Spanish Origin | 21 | Public Speaking |
| P5 | Cisgender Woman | Black, African-American | 19 | Conflict |
| P6 | Cisgender Man | Asian | 30 | Public Speaking |
| P7 | Prefer not to say | Black, African-American | 29 | Conflict |
| P8 | Cisgender Woman | Asian | 19 | Conflict |
| P9 | Cisgender Man | Black, African-American | 21 | Conflict |
| P10 | Cisgender Man | Asian, White | 22 | Social Party |
| P11 | Cisgender Man | Asian | 42 | Social Party |
| P12 | Cisgender Woman | Asian | 24 | Conflict |
| P13 | Prefer not to say | White | 41 | Social Party |
| P14 | Cisgender Woman | Asian | 29 | Social Party |
| P15 | Cisgender Woman | Asian | 20 | Conflict |
| P16 | Cisgender Man | Hispanic, Latino, Spanish Origin | 22 | Conflict |
| P17 | Cisgender Man | White | 38 | Public Speaking |
| P18 | Cisgender Woman | Black, African-American | 19 | Conflict |

| P19 | Cisgender Woman | Asian | 23 | Public Speaking |
|---|---|---|---|---|

*Exposure Condition*

Participants were asked to report their expected stress and anxiety levels for the three scenarios: (1) answering questions from an audience after their public speaking, (2) attending a crowded social party, and (3) having an interpersonal conflict with a roommate. Participants rated their stress levels ("On a scale of, how stressful or anxious do you anticipate yourself to be when...") for these three situations using the Subjective Units of Distress Scale (SUDS) scale from 0 to 10, which is an established rating system in psychology (Kiyimba & O'Reilly, 2020; Wolpe, 2018). After participants finished this survey, we recorded participants' highest stress situation and asked participants if they felt comfortable experiencing simulation for it. All participants said yes; we informed participants that at any point they could ask us to change scenarios or stop the interview altogether. For participant safety, we also automatically excluded showing them scenarios for which they rated their anticipated stress 9 or 10, and also provided all participants with the local county mental health hotline number for extra precaution.

*Interview Protocol*

We showed the eight prototypes on the participant's highest stress topic given their initial survey. In addition, we also showed the "onboarding" guide to the breathing guidance system before the participant saw **`VR-Guided`** or **`AR-Guided`** to help users engage with the exercise (Patibanda et al., 2017); in the onboarding guide, the participant saw only the breathing orb and no other simulation (see Figure 13), during which the participant could take as long as they needed to practice the box breathing pattern before proceeding. For each prototype, we provided an introduction blurb for participants to imagine. For example, before letting participants interact with the VR prototypes for interpersonal conflict with a roommate, we stated: "*Imagine that you recently moved in with a new roommate. Things have been tense, though, and you plan to confront your roommate at home later today after work about an ongoing source of disagreement between you two. Nervous about how this conversation will go, you put on a virtual reality simulation to practice* *what you are going to say to them. In this scene, you'll be met with your roommate who will speak with you about this ongoing conflict. You'll be able to choose the topic of the conflict*." For the public speaking and interpersonal conflict scenes, participants could choose their own topic of choice for their speaking topic and conflict reason, respectively, or choose from pre-sets as shown in Figure 14.

We aimed to understand people's needs, whether this technology met those needs, and understand the design opportunities to improve self-care technology. The interview protocol consisted of participants openly interacting with each of the eight prototypes and asking participants questions revolving around four main concepts: (1) **initial impressions**, (2) **usefulness** ("*To what extent do you feel this is useful?*", "*How does practicing stress relief in this way compare to your current practices?*") (3) **realism and natural interaction** ("*To what extent did the interactions feel natural or unnatural?*", "*What aspects felt realistic or unrealistic?*") and (4) **general feedback** ("*To what extent did you like this design?*", "*What additional features would be useful?*"). For the prototypes involving the breathing guidance, we additionally asked participants what their thoughts were on using guidance or, if they didn't use guidance, their reasoning for not using the breathing guidance system. Lastly, we asked participants about their experience as a whole with the simulation as well as their general thoughts on applying simulation to their everyday stress relief practices. Given the semi-structured nature of the interviews, our research team's interviewer would also ask follow-ups, clarifications, and ask participants to expand on thoughts they expressed in an evolving nature during the course of the interview. The full list of interview questions are in our appendix.

### *Data Analysis*

After completing all interviews, we anonymized, transcribed, and coded all 19 interview transcripts. Based on Braun and Clarke's method of thematic analysis (Braun & Clarke, 2012), our iterative analytic cycle consisted of: (1) recording and transcribing interviews (2) coding transcripts (3) amalgamating codes (4) discussing codes (5) highlighting themes (6) writing and revising memos. Our interviews and analysis continued until saturation, where new interviews gave no new insights. We first generated the lowest-level, granular codes through each transcript, then organized these codes into axial-codes. Afterwards, we arranged these findings into five high-level themes.

## 6.2 Results

We organize our findings around five high-level themes identified through our interview analysis, including people's current experiences of stress, overall perspectives and experiences using simulation for self-care, and findings around our three high-level dimensions for designing simulation for self-care: modality, interactivity, and guidance.

### *Current Experiences of Everyday Stress*

Participants experienced a range of different stressors in their everyday lives, most of which were in the themes of social anxiety and work. 12 out of 19 participants described their everyday experience of stress as related to social anxiety, including being in social settings with people they don't know (P3, P12), being in crowded spaces (P10, P12), giving presentations (P2, P14, P15), and talking on the phone (P19). Other sources of stress include taking exams, meeting deadlines, and generally having a lot of work or personal tasks to complete.

*Current methods to handle stress.*

In addition to asking participants what types of everyday stress they experience, we also asked them how they dealt with stress. While participants did mention a range of coping mechanisms, such as relying on social connections (P11, P14) and problem-solving (P3, P10, P14, P15), the most common theme across participants was using distraction.

> "*talking to people about lighthearted things...so I can* ***distract*** *myself*" (*P*5)
> "[*I use*] *messaging applications...talk about random stuff and* ***distract*** *myself from any conflicts going on.*" (*P*16)
> "***distract*** *myself a bit, browsing social media*" (*P*1)
> "*I usually* ***distract*** *myself*" (*P*12)
> "*Not a technique, but I do* ***distract*** *with hobbies that are not related*" (*P*16)

Regarding strategies targeted towards mitigating their own stress, many participants had tried habits like journaling (P14, P17), meditation and meditation apps (P3, P12, P14), and breathing techniques (P1, P2, P4, P5, P11); however, almost all did not sustain these practices over time with the exception of regular exercise, which was cited as effective at keeping chronic stress low (P4, P7, P8, P9, P13). Participants' reasons for not continuing use of self-care techniques during stressful situations ranged from finding some techniques **difficult** ("*Journaling was a thing that my therapist said, but I don't have patience for it…I wish it would be more straightforward somehow*" (P17), "*I have briefly tried meditation apps... I just kept not focusing. So* [*it*] *didn't really go that well.*" (P8)), a **lack of enjoyment** ("*I downloaded a few apps, but again, I find it annoying.*" (P17)), **accessibility issues** such as cost ("*I used Headspace* [*a meditation app*], *but I don't really do it anymore because I didn't want to pay the subscription*" (P14)), and **start-up work** (*"it's a lot of work that you need to get in touch with* [*a*] *therapist, and you need to experiment with them to see if you work well together."* (P19)). Others had no particular strategies at all to deal with stress in their everyday life, such as P19 who said they "*just wait till* [*the stress*] *is gone*"**.**

*Perceived ineffectiveness of current stress reduction habits.*

When describing their current experiences of everyday stress, participants generally did not feel that they had good techniques for stress relief (e.g. "*I don't think I've had a ton of success*" (P3)). In particular, **participants felt they lacked self-sufficient skills.** For example, P16 said that their tendency to use distraction to deal with stress was "*evasive and destructive, which has worked enough, but I feel like it relies on something else... And I do think I don't have the techniques or knowledge to just by myself meditate or calm down*" (P16). When speaking about how she copes with making calls to customer service, a source of stress for her, P19 expressed, "*If I need to make the call by myself, I think I don't have a really good way to do that*" and this caused even more stress around not having a good way to deal with that stress: "*I think I feel stressed by not having a good way to cope with* [*the stress*]" (P19).

Additionally, heightened emotions were cited as a barrier for handling acutely distressing or stressful situations, such as interpersonal conflict. For these participants, calming their emotions was often seen as a necessity for them to properly destress or self-soothe: "*I have to calm down a little bit. My reaction is a lot* *of anger. I clench my teeth, then I start sweating [and] my eyes become a little hazy...I do nothing until I've calmed down*" (P6). Another participant expressed, "Anger really makes [it] hard to get out of [stressful] situations... it might be good for me to practice how to control my feelings by practicing breathing" (P11).

### *Overall Perspectives on Using Simulation for Self-Care*

After understanding participants' current experiences of stress, we proceeded to gather their thoughts on our prototypes and overall perspectives on the role that social simulation may take in their self-care.

Overall, participants responded very positively to some or all of the prototypes shown as well as the general idea of using virtual simulation to practice stress relief. **17 out of 19 participants said they would use one or more of the shown prototypes** (VR, AR, or text-based) in their current lives if they had access to the technology. Participants felt using simulation would **improve the outcomes** of situations ("*it would easily complement what I would normally do in reality, and just make all these situations work out better for me*" (P9)), help **explore different responses** ("*Just being able to articulate it beforehand and think, 'okay, maybe I should try saying this instead.' I think it would be helpful for me*" (P18)), make them **feel more prepared** ("actually having this scenario completely play out before it plays out" (P5)) and **mitigate negative emotions** ("*you can use it to mitigate some anger, stress*" (P9)). People even re-imagined the use of this technology, such as P16 who mentioned that they feel it would be useful in professional therapeutic settings rather than just for personal self-care, saying, "*If a therapist had this available during a session, it could be quite useful to test-drive situations or problems that you're addressing during therapy*."

#### *Benefit of Realistic Self-Care Practice.*

Participants felt that the main benefit of social simulation was to practice stress relief techniques they could use "*in-the-moment*" of real-world scenarios. Doing so within a realistic environmental setting was praised by participants and seen as a step up from their current ways of coping with stress, as P16 notes, "*The closest that I'd imagine doing before I came into the study would have been practicing something in my head or in front of the mirror, which compared to this, is way less evocative and* *doesn't put you into the same mentality. So it wouldn't be as useful or as effective as something like this*." P5 also appreciated the practice "*of tackling the stress by just being in the situation itself*" as it helped them "*mitigate the fear*" when approaching an interpersonal conflict. In contrast to retroactively reflecting, the simulation helped surface emotions in-the-moment of stress, which could result in a more accurate and reflective picture of one's emotional state. For example, one participant noted, "*I feel like this one, I can do it in the middle of whatever I'm experiencing. I definitely feel more in control because other techniques, you just reflect on it afterwards. With this one, I feel more like reflecting in that moment, trying to observe my reaction. In that sense, it was more helpful*" (P14). Moreover, participants noted that doing the breathing guidance served as a reminder to calm themselves "*in the middle of a situation*" when they expected to be too emotionally elevated and "*may not really come up with that idea* [*to*] *breathe at this moment*" (P11). Our simulation was also seen as fostering better transferability of these skills to real-life situations compared to interventions that remove individuals from the context of stress. Participants enjoyed the practical aspect of practicing deep breathing while "in" the stress-inducing environment, versus other technological interventions for self-care that either have no environmental immersion or use abstract or "escapist" (e.g., forest, beach) visuals:

> "*There's two types of things, right? There's one thing that you do on a daily basis just so that you don't get triggered as much... I feel like meditation falls in one of those, whereas deep breathing or stress balls or whatever people do – those fall into in-the-moment stress. And I feel like this is great for the second part... I could see myself using this, where I know a situation is probably going to trigger some stress for me. I would like to take all of those layers out as much as I can* [*by*] *using something like this before I actually enter into that situation*." (*P*12)

*Building Confidence and Preparedness.*

Participants also articulated that the non-guidance versions (e.g., **`VR-Interactive`**, **`AR-Interactive`**, **`Text-Interactive`**) felt useful for simply preparedness and confidence having already had (virtual) experience with a situation beforehand. P18 articulated that this practice "*has the use of getting over the initial nervousness of confrontation*" and helped "[*get*] *over anxiety and trepidation*."

P15 mentioned feeling more prepared after just using **`VR-Interactive`** without any mental health guidance: "*I would feel more prepared to actually confront them. The* [*avatar*] *showed me some attitude and I was kind of caught off-guard. If my roommate actually did that, I would have been prepared*." The technology also helped de-sensitize anxious situations as one participant notes regarding her experience coping with social anxiety through avoidance: "*you start avoiding places that make you uncomfortable... but some stress you can't avoid*" and described using the simulation as a "*choose-your-own-adventure*" tool to "*practice stepping out of your comfort zone*" and to help her "*reintroduce to social situations*" (P13).

### *Modality: Benefits and Potential Risks of Immersion and Realism*

We summarize below how participants responded to different modalities and allowances for immersion in our prototypes. Overall, participants expressed a desire for increased realism in the prototypes but also raised concerns about emotional safety.

*Preference for Augmented Reality*.

In terms of which modalities were preferred among the eight prototypes, participants generally preferred AR over VR or text-based prototypes. However, rather than a strict favorite, some saw value in multiple designs depending on context:

> "*If I'm out in a public scenario, and putting on the VR headset is a little much...then I started typing away. I think that's the best use for* [*the text-based chatbot*]. *For me, the AR one* [*if*] *I'm in the area in which I'm going to have the talk... The VR one* [*if*] *I'm inside, I'm alone*." (*P*9)
>
> "*Using the VR is like the initial part, just to only be able to focus on the conversation. The AR is the next step where you're in the space where the issue is...so* [*first*] *being able to practice in a controlled environment, which is the VR, then move to the AR*." (*P*18)

One notable consensus among participants regardless of their overall preferences was that AR was most useful when available in the actual locations where they anticipated stressors, such as the classroom they expected to give a talk in or their apartment if expecting a confrontation to occur in their home. For example, participants said that they'd "*use AR when it is close to where I will actually give a speech*" (P19) and practice stress-relief for a roommate conflict "*in my actual room...maybe I can breathe and walk around while I breathe*" (P9). Indeed, "*a lot of the confidence comes from just being comfortable in your space*" (P12) and practicing situations in the expected environment was already what some participants said they would do now if preparing for something like public speaking. This would also allow for people to practice their own physicality. For example, P9 felt AR would "*encourage me to move around a little bit, do hand gestures*" and, more than the dialogue, P17 saw the simulation as more useful for his "*physicality, you know moving around...engaging with people, which approaches I would do. I could train that.*"

Experiencing AR also opened up suggestions of different mental health guidance options according to people's personal needs, such as navigating their surroundings to reduce stress. For example, P11 felt he would use AR to practice in his real home being able to retreat to a different room to do deep breathing if overwhelmed at a social party. Similarly, people could also practice doing guidance rooted in visual stimuli in the environment as P12 mentions: "I know a few people who do struggle with social anxiety or confrontation. I remember somebody had mentioned to me that...there's a point in their room that they look at and that helps them just go into zone of 'I need to breathe three times'. [AR] could translate to that...you can just look at the same spot, and it should prompt the subconscious breathing."

*Environmental Realism*

Although past work in social simulation has focused on agent interactions to simulate a "real-world" setting (Park et al., 2022; Ren et al., 2014), we found that environmental factors are critical in participants' experience of social simulation; immersive simulation goes beyond just accurate interactions or behaviors. In particular, ambient sound and visual immersion were noticed and desired by participants. For example, multiple participants felt that the background sound in our prototypes was key for their feelings of presence ("*listening to the sound [of the party] made me anxious*" (P11), "*the sound was very important [for immersion]*" (P17)) and wanted even further sound immersion, such as emphasizing particular sounds (*"In real life, when you're stressed or anxious, certain sounds are emphasized, and it adds to the feeling of stress*" (P17)).

We saw the importance of environmental factors for realism when comparing users' reactions between AR/VR to the text-based simulations, which participants described as "*artificial*" (P11) and "*more like a tech simulation*" (P10) despite GPT prompting and dialogue style being consistent across VR, AR, and text-based interactions. P16 pointed out the stark difference between the visual experience of AR/VR versus text: "*I'm comparing it now to what I was experiencing during the AR/VR experience...it feels a little more clinical, removed. Obviously it's the same engine and it's the same conversation model, but the act of either being in AR or VR space and having, albeit rudimentary, a representation of another person does help the believability and the immersion.*" The disconnect participants felt in text-based simulations despite the same dialogue prompting as in VR/AR showed the importance of creating rich multisensory environments.

Participants consistently expressed a need for customizable avatars and environments to better reflect real-life scenarios: "*This scenario could be more personalized. For instance, if I could say, 'I'm attending this party tonight and meeting these specific people, and I'm nervous about it,' then maybe the conversations could reflect that*" (P14). Participants desired for simulations that not only felt plausible but were specifically relevant to their own lives. Even smaller physical differences, like positioning of avatars, could break immersion. For example, P12 pointed out that the roommate conflict simulation felt unnatural because the avatar stood facing across the user. She commented that she would not imagine this positioning for a conflict situation and desired the ability to customize this, stating, "'*do you want to be standing up or sitting down?' – maybe just as simple as that. The room that you have right now, I would choose to sit on the couch.*" Generally, we found that participants desired modification features for avatars and environmental details to match their real-world expectations.

*Risks in Realistic Simulation of Stress*

While participants believed our prototypes to generally be immersive, they wanted the simulation to produce higher stress levels to more closely emulate real life. Participants saw this as giving "*insights* [*on*] *how to handle or how I would*

*feel...Basically, it's a way to anticipate my stress or my anxiety in a real setting*" (P17) and avoiding getting false security since "[*if you*] *go to practice with this, and then you go for a real environment with real people – maybe they* [*will not*] *be so gentle with you*" (P7). P10 echoed this and surfaced a potential limitation with using GPT for interactive dialogue, saying, "*I went in with the impression that no matter what I responded with, there would be people bringing up other new topics and things that I could respond to. In real-life conversation you're also obligated to keep the conversation flowing.*" In fact, many participants correctly identified our dialogue engine as GPT-generated, and commented on GPT being unrealistically polite. Indeed, several participants chose not to use the available guidance during **`VR-Guided`**, citing a lack of stress. Although some felt guidance would be useful, they did not need it during the simulation: "*If the roommate would have persisted a little more, I would have probably* [*selected the breathing guidance*]*...maybe I just wasn't stressed enough in this situation*" (P12).

However, the risk of simulating high levels of stress is potential risk of trauma. For example, when P19 was asked whether she would like the simulation to be more realistic, she expressed hesitation and stated, "*I sometimes feel like I would be traumatized when I'm having conflict with somebody. I just want it to happen once. And I think practice will help to reduce the stress, but I kind of feel like maybe I'll get traumatized.*" Another participant noted, "*If somebody has a very deep fear of public speaking, this might be too realistic...confronting the fear head on, depending on the person, might help or might not*" (P5). Some participants felt lacking some realism created a safer environment for practice. P8 expressed that greater realism "*kind of scares me*" and P5 appreciated the current balance as helpful but safe: "*I do like that it's not completely realistic. It helps me remember that it's just a simulation and that it's more for mental health.*" The current level of (or lack of) realism allowed users to feel safer in verbal expression and more willing to engage in difficult scenarios. This allowed participants to feel it was "*okay to fail in the conversation*" (P14) and "*it's a safe environment... I know it's virtual* [*so*] *I am more relaxed...it will not have a repercussion*" (P17).

*Value in Low-Immersion Designs*

Despite a preference for immersive features, participants expressed value in low-immersion, text-based prototypes due to their accessibility and ease of use. The ease with which users could access text-based simulations made them a practical alternative for stress-relief practice in everyday settings. Participants mentioned that accessing chatbots "*might be more accessible for most users*" (P15), and "*in your everyday, I think this is way easier to pull up on your computer or your phone...like VR is something that most people either don't have or don't have on them always*" (P8). As a result, while environmental immersion was an important factor to people finding simulation helpful for practicing stress relief, participants still valued non-immersive but accessible systems.

## *Interactivity*

Interactivity through dialogue was preferred across the board among participants. When asked, all participants explicitly said they preferred **`VR-Interactive`** and **`AR-Interactive`** over the static counterparts due to the dialogue. However, despite people preferring interactivity, non-interactive designs were still seen as more valuable than current means of practicing stress relief.

*Interactive Dialogue*

Having interactive, realistic dialogue was key for participants in evoking accurate reactions and emotions. The interactive nature evoked genuine emotional responses, which participants felt surprised but pleased by, that fostered deeper engagement and self-realizations. For example, P11 said, "*I didn't really think the headset [could] make me very stressful...I was very surprised*", and the dialogue even evoked hesitancy akin to how participants would feel in real life:

> "*Even though it's a simulation, you could hear [my] stuttering and the hesitation. I was trying to still not go in with too much anger or irritation, but then also trying to be as direct with the problem.*" (*P*18)

> "*Even now, I was stumbling over my words and I wasn't really sure what I would say. So this is a perfect scenario, where I could think of how to approach the situation delicately.*" (*P*5)

Participants commented on enjoying how they were prompted to give realistic responses. P12, who engaged in an interpersonal conflict situation, stated that they enjoyed how "*it was not like they suggest the solution upfront because I feel like most of the times, it doesn't go that same way,*" and P10 felt questions received during the public speaking scenario were "*in-dept*h" and "*elicit a very thoughtful response*."

Beyond just realistic dialogue, participants emphasized that they desired more nonverbal cues such as facial expressions and body language due to its relevance in real life. P14 pointed out that "*nonverbal behavior is what's causing a lot of the anxieties as well,*" and P19 also felt so regarding public speaking, stating, "*I would prefer* [*the avatars to*] *have more facial expression when I answer some questions...some* [*audience members*] *will make some faces that make me really nervous...it also adds more stress to myself, but I would say it will make the scenario more realistic*."

All 19 participants favored dictation using our speech-to-text feature over typing on the VR keyboard due to ease of use and its similarity to real-life interactions. However, since our designs did not implement text-to-speech, nearly all participants expressed wanting audible voices from avatars. As P19 notes, avatars speaking out loud rather than our design of showing dialogue in avatars' text boxes would help give "*a shorter time to process the information*" to create "*more realistic interactions*" akin to real life.

*Movement and Interaction with Environment*

Movement and interaction with the environment were interactions that participants expressed wanting more of. As we reviewed previously, people enjoyed the thought of AR due to their closeness to the real environment they expected to encounter stress. Some of this desire came from expecting physical movement to play a significant role when feeling discomfort. People therefore felt that the fact they were stationary in our VR/AR prototypes was limiting. For example, P13 noted that their natural response to feeling socially anxious would be to move around, "*my impulse would be to walk to a new location within the space. A lot of people in a social situation, if they see a snack table will be like,* '*look, snacks*!' *and you run over there*." The ability for greater levels of interaction with the environment could help users also simulate different ways of handling stress, such as finding more comfortable spots in a room or occupying themselves with other objects. Additionally, allowing interaction with the avatars based on the simulation's environmental factors could help people practice icebreakers and conversational skills in a more natural way, as P13 points out about how socially anxious people may choose to converse when feeling nervous in social settings: "*it's the cat, or these plants are nice. You look for some shiny objects and start a conversation based on artwork in the room or something.*"

*Value in Non-Interactive Simulation*

Users enjoyed the realism and interactive aspects of the prototypes, as reviewed above. However, interestingly we found that participants still perceived value even in the prototypes without any interaction (**VR-Static** and **AR-Static**) as they were still perceived as more helpful than participants' current means. For some, this was due to the fact that they added any kind of visual to an already existing simulation people currently do in their heads to prepare for difficult situations ("*mirror the act of pulling it up in your head*" (P16)). Similarly, regarding **VR-Static**, P12 felt, "*I'm comparing it to practicing in the mirror...it's the same fundamentals. You're trying to get used to seeing things that you're not normally used to saying...it is good.*" The lack of realistic graphics was also not seen as necessary to still gain value from the technology for most participants: "*I don't think it's a problem that it's low poly or not necessarily realistic graphics. The act of being in this space and having a roommate character in front of you makes it into the psychology of being in that situation*" (P16).

### Guidance Needs

Although participants still felt that the prototypes without mental health guidance were useful, breathing guidance in **VR-Guided** and **AR-Guided** prototypes yielded an opportunity for **emotional regulation** and **more productive responses.**

Breathwork helped participants calm themselves and self-reflect on their emotional state. For instance, participants noted that breath work "*definitely made me calm down*" (P14) and helped "*mitigate some anger, stress*" (P9). Participants usually triggered breathwork when dialogue became more uncomfortable, such as a roommate becoming defensive or receiving a pressing question from an audience member. In doing so, participants were able to be "*more aware of the situation and* *gives me some time to actually watch what kind of emotions that I have in myself*" (P14) and that breathing during a more stressful encounter "*actually did a good job in helping manage mostly anger*" (P9).

Additionally, many participants also mentioned its usefulness for helping them respond more productively. For instance, participants stated "*with that breathing exercise, it will help me to be calm and think about the answer*" (P17) and "*being able to articulate it beforehand with a response and think, 'okay, maybe I should try saying this instead'*" (P18). Pausing during stressful moments gave a moment to "*recenter and get back to what I practiced*" (P18) as a healthier response to unexpected stress arising. Additionally, given participants' current methods of stress relief revolving around avoidance and distraction, it was particularly notable that participants felt that practicing breathing would help them to not immediately leave situations in unproductive ways. For example, P9 commented "*if I didn't press the breathing bubble, I would have just been like, 'okay, it's whatever' and then left when in reality, that's not going to resolve a situation. I think taking that extra time to breathe there was better.*"

*Transferable Skills to the Real-World*

While some participants felt the practice was good for habit-building for applying to real life (e.g., "*you can make it a practice...do it subconsciously so that it calms you down*" (P12)), other participants expressed that deep breathing was not necessarily a transferable practice due to its **impracticality during real-world situations.** As a result, we discovered that **transferability, practicality, and even social acceptance** were key dimensions that affect whether mental health guidance is perceived as useful. For instance, participants expressed hesitation on doing box breathing in real life:

> "*I do kind of worry that in real life people are not gonna wait for me to breathe and breathe out.*" (*P*14)
> "*People will think*: *why the long pause?*" (*P*6)
> "*I can't stop this conversation anytime without worrying about* [*the avatar's*] *feeling. So I can't just roughly* [*and*] *abruptly stop talking with him without answering his question.*" (*P*11)

As a result, people wanted and suggested different types of guidance that they thought were more relevant to their specific lives. For example, P19 suggested that her real-life calming mechanism during public speaking was seeing encouraging facial expressions from the audience and thus recommended simulating this. P18 mentioned techniques that felt easier to do in the middle of a conversation, such as "*closing your eyes and counting to a certain number and then gather your thoughts and recenter yourself*" (P18). Other participants simply wanted options, such as "*not only the breathing, but maybe like a couple other options I could click for how to relieve the stress*" (P9), or what they felt would be more relevant to the context: "*The assistant was being sarcastic and the guidance said to practice box breathing, but I don't think I needed box breathing for that. Maybe if they were being a little bit more aggressive, I would need box breathing*" (P15).

*Manual Guidance: Empowering Users' Agency and Self-Reliance*

We also investigated participants' preferences between GPT-timed breathing guidance and self-triggered guidance. One reason people preferred manual control was their confidence in recognizing their own stress. For example, one participant commented on how she understood her needs when the moment arose: "*It was kind of vague in the beginning whether I should turn [breathwork] on or not. But when the moment came that I felt, 'oh, maybe this is a good time to take a break'...I just made more [of a] conscious decision on when I need[ed] that*" (P14). Participants also felt manual control better reflected real-life decision-making, encouraged habit-building, and facilitated learning. As P9 explained, "*It's more realistic and applicable if I control it, because in real life I would be controlling when I do something like that,*" and P16 echoed, "*Realistically, you would be the one trying to execute your own calming techniques...I think you yourself are the best arbitrator of when something's stressful.*"

More explicit concerns over **intrusiveness** also arose. P9 commented, "*I wouldn't like it if...the simulation made me do the breathing thing, because maybe I didn't feel like I needed it. Or I felt like I could handle the situation without doing the breathing,*" while P7 similarly felt "*forcing someone in middle of the conversation might make things much worse...what if I don't want to take a break?*" These automated recommendations could even cause people to feel self-conscious; P14 encountered a suggestion for breathing guidance during **`Text-Guided`** and stated, "*I felt like I wasn't feeling stressed, but I think I got direction that I'm stressed...Like, am I stressed? When I'm not stressed, do I look stressed? I don't really like being told that I look stressed*."

Interestingly, participants also expressed a general distrust in the accuracy and universality of AI-driven mental health guidance. Even participants who felt GPT's recommendations accurate during the study remained skeptical about future scenarios. For example, P9, who believed the breathing suggestion would be useful, still had reservations: "I actually thought the system did pretty good on this one, I thought the situation was getting a little stressful...but I don't know if it will do that every time. But for this time, it did good."

*Automated Guidance*: *Enhancing Self-Awareness*

On the other hand, automated guidance helped users reflect on their emotional state and thus could lead to gaining self-awareness. While some participants expressed concerns about system-determined guidance, others appreciated its value in moments when stress might go unnoticed. P19 shared that she preferred LLM-powered guidance suggestions due to a lack of anticipation of her own stress, stating, "*When I was answering some of the questions previously, I did not feel like I'm stressed until I was stuck. Then I started to feel stressed, and that's something that I did not anticipate*." For these participants, automated guidance also was more intuitive to include in a training system such as ours in order to develop further self-awareness skills and allow users to consciously notice moments of stress they may have otherwise missed.

Overall, we found that users feel automated guidance can play a supportive role in helping users develop better self-awareness and emotional regulation skills. However, serious concerns over accuracy and agency also emerged. Perhaps due to the distrust of LLM-powered recommendations and its universality, multiple participants surprisingly suggested using **biosignals** to guide automatic recommendations instead. In subsequent interviews, when probed with the idea, all participants reacted positively to this idea. For example, P7 commented that "*I prefer to take charge. But I guess for such a system, tracking your heart or something for it to be accurate? Or maybe your voice tone*." and P16, another participant who suggested this idea, said, "*I like the idea of monitoring your vitals and signs of stress, maybe especially in a VR/AR setting. Having like a non-intrusive pop-up or gentle reminder*" when the user showed a signal like high heart rate. Integrating biosignal monitoring could also provide an alternative route to an accurate and personalized approach that still facilitates practicing self-awareness around stress management.

## 6.3 Discussion

This study explored user perceptions of applying social simulation technologies for practicing self-care techniques for their everyday situations. Through prototype-driven interviews with 19 participants using eight built prototypes, we gathered valuable insights into the benefits, risks and challenges, and future design implications for applying these technologies to self-care.

### *Designing Simulation for Everyday Self-Care*

As Nunes et al. note in a review of HCI technology for self-care, the concept of technological support for people's mental health should be expanded to center on people's everyday life experiences rather than just patients of chronic conditions (Nunes, 2015; Nunes et al., 2015). Our study contributes to the field of HCI technology for everyday self-care and explores what design needs exist in this relatively unexplored space.

*Modality*. One of the most significant findings was the clear preference for AR among our prototypes. Participants envisioned using AR in real-world environments to overlay both practice scenarios and mental health guidance, allowing for the direct application of coping strategies in the settings they encounter in daily life. In the context of everyday stress, where individuals are frequently exposed to environments that trigger emotional or mental distress, our interview findings suggest that AR lowers barriers in transferring skills to real-world settings and individuals prefer to practice coping techniques in spaces that closely resemble real-world settings. While high levels of immersion in VR/AR can foster engagement (Grinberg et al., 2014; Shin, 2019) and be beneficial for simulating scenarios that aren't immediately accessible, its separation from real-world contexts may limit the practical application of skills learned in the virtual space. Text-based prototypes, unsurprisingly, were considered the least immersive given absence of visual and auditory feedback. However, many participants said they would use text-based systems because of its availability throughout the day, suggesting that even non-immersive tools have a valuable role in the self-care space.

*Interaction*. Participants clearly favored features like dialogue and environmental interaction, as these allowed them to practice communication skills in a system-responsive manner. This interactivity provided more opportunities to experience realistic stress and practice self-soothing techniques in response. Notably, even non-interactive prototypes like **`VR-Static`** and **`AR-Static`** were considered more helpful than participants' current methods of stress relief, indicating that while interactivity is desired, it is not essential for perceived usefulness.

Participants' reactions to our prototypes echoed past findings (Abbasiantaeb et al., 2024; Farrahi & Clare, 2024) that LLMs are capable of generating natural, context-aware dialogue that participants said were realistic compared to real-world interactions. In the future, given the success of speech-to-text in our study, LLMs could be used to also dynamically design environments rather than just provide dialogue as it did in our study. Given the accuracy of our speech-to-text engine using WhisperAI, it is plausible that users could verbally describe anticipated situations, and the LLM could then generate corresponding dialogue, character personas, physical avatar appearances, and environmental features to create a VR or AR scene for the user to practice with. Since participants consistently expressed a desire for customization features to adjust simulations to reflect their personal environments and stressors, using LLMs for designing more aspects of the simulation could facilitate this capability while reducing the time, effort, and technical skill needed on the user's part.

*Mental Health Guidance*. As seen through our interviews in how people responded to box breathing and their original approaches to self-care in the initial parts of our interview, there is not a one-size-fits-all technique for in-the-moment stress relief. Overall, designs should opt for flexibility or adaptability in guidance. In terms of our specific system, though, participants believed the guided interventions to be useful during moments of heightened stress. However, an important theme that emerged was the transferability of such techniques to real-world interactions. Participants expressed concerns about the practicality of taking a moment to themselves during a heated conversation or stressful social event. Although one method of addressing these concerns is to offer techniques that are more easily transferable, this feedback might also suggest a greater opportunity to also teach people how to advocate for their own self-care needs in social settings and/or learn to transition into self-soothing skills during social settings. For example, instead of social simulation pausing the scene when the breathing guidance was triggered as in our study's designs, social simulation could also continue the interactions while guidance appeared and even simulate realistic ways that people may respond during these moments (e.g. facial expressions, asking a question again if the user does not respond in a timely way). By doing so, users can practice the interaction necessary to pause, transition into self-soothing techniques, or step away from a conversation to manage stress. While our prototypes focused on the concept of individuals learning these skills to incorporate into their own daily lives, our interviews also surfaced the significant role that social dynamics play in the application of self-care strategies and even the interpersonal consequences of doing so.

### *Ensuring Safety and Agency through Customization*

Although our simulation is designed as a safe environment to help users practice difficult situations, simulation can also carry risk in itself. The most apparent risk that emerged through our interviews was a tension between people wanting more realism, but also having concern over the potential of discomfort and trauma. Balancing realism and virtual imagery is crucial, as it impacts both user comfort and the risk of trauma. As one participant had noted in our study, stressful situations like interpersonal conflict can be difficult enough to deal with already just once in the real-world.

Our recommendation for giving users more control, minimizing surprises, and preventing stress responses beyond users' boundaries is to implement several **customization features** in the simulation. Customization features were mainly suggested by participants as contributing to realism, but that same control can ensure safety for users' boundaries as well. Customization of avatar and environmental features, for example, can be made more real or artificial according to users' preferences. Since sound was a key immersive element as well, simple features like volume control or the option to mute should be available at any time during simulation in order to reduce auditory immersion if needed. Additionally, a theme that emerged from our findings was *tailoring of difficulty level*, given that some participants wanted more challenging situations to practice with while others had concerns over trauma. Adjustments in mental healthcare to gradually lead up to more challenging situations are well-established in traditional therapeutic settings already, such as "gradual exposure" in Cognitive Behavioral Therapy (J. A. Cohen et al., 2012) and traditional Exposure Therapy (Craske et al., 2014). Challenges in dialogue, such as a roommate being more defensive or combative in our simulation, may be tailored to GPT, and further work may need to be done to find prompting techniques that generate appropriately easy versus difficult output that is appropriate and consistent for users. As a result, we strongly recommend simulation for self-care to provide control (and thus more safety) over specific features that may affect one's stress levels (e.g. amount of people at a party, volume levels) as well as difficulty when prompting LLM-powered dialogue. Overall, we find customization not only gives users agency, but also mitigate risks as users navigate their own boundaries.

As one participant suggested in our study, social simulation has potential to go beyond personal care and could be deployed in a clinical or therapeutic setting for supervised observance. Doing so may help therapists or clinicians see first-hand how their clients react in situations they encounter in life outside therapy, suggest coping techniques, and allow clients to reflect more accurately on feelings during situations they bring up in therapy. Although our work was originally built with a user's own individual, non-supervised learning and practice in mind, it is plausible that learning unsupervised may be best for people who already have personal understanding of their stress (Nunes et al., 2015). Different autonomy levels may be required for different "treatments" – while some care decisions such as those people make in our study's scenarios (e.g. public speaking, conflict) are done on a somewhat regular basis, other situations causing higher risk distress decisions should be supported and supervised by professionals.

We also note that participants had very positive reactions to incorporating biofeedback into the simulation system. Using biofeedback through heart rate monitoring or skin conductance measurements (Hickey et al., 2021) could inform the system when the user is becoming stressed (and even adjust the simulation accordingly) but also generally make the user more informed of their physiological state for them to gain more self-awareness. Past work has often applied HCI technology, including wearable devices and biofeedback, just for informative purposes for users' health as symptoms and trends can often go unnoticed (Nunes et al., 2015) and have become popular inclusions for mental health technology in recent years to help guide people to understand their habits and responses as well as the characteristics of their environment that contribute to their status. Given people's concerns over LLM-powered recommendations' accuracy but their desire for more self-awareness and reflection opportunities, future iterations of social simulation should consider using people's biofeedback to calibrate system responses and generally connect the user more with their own body's signals during mental health and self-care interventions.

### *Shifting Paradigms in HCI for Mental Health*

*Designing for Self-Care in the Real-World*. Prior literature that has tackled immersion for mental and emotional well-being have placed the user in the beach (Chandrasiri et al., 2020; Yildirim & O'Grady, 2020), gaming environments (Sra et al., 2018), abstract scenes (Ng & Caires, 2023), and outer space (Miller et al., 2023). Escapism has so far been the traditional way that HCI research has approached helping people find emotional relief, transporting users to a different landscape as a temporary escape from their current mental or emotional difficulties. Escapist approaches certainly have their place in stress reduction – studies about escapism in gaming for mental health have found that these approaches can indeed help with emotional regulation if not used to an extreme (Kosa & Uysal, 2020). However, there is also evidence to suggest that distraction, which was our participants' most common method of coping, can be effective as a short-term strategy for mood regulation but is not sustainable for long term well-being nor better for well-being than more active emotional regulation strategies (Broderick, 2005). Using escapism as coping can also cause more feelings of isolation, while methods to emotionally regulate directly (such as assessing emotions while in distress) can lead to better well-being (Kosa & Uysal, 2020). Our findings also suggest a significant gap and need for reflecting the real-world and its stressors in order to create applicable, transferable, and sustainable skills that we can apply without technological involvement in our everyday lives. In this way, our work attempts to shift the conversation from how to apply technology so it can **make people feel better**, to how to apply technology so it can **teach people how to feel better.** Realism played a crucial role in participants positive experiences during our study, and participants consistently favored tools that replicated real-world environments (e.g. AR, desiring more realistic body language interactions). Immersion enabled users to apply coping mechanisms in settings that mirrored the actual environments where stress arises, such as their homes or workplaces; participants felt that this design ultimately brought more confidence to encountering stressful situations as well as better transfer of emotional regulation and reflection skills to real-life situations.

*Reframing HCI for Self-Care*. Lastly, we discuss how we hope our work contributes to the broader landscape of HCI for mental health by framing self-care — and mental healthcare in general — as **fundamental needs for everyone**, not just those with diagnosable, chronic conditions. Traditionally, self-care technologies in HCI have been positioned within the framework of clinical interventions or therapeutic support, often focusing on structured tools for chronic conditions like depression or anxiety (Lucock et al., 2011; Nunes et al., 2015). However, as our findings highlight, there is a growing need to reframe self-care in HCI to create tools for everyday practice; in particular, we aimed to use HCI methods to equip users with the skills to manage stress in real-time for the contexts where it occurs. Our work's findings and implications also emphasize the idea of people's own **self-efficacy** in taking care of themselves. As seen in our interviews, people want to be self-sufficient but do not currently engage in means to do so when it comes to mental health, despite it being crucial for our well-being just as much (and maybe even more) as social or professional support needs. In our study, participants were able to engage in a simulation that they felt gave agency to learning self-care skills in the long-term to enact on their own in the real-world.

In addition, self-care in HCI has traditionally focused on reactive strategies to help people manage stress after it occurs or self-management and tracking for specific conditions and diseases (Hilgart et al., 2012; Nicholas et al., 2017; Nunes et al., 2015; Søgaard Neilsen & Wilson, 2019). However, our study suggests that HCI research can go beyond these framings to instead foster **proactive self-care** by allowing users to practice personal coping skills before they enter stressful situations. This shifts self-care from a purely reactive measure to an ongoing skill-building process, where users can refine their emotional regulation techniques over time. HCI's role in supporting this proactive approach is critical, as it positions technology not just as a fix but as an integrated part of daily life that empowers users to actively manage their mental health on their own terms.

### *Ethics Considerations*

In this project, we strive to keep in mind the risks and unknowns surrounding LLM use in mental health technology. Given the recent rise of LLMs and simulation capabilities relevant to the mental health domain, we also include discussion here on three highlighted risks: bias in LLMs, misinformation in health information, and data safety.

First, we acknowledge the potential bias that could occur in simulations such as ours, especially given known biases present in LLMs. As a result, LLMs may inadvertently reproduce or even amplify stereotypes or harmful assumptions, posing risks to users, especially those from marginalized or underrepresented groups. In the context of our study, these biases could manifest in the simulated conversations, potentially leading to interactions that are inappropriate, distressing, or counterproductive for the intended stress-relief purposes. These challenges highlight the importance of ongoing scrutiny, testing, and calibration of LLMs to minimize potential harm, as well as the need for user-centered designs that allow participants to flag or opt out of problematic interactions. As a result of many unknowns and known risks with LLM applications to mental health, we have chosen not to implement features that participants in the previous design study expressed wanting of – for example, specifying demographic qualities of avatars (e.g., gender, race). Instead, avatar appearances are randomized using URM (explained later in this chapter) and no demographic information regarding the avatars or user are fed into LLM prompts for generating the simulation dialogue. While this is not a perfect solution, we believe this mitigates some of the issues that may appear around stereotyping and incorrect assumptions generated using large language models when they are fed demographic information \cite{}. Although this means less customization on users' parts, we believe that further investigation to mitigate these biases are required to be able to allow these features responsibly. Currently, we as scholars do not understand the short- and long-term risks of these types of features, including on the primary user themselves but also on others around them that may be simulated through the users' choice. However, preventing demographic information being fed into GPT still does not address the fact that GPT4 can still output biased content in its reference to people, events, etc. even without any particular information input by the user. Given that we are only developing this system for public speaking for which LLMs are used to ask questions regarding the topic at hand, we believe that this choice of scenario offers less opportunity for LLM avatars to output biased conversational content that may be more present in, for example, more personal scenarios like parties, friendship, and conflict situations. That being said, any future development of this technology to a larger scale will absolutely require more safeguards and research around how LLMs – even in scenarios like public speaking – can amplify or generate harmful bias.

Second, both real-world events and research highlight the lack of safeguards surrounding LLM-driven mental health recommendations. For example, direct interactions with LLMs for mental health interventions have resulted in inappropriate and harmful advice (Wells, 2023) and may fail at providing nuanced advice needed for many mental health-related topics \cite{}. There are significant unknowns—as well as documented harms—associated with using LLM-driven interventions, which demand extreme caution when integrating LLMs into the sensitive context of mental health. Consequently, consistent with our previous work (Fang et al., 2024), we avoid using LLMs to provide mental health guidance to participants. Instead, our focus is on leveraging LLMs solely for generating realistic dialogue to simulate authentic, real-world interactions, an area where LLMs have demonstrated effectiveness \cite{}. Mental health guidance in our study is not derived from GPT or any LLM. Instead, we rely on a research team to carefully curate and hard-code self-soothing techniques, informed by a comprehensive literature review of evidence-based practices in traditional therapeutic settings.

Lastly, we would also like to discuss the misuse and potential data privacy issues that can occur with simulation regarding mental health. Our study does not record or log any user responses within the simulation given the sensitive and vulnerable nature of simulation for personal mental health practices; we only record the *time* that the participants use the system rather than any content recording. We recommend that researchers in this space continue to not engage in data collection regarding users' activities during simulation if not necessary, or do so only with appropriate, thorough, and informed consent from users. We also note that analysis and collection of user experiences goes beyond just direct collection; implicit inference given analyzing behavioral patterns or linguistic choices by users could also offer sensitive, vulnerable, and undisclosed insights about a person's mental state. Applying simulation to the especially sensitive context of mental health highlights the need for prioritizing consent, data anonymization, transparency, and continuous attention to ethical research design.

# *Chapter 7*

# Measuring the Effects of Contextual Environments in Augmented Reality for Mental Health Practice

In the previous chapter, I described my work in "*Social Simulation for Everyday Self-Care*: *Design Insights from Leveraging VR, AR, and LLMs for Practicing Stress Relief*". We found that while most individuals experience stress regularly, they feel that they lack effective strategies and preparatory means in order to manage this stress. Our design prototypes using XR simulation to embed self-care strategies into emulating real-world stress situations garnered overwhelmingly positive reactions for its perceived ability to address these challenges, providing a virtual space for people to practice navigating stressful situations and in-the-moment soothing techniques like breathwork.

However, it remains unclear how social simulation actually quantifiably affects how people learn, apply, and transfer the learned skills from the training application into the real-world. While participants feel that embodied social simulation is a useful tool for helping them feel safe in experimentation, the measurable effects of using social simulation compared to interventions without social simulation remain unknown. In this chapter, I describe my follow-up study that explores the effects of using social simulation to train mental health skills.

## 7.1 Introduction

As we previously reviewed, the past several years has seen immense growth in HCI designing immersive, embodied, and interactive tools to support mental health and well-being outside the traditional medical setting. For example, virtual and augmented reality (VR/AR) applications provide embodied environments that can help people experience calming or therapeutic settings (Freeman et al., 2017; Riches et al., 2024), while advances in sensing and biofeedback have enabled the integration of users' real physiological and behavioral cues into mental health interventions (Daudén Roquet et al., 2022; Mitsea et al., 2024; Pera, 2022). This trend in HCI is informed by research in psychology, education, and health, which has all demonstrated that embodied and immersive interaction can foster more engaging and adaptive skill development (Marshall et al., 2003; L. Xie et al., 2008; Z. L. Xie, 2008). Building on these foundations, HCI research for the mental health space has used a range of interactive visual, auditory, and tactile stimuli in order to enhance presence, support behavior change, and improve users' well-being.

However, despite the numerous benefits of existing immersive technology for mental health such as greater engagement, better learning outcomes, and more enjoyment from users (Freeman et al., 2017; X. Wang et al., 2022), these interventions largely remain focused on short-term relief and aiding users by presenting calm (Chandrasiri et al., 2020b; Miner et al., 2024), otherworldly spaces (Miller, Stepanova, Desnoyers-Stewart, Adhikari, Kitson, Pennefather, Quesnel, Brauns, Friedl-Werner, Stahn, et al., 2023), abstract (Ng & Caires, 2023), and gamified environments (S.-Y. Wang et al., 2023). These settings, though effective for immediate stress reduction, are almost always disconnected from the everyday contexts in which mental distress arises — e.g., at work, in our personal relationships, at social events. As a result, users may struggle to actually apply these strategies when confronted with real-world challenges, where distress may be inevitable but ongoing access to professional or non-professional support is rarely available (Fang et al., 2025; *Stress in America* 2023: *A Nation Recovering from Collective Trauma*, 2023).

This highlights a key gap between interventions that teach self-care skills through calming experiences and the practical challenges of day-to-day stress in socially situational contexts. Research in psychology (Tulving & Thomson, 1973) and applications across domains like nursing and healthcare (Drews & Bakdash, 2013; Rutherford-Hemming, 2012), navigation (Ferguson & Rice, 2001; Wrisberg & Liu, 1991), and sports (Michalski et al., 2019) using have all supported the notion that learning techniques in realistic contexts improves skill transfer, recall, and confidence in applying skills during real-world contexts. The contexts in which we navigate mental health distress similarly involve various constraints of social acceptability and environmental limitation that immersive interventions rarely capture (Fang et al., 2025; Ross & Mirowsky, 1984). Some growing HCI and design literature have begun to now propose the importance of situating mental health learning in direct, real-world stress contexts — thus making the transfer of these skills to the real-world less difficult and nebulous (Fang et al., 2025; Nunes et al., 2015a; Nunes & Fitzpatrick, 2018; Wagener et al., 2025). This work has primarily been theoretical through conceptual frameworks of designing future self-care technology (Nunes et al., 2015a; Thieme et al., 2015; Wagener et al., 2025) and related qualitative studies have found people to indeed perceive contextual practice as important for their real-world coping (Fang et al., 2025). Yet, an empirical and systematic basis for whether practicing in what we will refer to as a social simulation — the simulation of real-world conditions in which a skill, behavior, or activity is typically performed — actually affects people's ability to learn, apply, and transfer well-being skills remains unknown.

As a result, our work builds upon this historical gap and recent prior work to empirically measure how situating practice in social simulations actually affects users' ability to apply and sustain mental health skills. Our study presents the first empirical evidence of the effects of practicing mental health techniques within an embodied and social simulation, as compared to conventional isolated or decontextualized environments. In this paper, we explore:

**RQ: How does practicing mental health skills in a virtual social simulation affect the application and transfer of skills to the real-world, compared to practicing in a non-social simulation?**

We present findings from a two-week in-the-wild experiment study with 43 participants who use an augmented reality application that teaches calming techniques for the scenario of public speaking. Our results show that participants who practice mental health techniques using a contextual simulation environment not only showed a slower increase in heartbeat stress responses to a mock in-person speaking task, but also both quantitatively and qualitatively show greater propensity towards using self-care techniques in their daily lives outside the intervention. Our findings suggest that situating virtual self-care practice in realistic contexts can enhance various dimensions of effectiveness and longer-term incorporation of mental health skills.

By presenting this work, we build on HCI research for developing mental health technologies by exploring the potential benefits and limitations of contextual simulation in VR/AR applications for teaching coping strategies. We provide the first empirical evidence that practicing self-soothing in a simulated stress environment improves both physiological and experiential outcomes, and provide various opportunities and limitations for the future design of virtual self-care systems.

## 7.2 Hypotheses

Given our research question and prior study (Chapter 6), we formed five hypotheses to measure how realistic and contextual simulation in an embodied environment affects people's (1) **engagement**, (2) **change in stress**, both state and longer term anticipated stress, (3) **effectiveness** of applying skills, and (4) **propensity for transferring learned techniques** to the real-world. Below, we review these hypotheses along with relevant prior work that guide their creation.

First, a growing body of HCI research suggests that the effectiveness of training or care tools depends heavily on how relevant and meaningful they feel to users (Jardine et al., 2024; Rapp & Boldi, 2023; Slovak et al., 2023). Interventions that remain too abstract or detached from everyday life risk being perceived as less engaging and less useful (Fang et al., 2025), reducing adherence and long-term impact. In contrast, simulations that mirror the types of stressors people actually encounter could increase users' motivation to practice, given urgency and need in their real-world lives. Building on this foundation, we investigate whether practicing self-care skills in the context of simulated real-world stressors leads to greater engagement with the practice system. We hypothesize that:

**Hypothesis 1** *People who practice self-care skills in a virtual social simulation will use the system more frequently and for longer, compared to those who practice in a non-social simulation.*

Second, we evaluate how using social simulations influences participants' stress levels. State, rather than trait, anxiety is defined as "*dependent upon both the person* (*trait anxiety*) *and the stressful situation*" (Endler & Kocovski, 2001) and is often the assessment measure for fear triggered by specific situations such as public speaking (Poeschl, 2017). Thus, we hypothesize on how social simulation affects state stress measures. Intuitively, adding a stress environment like in contextual, embodied environments could lead to even greater stress in the immediate aftermath of the exposure, even without particularly realistic representation (Fang et al., 2025); on the other hand, the system may lead to increased confidence for the user in that they can practice self-soothing themselves in the face of the stressor. Given the intervention is episodic and studies generally do not show significant results in a short-term period immediately after a simulation intervention, there is no evidence to suggest that there will be state stress differences (Brasil et al., 2021). However, given previous work that realistic environmental stressors when practicing will result in better preparation and skill acquisition, we hypothesize that participants will indeed report lower anticipated stress over time towards public speaking, compared to their baseline stress when starting the study.

**Hypothesis 2** *People who practice self-care skills in a virtual social simulation will not show any significant differences in their state stress measures on a day-to-day basis, compared to those who practice in a non-social simulation.*

**Hypothesis 3** *People who practice self-care skills in a virtual social simulation will report significantly lower anticipated stress towards public speaking compared to their initial stress levels, compared to those who practice in a non-social simulation.*

Next, we investigate more distal effects such as effectiveness when applying skills and propensity for transferring skills to the real-world. Prior work has found that realistic simulation (both in digital and non-digital ways) is valuable for both immediate and longer-term learning outcomes for applying skills in the real-world (Lateef, 2010; Nestel et al., 2011). We build on this to measure whether practice in a social simulation leads users applying these skills more effectively in real life situations. We hypothesize that:

**Hypothesis 4** *People who practice self-care skills in a virtual social simulation will display lower levels of stress when enacting the mental health skills in real life, compared to those who practice in a non-social simulation.*

Research in psychology and learning sciences has consistently highlighted the importance of situated practice for promoting the transfer of skills from training environments to real-world contexts (Drews & Bakdash, 2013; Michalski et al., 2019; Rutherford-Hemming, 2012; Tulving & Thomson, 1973; Wrisberg & Liu, 1991). When learners practice skills in contexts that approximate the physical, social, or emotional constraints of real-life situations, they are more likely to encode relevant cues and adapt strategies effectively when faced with similar challenges outside of the training environment. Building on this, we hypothesize that:

**Hypothesis 5** *People who practice self-care skills in a virtual social simulation will show higher propensity for practicing and applying these skills in their real-world lives, compared to those who practice in a non-social simulation.*

## 7.3 Methods

In this section, we describe our methods in three stages:

First, we describe our development of an augmented reality application consisting of two versions: control and treatment}, which participants are randomly assigned to. We start by overviewing the two systems that form our experiment groups to test the above hypotheses and review key design principles that inform how we developed the application. (System Design)

Next, we review the experiment design including the process by which participants interacted with their assigned version of the application for a required period of 7-days, followed by an optional "in-the-wild" period of 7-days. We detail the participant recruitment criteria, stages of the user study, and participants' required tasks. During and after the participants' 14-day period with the system, we collected day-to-day insights on their self-reported stress levels, use of the AR system, and transfer of the included mental health skills in the real-world. (Experiment Design)

Finally, we review the analysis process including an overview of measures collected and our analysis methods used. (Analysis)

### *System Design*

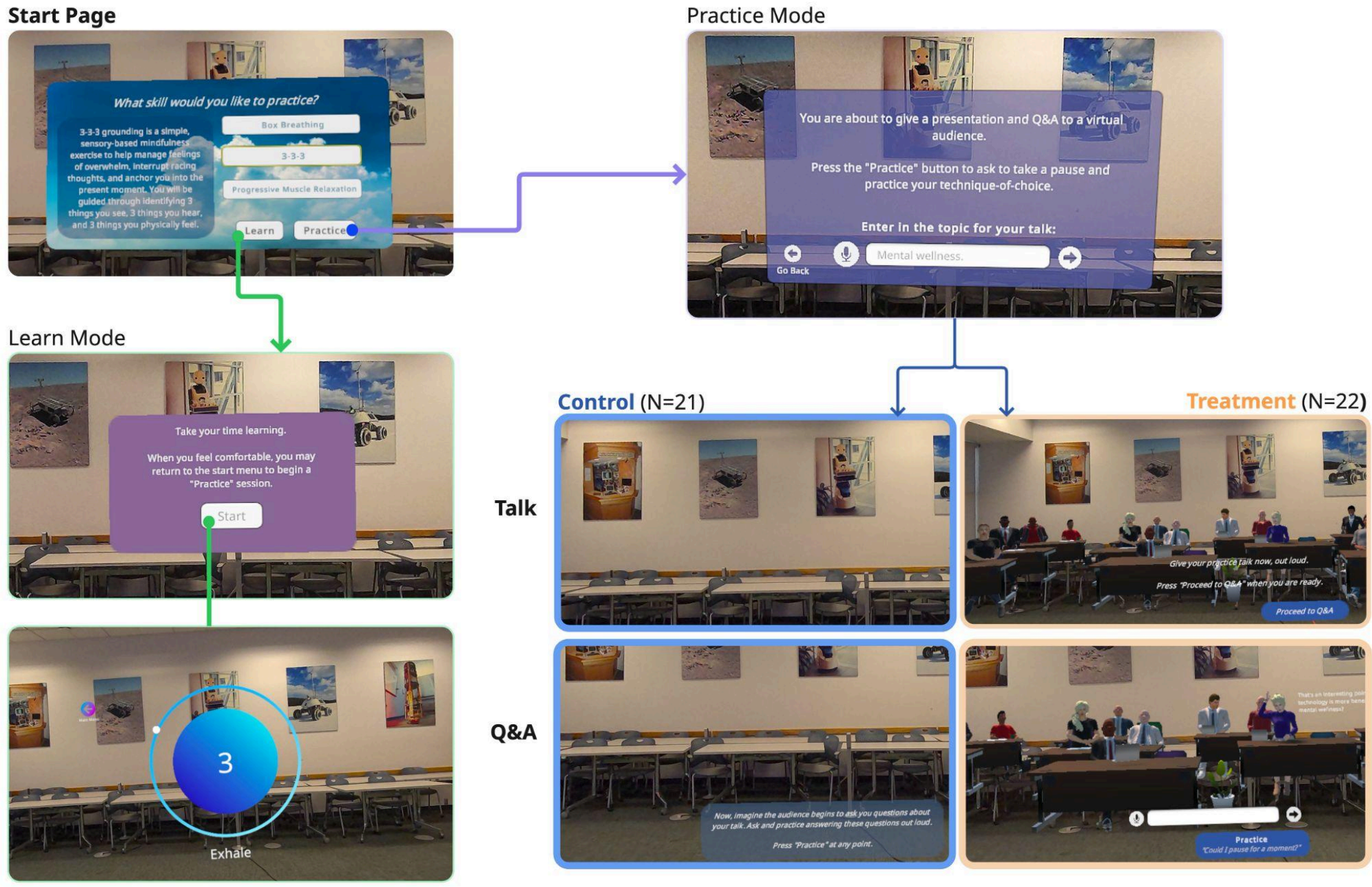

**Figure 15.** *User flow through the control and treatment versions of the augmented reality system. Screenshots are taken from a classroom on our research team's campus.*

To examine the influences of realistic environments on the practice of self-care, we built an augmented reality application for the Meta Quest 3. Although distress triggers vary from person-to-person, we designed the application to focus on just one type of scenario in order to control across participants: public speaking}.

Public speaking is the most common fear (Dwyer & Davidson, 2012) and especially relevant to our recruited population; for example, 64% of college students experience fear of public-speaking (Marinho et al., 2017). Public speaking is also a context that has been of particular interest for embodied technology, and evokes similar levels of fear and arousal from people as in-person engagements (Chanwimalueang et al., 2016; La et al., 2025; Slater et al., 1999). As a result, we designed the application to act as a practice tool for people to tackle mental health distress while giving a public talk as well as a following Q&A session from an audience.

To evaluate our hypotheses, we developed an augmented reality application. Below, we describe the full system followed by the distinguishing factors between two versions of the system for our experiment: control and treatment. See Figure 15 for a complete user flow with snapshots of a typical user interaction in the application.

### *Overview of System*

We begin by describing the user flow for the application.

Users begin by selecting one of three mental health skills they would like to practice: (1) Box Breathing, (2) 3-3-3, or (3) Progressive Muscle Relaxation. Upon hover, users see a brief description of each of these techniques as well as its intended outcomes. For example, hovering over Box Breathing shows: "Box Breathing is a simple and effective breathing exercise that promotes relaxation and calms the nervous system. You will be visually guided with an expanding and contracting orb through a controlled, rhythmic pattern.". All methods (i.e. box breathing, 3-3-3, progressive muscle relaxation) that our team chose to be included in the application have been found to be effective at reducing stress, anxiety, and/or panic as well as are fairly easy for beginners to learn (Carlson & Hoyle, 1993; Gan et al., 2022; Nunes-Harwitt, 2018; F. F.-Y. Tan et al., 2023b).

After the user selects one of the three methods, they may select one of two modes: Learn and Practice.

**Learn Mode.** For the selected method, Learn Mode allows the user to familiarize themselves with the method without any stressor stimuli. Learn Mode superimposes visual and auditory guidance for practicing the technique, at the center of the attention of the user. For Learn Mode}, Box Breathing consists of an expanding and contracting orb that guides the user through a countdown timer for inhaling, holding the breath, exhaling, and holding at the bottom of the breath for four seconds each; 3-3-3 contains text and audio guidance on identifying objects, sounds, and feelings in the user's physical world; and progressive muscle relaxation guides the user through text and audio for relaxing different parts of the body from feet to the head while the user simultaneously follows an inhale-exhale breath pattern through a soundbite guiding their breath through audio. In order to help the user focus and feel the calming nature of these techniques in a non-distracted atmosphere, all Learn Mode techniques also have the same soft music playing in the background that is sourced from free online media. Learn Mode intentionally emulates the design of current self-care applications on market, such as meditation apps.

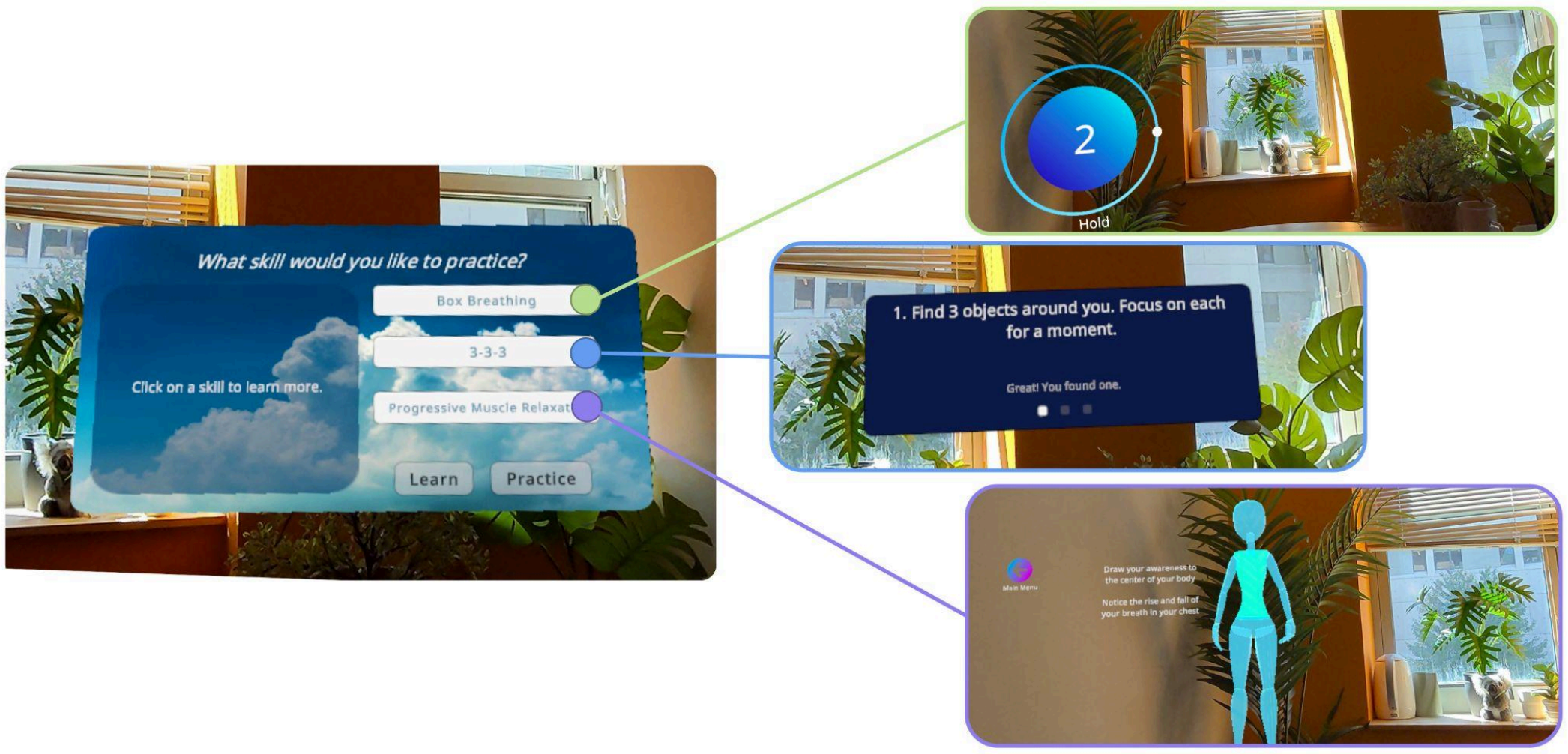

**Figure 16.** *Screenshots of the three mental health skills present in our system: Box Breathing (top), 3-3-3 grounding (middle), and Progressive Muscle Relaxation (bottom).*

**Practice Mode.** In Practice Mode, users are prompted through both text and spoken voice to give a talk out loud. Since the application is in augmented rather than virtual reality, users can also refer to their own notes and presentation slides if wanted. In the treatment condition, the system begins a visual and auditory simulation of a classroom/meeting room environment filled with audience members. Avatars are seated in the space, accompanied by soft background noises (e.g., shuffling and typing). After the talk, users proceed to a Q&A session in which virtual audience avatars take turns asking questions, and participants can reply verbally through a speech module (Figure 15). Throughout Practice Mode, participants can press a button to bring up their selected mental health skill (box breathing, 3-3-3, progressive muscle relaxation) in an abbreviated version that appears in the corner of the screen if they wish to practice during their talk or Q&A.

**Control vs. Treatment Versions.** We create two versions of the above system. What we will refer to as the control version of our system is the same as the treatment and full version of the system, but with the social simulation (virtual audience, augmented classroom objects) removed. Instead of facing avatars and environmental audio, control participants were prompted to imagine an audience during both the speech and Q&A. On the other hand, treatment does include the full system features as described above including the contextual audience. Both versions share the same Learn Mode, the speaking Q&A structure of Practice Mode, and the ability to deploy self-soothing skills mid-task, but differ in the presence of a simulated audience environment.

### *Design Principles and Implementation*

Below, we overview implementation details guiding design of both control and treatment systems.

**Modality.** We built an augmented reality system to create an immersive environment that could validly examine how users differ in practicing in their normal surroundings versus a virtually simulated social simulation. We built the application using the Unity game engine and deployed on a Meta Quest 3 headset. We decided to build the environment in augmented reality rather than virtual reality due to previous work that has found it to be effective in simulating socially-related stress as users can remain situated in physical surroundings where stress might arise, such as classrooms or meeting rooms (Fang et al., 2025).

**Environment.** For treatment, we present participants with a simulated audience. All avatars were created and customized using Unity Multipurpose Avatar 2 (UMA 2), given their relatively greater realism compared to other avatar packages and customizable features that allowed our team to create unique and diverse characters with appropriate clothing for the scenario (e.g., business to business casual). Additionally, we looped the sound of soft classroom and hallway noises to add audio realistic to a typical speaking engagement environment.

**Interaction.** Dialogue for the avatars was generated using GPT-4o. We chose to power avatar dialogue using LLMs given its ability to generate realistic, responsive, and context-specific replies to the user that more closely emulate real-world interaction. For the treatment condition, we used a simple prompting pattern (see Appendix) for avatars to switch turns asking the participant questions on their talk. To ensure natural exchanges, the first and third authors iteratively refined prompting patterns, eliciting diverse tones and question types including those that were less deferential. Dialogue was presented as both on-screen text bubbles and voices, with voices generated through OpenAI's AudioAI text-to-speech system. Participants interacted verbally using a dialogue system that we designed with simple microphone input, which we transcribed using OpenAI's Whisper. Avatars were assigned behaviors including head nods, posture shifts, and sitting positions.

**Mental Health Skill Development.** Across both versions, participants could select from three techniques: Box Breathing, 3-3-3 grounding, and Progressive Muscle Relaxation. These methods were chosen as beginner-friendly, evidence-based strategies for reducing distress (Carlson & Hoyle, 1993; Gan et al., 2022; Nunes-Harwitt, 2018; F. F.-Y. Tan et al., 2023b). Their design drew on prior literature—for example, breathing guidance used cool colors (Lukic et al., 2021) and abstract expansion imagery (F. F.-Y. Tan et al., 2023b). Participants in both control and treatment could access their chosen skill at any point during Practice Mode, ensuring strategies were available during stressful moments. This mirrored real-world coping, allowing us to study how users incorporated techniques under stress — either with a simulation of a social simulation (treatment) or an imaginative environment (control).

### *Experiment Design*

We conducted a 43-person between-subject user study. Participants were randomly assigned to the control or treatment systems described above. Below, we describe participant recruitment and criteria, followed by each of the four stages of the study.

#### *Participants*

To ensure safety of our Meta Quest 3 equipment given the large number of participants and in-the-wild nature of the study, we recruited out of our academic institution in an urban area of Mid-Atlantic USA. All participants had to sign up via their institutional email address, and so are either undergraduate or graduate students, staff, or faculty at our research team's institution. We recruited participants using physical flyers, social media, institutional mailing lists, and snowball sampling.

**Table 15.** *Participant demographics by condition. Initial SUDS is on the self-report scale of* 1 *to* 10.

| Treatment Group | | | | | Control Group | | | | |
|---|---|---|---|---|---|---|---|---|---|
| **ID** | **Gender** | **Race** | **Age** | **SUDS** | **ID** | **Gender** | **Race** | **Age** | **SUDS** |
| 1 | Woman | Asian | 25–29 | 7 | 3 | Woman | Asian | 18–24 | 5 |
| 2 | Man | White | 18–24 | 6 | 4 | Woman | Asian | 18–24 | 4 |
| 5 | Man | Hispanic | 18–24 | 5 | 6 | Man | Asian | 18–24 | 4 |
| 7 | Woman | Asian | 25–29 | 7 | 9 | Woman | Asian | 18–24 | 6 |
| 8 | Man | Asian | 18–24 | 4 | 11 | Woman | Asian | 25–29 | 5 |
| 10 | Man | Asian | 18–24 | 3 | 12 | Woman | Black | 18–24 | 5 |
| 17 | Man | White | 18–24 | 5 | 13 | Man | Black | 18–24 | 3 |
| 18 | Woman | White | 18–24 | 3 | 14 | Man | Hispanic | 25–29 | 8 |
| 19 | Man | Asian | 18–24 | 4 | 15 | Woman | Asian | 18–24 | 5 |
| 20 | Man | Asian | 25–29 | 3 | 16 | Man | Asian | 18–24 | 6 |
| 22 | Woman | Asian | 18–24 | 5 | 21 | Woman | Asian | 18–24 | 7 |
| 26 | Woman | Black | 18–24 | 7 | 23 | Woman | Asian | 18–24 | 5 |
| 28 | Woman | Asian | 18–24 | 8 | 24 | Woman | Asian | 25–29 | 3 |
| 30 | Woman | Asian | 25–29 | 2 | 25 | Man | Asian | 25–29 | 3 |
| 32 | Woman | Asian | 25–29 | 2 | 27 | Woman | White | 25–29 | 5 |
| 33 | Man | Asian | 25–29 | 2 | 29 | Woman | Asian | 18–24 | 4 |
| 34 | Woman | Asian | 18–24 | 5 | 31 | Man | Black | 25–29 | 5 |
| 36 | Man | Asian | 25–29 | 7 | 35 | Woman | Asian | 25–29 | 3 |
| 37 | Man | Black | 30–34 | 5 | 38 | Nonbinary | White | 18–24 | 2 |
| 39 | Man | Asian | 35–39 | 7 | 41 | Woman | Asian | 18–24 | 7 |
| 40 | Man | Hispanic | 18–24 | 5 | 43 | Man | Asian | 18–24 | 3 |
| 42 | Woman | Asian | 18–24 | 3 | | | | | |

Given that we are not focused on self-care for chronic conditions, we recruited a non-clinical population. We define ‘clinical’ and ‘non-clinical’ populations based on prior literature [5,57,267] by recruiting participants who explicitly do not have any diagnosis of anxiety or stress disorders and are not engaged in any professional treatment (e.g., medication, professional therapy) to minimize confounding variables with our study's intervention. Upon filling out our interest form, participants are asked to self-score themselves on the Subjective Units of Distress (SUDs) ("*On a scale of, how stressful or anxious do you anticipate yourself to be when engaging in public speaking?*") from 0 to 10, which is an established rating system in psychology (Papapetropoulos et al., 2023; Wolpe, 1969). We recruited participants who reported themselves to have at least mild public speaking anxiety (a score of 2 or higher) on the 10-point scale; we do not recruit participants who report themselves to have extremely severe public speaking anxiety (9 or 10 points) on their self-report, for participants' psychological safety reasons. We also recruit individuals who confirm that they have at least one upcoming speaking engagement in the next two months. This approach allows us to better understand practice and use of self-soothing techniques for participants who find the public speaking application relevant and generates some amount of real-world stress.

In the end, we recruited 44 participants with 1 drop out (due to unrelated personal circumstance). Participants were compensated $50 USD through gift card or Zelle. This study was approved by the appropriate Institutional Review Board (IRB). All studies were conducted from June through August 2025.

*Stage 0: Onboarding (Day 0)*

Recruited participants visited our research site to receive study details, provide baseline data, and be randomly assigned to the control or treatment condition. During the 30-minute in-person onboarding session, participants were asked to fill out a consent form and provide self-report ratings for their familiarity with VR/AR and various self-soothing techniques (*"Rate your familiarity with (virtual and augmented reality technology / box or deep breathing techniques / 3-3-3 or grounding techniques / muscle relaxation or body scan techniques)"*) on a Likert scale of 1 (Not at all familiar) to 5 (Extremely familiar) (Vagias, 2006). Participants were given a Meta Quest 3 headset, which had the appropriate (control vs treatment) system according to randomized assignment. Finally, our research team members engaged in a 5-10 minute tutorial session to familiarize participants with the system controls.

*Stage 1: Free Use Period (Days 1 through 7)*

Participants were instructed to use the system every day for the next seven days for at least 10-minutes each day. We did not record the microphone or input content of the participants for privacy reasons, but did record local logs of button presses during their session so we could verify and analyze their daily use. Each day, the participants were instructed to fill out a survey, consisting of the same questions for before and after the participant's application use for the day:

- *Have you used any self-soothing skills learned from the AR application in your real life (outside of the headset) in the past 24-hours?*
- *(Before/After using the application) What is your stress level at the current moment?*

- (*Before/After using the application*) *What is your anticipated stress level, if you had to conduct public speaking in real life*?

*Stage* 2: *Free Use Period* (*Days* 8 *through* 14)

For the second week of the study, in order for us to understand participants' free use of the application and incorporation of the application into their daily lives in a non-study context, participants kept the Quest 3 and were told they could optionally use the system to their own volition. Participants did not have any required daily task during this stage but, similarly to Stage 2, on days of their use we again record logs of their use and survey responses for before/after use.

*Stage* 3: *Assessment and Semi-Structured Interview* (*Day* 15)

After two weeks, participants returned to the research site for a mock public speaking task in order for us to measure differences in physiological responses between participants when doing tasks in real life. We also conducted semi-structured interviews to gain qualitative insight.

We drew inspiration from prior literature (Douglas et al., 2022) and asked participants to engage in a public speaking task to our research team. All Stage 4 tasks were conducted by the first two authors. Upon first entry to our room on campus, we asked the participants to wear a FitBit Versa for heart rate measurement. We then let the participant settle into a chair while we recorded participants' resting baseline BPM. After the researchers and participant settled, we then asked the participants to imagine being interviewed in a room of executives. We asked the participant to then speak for roughly five minutes to answer the question: "*What are your best and worst qualities*?". We recorded their heart rate before they began answering the question, during the course of their answer, and after they finished speaking. We then asked participants to conduct a second task of conducting a mental health skill taught in the AR system (box breathing, 3-3-3, progressive muscle relaxation) for roughly two minutes. We recorded their heart rate before, during, and after their completion of the self-soothing task.

After the two tasks, we conducted 30-minute semi-structured interviews with all participants. The interviews were guided by a list of questions but allowed to deviate depending on topics participants may introduce. Questions included probing on participants experiences of stress over the last few weeks, methods used to tackle these situations, and transfer of skills (e.g., "*What did you do, if anything, to deal with the feelings of stress or anxiety in that situation?*", "*Since using the AR program, have you noticed yourself being reminded of or using any of the mental health skills in real life?*"). We asked about participant experiences with the application itself and their experiences navigating practice sessions (e.g., "*How did you decide, if at all, to pause during your speaking or Q&A session?*"). We voice-recorded all discussions and analyzed transcripts using a thematic analysis.

### *Analysis*

We first summarize and review the measures and outcomes we gathered from the user study, as a preface for the analysis then conducted.

- In Stage 1, we gathered participants' demographics, self-reported familiarity with VR/AR technology and various mental health skills, and their initial SUDS score for public speaking.
- For Stages 2 and 3, we gathered their daily survey information and logs of use, including duration and frequency of application use.
- Finally, in Stage 4 we collected participants' heart rate data during a mock in-person speaking task and self-soothing task, as well as qualitative interviews.

For all measures relevant to our hypotheses, we ran statistical tests and regressions to understand how treatment — along with other covariates — affected (1) engagement with the system through frequency of practice sessions and application use, (2) effectiveness of applying skills through analyzing heart rate changes during the mock speaking task, (3) change in self-report state stress and (4) anticipated stress}, and (5) propensity for transferring learned techniques to the real-world through participants' self-report of using skills in their lives outside the application. We also analyzed qualitative data through a thematic analysis. Our research team conducted thematic analysis through anonymizing, transcribing, and coding all 43 interview transcripts. Based on Braun and Clarke's method of thematic analysis (Braun & Clarke, 2012), our iterative analytic cycle consisted of: (1) recording and transcribing interviews (2) coding transcripts (3) amalgamating codes (4) discussing codes (5) highlighting themes (6) writing and revising memos.

## 7.4 Results

To evaluate our hypotheses, we measured the effect of treatment and control assignments for usage statistics (H1: Usage), self-report measures of stress (H2, H3: Self-Report Stress), effectiveness of applying skills (H4: Physiological Stress), and transfer of learned skills to participants' real lives (H5: Transferability). We assessed baseline balance between randomized groups on demographics and prior familiarity measures; we found no significant imbalances between treatment and control on any baseline variable (all $p > .05$).

### *H1: Usage*

***Table 16.*** *Regression models for: (a) log-transformed total application uses, (b) raw count of practice sessions (subset of total uses), and (c) length of use in minutes.*

| Variable | (a) Total Sessions (log) *β (SE)* | (b) Practice Sessions *β (SE)* | (c) Length per Session *β (SE)* |
|---|---|---|---|
| (Intercept) | **1.629**** (0.462) | -4.940 (4.769) | **8.980***** (1.362) |
| Treatment | -0.512 (0.720) | 7.490 (7.428) | **5.997**** (2.093) |
| Initial SUDS | 0.107 (0.081) | **1.983*** (0.834) | 0.413 (0.229) |
| Familiarity w/ VR/AR | 0.009 (0.090) | 0.173 (0.930) | 0.465 (0.278) |
| Familiarity w/ Breathing | -0.225 (0.146) | -1.664 (1.506) | -0.610 (0.423) |
| Familiarity w/ Muscle Relax. | 0.028 (0.147) | -0.870 (1.512) | 0.328 (0.409) |
| Familiarity w/ Grounding | **0.451*** (0.164) | **8.227***** (1.691) | -0.113 (0.384) |
| Treatment × SUDS | 0.011 (0.104) | -1.137 (1.077) | -0.190 (0.313) |
| Treatment × Fam. VR/AR | 0.070 (0.160) | -0.000 (1.647) | **-1.015*** (0.447) |
| Treatment × Fam. Breathing | **0.462*** (0.218) | 3.215 (2.253) | 1.003 (0.694) |
| Treatment × Fam. Muscle | -0.261 (0.223) | -1.025 (2.296) | -0.982 (0.751) |

| Treatment × Fam. Grounding | -0.435 (0.265) | **-7.958**** (2.733) | -0.444 (0.673) |
|---|---|---|---|
| **Multiple $R^2$** | **0.404** | **0.601** | **0.110** |
| **Adjusted $R^2$** | **0.382** | **0.455** | **0.077** |

**Hypothesis 1** *People who use a virtual simulation of a realistic environment to train self-care skills will use the system more frequently and for longer (including voluntarily).*

**Overall, we found H1 to be partially supported.**

We calculated the number of sessions that participants completed across their 14-day period. We define a session as starting when the participant launches the application and completes at least one Learn or Practice session. If the application enters sleep mode (either due to inactivity or because the participant manually puts the headset to sleep), subsequent use is deemed as a new session.

Upon comparison to the log model residuals, the residuals of the raw count of sessions already appeared to be more normally distributed, but its model explained a significantly smaller fraction of variance (adjusted $R^2 = 0.07$) compared to modeling the log-transformed outcome (adjusted $R^2 = 0.24$). Given that log-transforming the outcome improved model fit, although residuals were slightly less symmetric, we report results using the log-transformed outcome as the primary model.

As seen in Table 16, we found that the effect of treatment on the total number of sessions over the course of the study did not show significance ($\beta = -0.51$, $p = 0.48$). For the length (minutes) of use per session by participants, though, we found that treatment participants on average used the application for roughly 6-minutes more per day compared to control participants ($\beta = 5.99$, $p = 0.004$). As a result, we overall find that H1 is partially supported: results indicate that treatment alone did not meaningfully increase participants' number of application uses, but it did increase how long a user spent on a session. When we ran a similar regression model for predicting the number of days that participants continued to use the system during Stage 2: Free Use Period, we did not see any significant results.

Interestingly, though, we found that participants' previous familiarity with the technology and/or techniques were strong predictors of their usage behaviors. Familiarity with grounding techniques was positively associated with the number of sessions ($\beta = 0.45$, $p = 0.01$) as was the treatment by breathing familiarity interaction ($\beta = 0.46$, $p = 0.043$). Treatment effects are roughly 58\% higher for participants who had one unit higher familiarity with breathing techniques at the start of the study. This booster effect may be due to these users being already at higher propensity to use such systems given existing personal interest. Conversely, for session length, the treatment effect was smaller for participants with higher VR/AR familiarity. Intuitively, this may simply be a result of those less familiar with the technology needing to spend more time on interface navigation.

Additionally, upon a deeper look into participants' specific use statistics for Learn and Practice modes in the application, we found that results for the number of Practice sessions echoes these previous results. Only interaction effects between treatment and previous familiarity with mental health techniques showing significance while the main effect of treatment showing no significance ($\beta = 7.49$, $p = 0.32$). However, increased SUDS scores had a small positive effect ($\beta = 1.98$, $p = 0.024$); for a control participant, each 1-point increase in SUDS at baseline is related to an increase of ~2 additional practice sessions, but treatment does not significantly change this effect. Higher familiarity with grounding techniques was strongly associated with more sessions ($\beta = 8.23$, $p < 0.001$); however, the interaction between treatment and grounding familiarity was negative and significant ($\beta = -7.96$, $p = 0.007$), suggesting that treatment effects were smaller for participants already familiar with grounding.

*Qualitative Findings*

In terms of learning and applying mental health skills from technological tools, participants generally found the application to be more effective for learning skills than other self-care technologies like meditation apps or video tutorials; however, both control and treatment participants described using Learn Mode only for the first days of the study given that techniques were easy to learn. Treatment participants consistently emphasized that Practice Mode was the more useful feature due to environmental immersion ("*The best way to evoke a certain emotion in* *a situation is to put the person in that situation*" (P5), "70% *I would use the app* [*in the future*], *because it's helped me a lot to present and reflect...it helped me a lot to get used to the atmosphere*" (P7), "*Before, I used to practice by just saying out loud to myself, but I feel like that's different. Simulating that experience presenting on the spot, I feel like that was good to practice with*" (P26)). Although mixed, control participants also expressed positivity towards the tool versus traditional methods of learning self-care techniques, such as P16 who commented, "*It's immersive, it's just easier for me to perceive* [*in AR*]. *I often find a lot of situations where I learned something from online, from a video, but it's hard for me to process those steps even if the content creator thinks they are already slow enough, detailed enough. It's not just enough...but VR put me in that position*".

However, some also felt that VR/AR was similarly effective to video or other tutorials while being less accessible given the inconvenience of the headset. Many control participants expressed that the VR/AR environment needed additional contextual features to feel useful, similar to the social simulation that treatment participants were in. For example, P4 (control) said, "*I'm not a good imaginative person, I can't imagine myself in this situation. But with AR, it could be helpful and able to put me in that situation*" and P25 (control) similarly expressed, "*I do wish it had more applicability ... I feel like, if it could collaborate with me, if it could give me,* "*okay, this is it in a public speaking pattern*". *I would use that application then*".

### *H2, H3: Self-Report Stress Measures*

**Hypothesis 2** *People who practice self-care skills in a virtual social simulation will not show any significant differences in their state stress measures on a day-to-day basis, compared to those who practice in a non-social simulation.*

**Hypothesis 3** *People who practice self-care skills in a virtual social simulation will report significantly lower anticipated stress towards public speaking compared to their initial stress levels, compared to those who practice in a non-social simulation.*

We ran a mixed-effect multi-level regression model to examine how treatment affected participants' daily self-reports of their feelings of stress. **We found that H2 was supported, while H3 was not supported by quantitative results.**

**Table 17.** *Linear mixed-effects models predicting session-by-session stress differences: (a) state stress differences and (b) anticipated stress differences toward public speaking. Both models include fixed effects for treatment, mean-centered day, baseline stress, and their interactions, as well as a random intercept for participant ID.*

| Variable | (a) State Stress Change β (SE) | (b) Anticipated Stress Change β (SE) |
|---|---|---|
| (Intercept) | **-0.589*** ** (0.115) | **-1.058*** ** (0.211) |
| Treatment | 0.170 (0.164) | 0.202 (0.302) |
| Day (centered) | 0.037 (0.025) | 0.080 (0.043) |
| Starting SUDS (centered) | -0.038 (0.074) | -0.214 (0.136) |
| Treatment × Day (centered) | -0.038 (0.035) | -0.074 (0.061) |
| Treatment × Starting SUDS | 0.080 (0.098) | 0.062 (0.181) |

***p < .001, *p < .01, p < .05

Results of the model are shown in Table 17, where we calculate the state and anticipated stress change as the outcome — i.e. participants' stress levels post-application use, minus their level pre-application use. The model included fixed effects for the treatment group (intervention vs. control), mean-centered day of the study for which the self-report was filled out, mean-centered baseline stress, and their interactions. A random intercept for participant ID accounted for repeated measures across sessions. We chose not to include familiarity variables, gender, race, or age to maintain parsimony and given preliminary models indicating that these demographic variables did not improve the model or significantly alter treatment estimates.

Our findings show support for H2 but not H3. For the state stress model (H2), participants in treatment condition indeed did not show any reliably significant differences in how their state stress changed before to after application use. No other variables showed significance either. On the other hand, the anticipated stress model — which measures how treatment affects how users change in their anticipated stress towards real-life speaking events after using the application — found that there were no statistically significant results for any predictors.

*Qualitative Findings*

To understand these findings, we turned to qualitative results from the participant interviews. Upon being asked about feelings using the application, treatment participants' indeed mentioned that engaging with the system in some cases either did not change or even raised their level of stress (both state and anticipated). Participants stated that their practice sessions in the headset surfaced their hesitancy or lack of preparation for their talk, or consisted of difficult questions from audience members; these resulted in greater awareness of weaknesses in their preparation and self-consciousness, although this was not necessarily brought up as a negative outcome. For example, P10 (treatment) reflected, "it has been a long time since I spoke in public at all. When I first used the app, I was like, whoa. I'm kind of rushing". Similarly, P8 noted, "*Due to the questions that got deeper and deeper, that would make me a little bit nervous. In terms of learning, it's a positive experience. In terms of self confidence wise, it was a little bit worse". Participants also described the simulation as physiologically realistic, sparking the same nerves they might experience in real-world public speaking: "I actually did feel my heart rate rise. And I was like, what's going on? This is fake. But in the moment, this is simulating what's actually going to be like in real life*" (P17, treatment).

Interestingly, though, we found that both control and treatment participants expressed that the system gave them tools and "resources to call on" (P36, control) to deal with stress when it inevitably arises in the future, even if it did not actually reduce their anticipated stress. For example, P37 (treatment) stated, "I have some good tools in my toolkit now to apply if I ever get too wound up" while P27 (control) stated, *“I guess I’m more confident now because I didn’t have any good tools before — I just accepted my fate*". Similarly, P10 (treatment) explained, *“I don’t know if I’m going to be less stressed about public speaking, but even if I do feel stressed, I’ll be able to manage that well with the techniques*". Echoing, P39 (treatment) stated: "*Because of the practice I got, I feel I will be slightly more confident. I still don't feel that I'm good at public speaking. But I would at least be able to give it without panicking about it*".

Although we did not find significant differences in anticipated stress measures between control and treatment, one key distinction that emerged through the interviews was how treatment participants described a reframing of stress specifically tied to social and judgment}-related aspects of public speaking, which we did not see mentioned by control participants. This may reflect treatment participants' exposure to a virtual audience, which gives them more opportunities to confront the contextual and social dimensions. For instance, P30 (treatment) explained that they felt less afraid to answer questions, stating, *“It made me approach the whole thing better because I would always dread the Q&A session post-talk... In one of the tries* [*in the application*] *I just said ‘I don’t know the answer to this’ just to see what happens...we moved on to the next question...and I was like, guess what? I can do this in the real world too. When I’m asked something, I can always say, ‘listen, I don’t have the answer to this today*'". Similarly, P26 (treatment) reflected on their anxiety towards any silence in their talk and said, *“I feel like using this app has made me more aware of the fact that I can pause, like during a poster presentation...the app kind of reminded me that I can take a pause and take a breath sometimes, if I get nervous when I’m speaking*".

### *H4: Physiological Stress*

**Hypothesis 4:** *People who use a virtual simulation of a realistic environment to train self-care skills will display lower levels of stress when enacting the mental health skills in real life.*

We analyzed how participants perform when doing a mock in-person speaking task for 5-minutes to our research team as described in Experiment Design. We recorded their heart rate at baseline, during the speaking task, after completing the task, and during a self-soothing activity immediately following. Overall, we found that Hypothesis 4 was supported by our results as treatment participants exhibited lower rise in heart rate during the stress task and faster recovery rate during the self-soothing task.

***Table* 18.** *Mixed-effects models predicting heart rate for (a) trajectories across the full speaking task, (b) pre-peak heart rate analysis during speaking, and (c) heart rate recovery during the calming task. All models include a random effect for participants and predictors: treatment (binary), time in seconds, resting heart-rate, and treatment X time interaction.*

| **Variable** | **(a) Heart Rate During Speaking Task** β (SE) | |
|---|---|---|
| (Intercept) | **52.31*** | **(17.31)** |
| Treatment | -0.037 | (3.451) |
| Time (sec) | **0.053***** | **(0.002)** |
| Time $(sec)^2$ | **-.0003***** | **(2.2e-5)** |
| Resting BPM | 0.537 | (0.219) |
| Total time (min.) of app use | 0.089 | (0.084) |
| Treatment × Time | **-0.034***** | **(0.002)** |
| Treatment x $Time^2$ | 0.0001* | (3e-05) |

***p < .001, *p < .01, p < .05

| **Variable** | **(b) Pre-Peak Heart Rate During Speaking** β (SE) | | **(c) Post-Peak Heart Rate During Speaking** β (SE) | |
|---|---|---|---|---|
| (Intercept) | 35.59 | (20.31) | **81.55**** | (2.787) |
| Treatment | 2.703 | (3.80) | -1.53 | (5.236) |
| Time (sec) | **0.131***** | (0.003) | **-0.061***** | (0.003) |
| Resting BPM | **0.623*** | (0.258) | 0.355 | (0.352) |
| Treatment × Time | **-0.033***** | (0.004) | **0.022***** | (0.004) |

***p < .001, *p < .01, p < .05

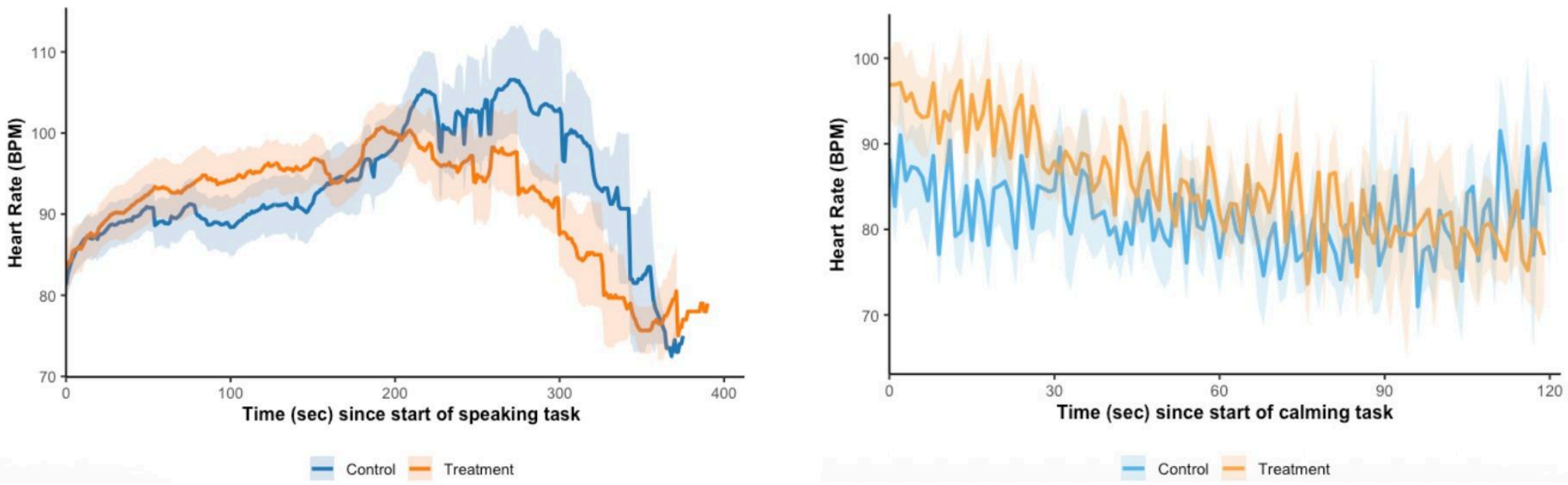

**Figure 17.** *Average heart rate trajectories over time for participants in the speaking task (left) and during a calming technique task (right). Lines represent group means for Control (blue and light blue) and Treatment (orange and light orange), with shaded ribbons indicating ±1 standard error. The plots show the difference in heart-rate rise between treatment and control groups across time.*

Using a linear mixed-effects model with participant-level random intercepts and accounting for heart rate measurement at every second over the course of the task, we examined the effects of treatment and its interaction with time (seconds elapsed during the 5-minute speaking task) as well as control variables of initial resting heart rate. We also ran various models using demographic variables and previous familiarities as covariates, but these resulted in minimal effect and no significant results for main or interaction effects with treatment. When analyzing participants' heart rate changes, we found that the treatment group exhibited a slower heart rate escalation during the mock speaking task compared to the control group (see Table 17). In particular, treatment and time interaction was significant ($\beta$ = -0.033 BPM (beats-per-minute)/sec, $p < 0.001$), indicating that **treatment substantially slowed the rise in heart rate during the task.** On average, treatment participants' heart rate increased by approximately 0.02 BPM per second, compared to 0.05 BPM per second for control participants over the 5-minute task. Given results around the slope of heart rate, we also analyzed how participants' heart rate changed leading up to the peak of their heart rate during the task. In other words, we exclude any heart rate changes during a "recovery" phase, during which participants tended to become less stressed near the later part of the task. On average, the peak time of heart rate occurred at 143 seconds for control participants and 141 seconds for treatment participants, with a t-test showing no significant difference in time to peak ($p = 0.918$). Using a linear mixed-effects model similar to previously, results indicated a significant main effect of time and a significant treatment and time interaction, suggesting that treatment indeed slowed the rate of heart rate increase. Resting heart rate was also a significant covariate, indicating a higher resting rate was associated with a greater rise in pre-peak heart rate trajectory as well. Overall, treatment significantly moderated the slope of heart rate rise both throughout the speaking task and leading up to participants' greatest physiological stress point, even though peak heart rate itself did not differ between groups.

However, analyses of heart rate variability measures (RMSSD and SDNN) showed no significant differences between treatment and control groups throughout the stress task, suggesting that the treatment effect is primarily on the rate of heart rate escalation rather than variability. Model results are shown in Appendix. Given that HRV is typically done with millisecond-level recordings, this finding likely reflects some limited resolution of our heart rate recordings.

***Table* 19.** *Linear models predicting heart rate recovery during the calming task.*

| Variable | Heart Rate During Calming Task β (SE) |
|---|---|
| (Intercept) | **99.08*** (21.14) |
| Treatment | 5.490 (3.96) |
| Time (sec) | **-0.043***** (0.004) |
| Resting BPM | -0.187 (0.268) |
| Treatment × Time | **-0.020**** (0.007) |

***p < .001, *p < .01, p < .05

During the calming task, all participants' heart rate tended to decrease over time. The linear mixed-effects model (see Table 19) revealed a significant treatment and time interaction ($\beta$ = -0.020 BPM/sec, $p$ = 0.003), indicating slightly faster recovery for treatment participants. Neither resting rate nor the main effect of treatment were significant. These results indicate that while all participants showed physiological recovery during the calming task, those in the treatment group demonstrated a modestly accelerated decline in heart rate and suggesting that treatment helped participants also improve their ability to self-soothe in response to a real-world stressor.

For all above models, we found no statistical significance that length of use mediated treatment effect, and treatment itself didn't have a direct effect on heart-rate.

### *H5: Transferability*

**Hypothesis 5** *People who practice self-care skills in a virtual social simulation will show higher propensity for practicing and applying these skills in their real-world lives, compared to those who practice in a non-social simulation.*

**Table 20.** *Predicts total number of reported uses of mental health skills in real life, per participant. Random intercepts for participants included.*

| Variable | **Number of reported uses of mental health skills** $\beta$ (SE) |
|---|---|
| (Intercept) | 1.475 (0.824) |
| Treatment | -2.554 (1.520) |
| Initial SUDS (centered) | 0.314 (0.242) |
| Opportunities for use (binary) | -1.492 (0.918) |
| Treatment x Initial SUDS | -0.345 (0.304) |
| Treatment x Opportunities for use | **3.341*** (1.614) |

***p < .001, *p < .01, p < .05

**Table 21.** *Linear mixed-effects model for total number of application uses by participants during the optional use period in the second week of the study. Random intercepts for participants are included.*

| Variable | Number of voluntary application uses β (SE) |
|---|---|
| (Intercept) | 0.100 (1.859) |
| Treatment | -0.062 (0.838) |
| Initial SUDS (centered) | -0.698 (0.470) |
| Familiarity: Breathing | -0.167 (0.439) |
| Familiarity: Grounding | -0.464 (0.553) |
| Familiarity: Muscle | 0.086 (0.389) |
| Treatment x Initial SUDS (cent.) | 0.205 (0.549) |

***p < .001, *p < .01, p < .05

**We found both quantitative and qualitative evidence to support Hypothesis 5. Participants who used social simulation (treatment) showed a higher rate (~2.2 times) of using the mental health skills when presented with opportunities to do so, compared to control (~1 time) over the 14-day period.**

We ran a negative binomial mixed-effects model on whether participants used learned mental health skills in their real life, self-reported every day during the study. Participants could indicate each day whether they used a technique presented in the intervention (box breathing, 3-3-3, and/or muscle relaxation).

We centered numeric predictors (day, starting SUDS score) and initially fitted generalized linear mixed models for a binary outcome of whether a participant used a technique present in the intervention ("relevant technique") on a given day. Additionally, we coded given participant post-study interviews whether they indicated (yes/no) they had a stressful event in the previous two weeks of the study that was an "opportunity for use" for a stress-reduction technique. Since our day-by-day analysis showed no statistically significant effects of treatment likely due to high variability within days and limited statistical power, we analyzed aggregate data across the first seven days of the study to determine treatment's effect on the total count that a participant completed either relevant or irrelevant techniques during the study. We included covariates of baseline stress toward public speaking as well as initial familiarities with the techniques presented in the study, given that participants may be more likely to use the techniques if they came into the study with familiarity. Given the count outcome and preliminary analysis showing overdispersion, we modeled the data using a negative binomial regression model.

Results are shown in Table 20. We found that treatment participants showed significantly higher rates of using mental health skills in their real lives *when they report having opportunities to do so*. Specifically, treatment participants, aggregated over the course of the study, used 2.7x times relevant techniques in their real lives outside the headset, compared to the control group. Overall, these models suggest that the treatment increases technique transfer to participants' real lives for techniques specifically included in the intervention when presented with stressful in-person events, therefore supporting Hypothesis 5.

*Qualitative Insight*

Treatment participants' qualitative perspectives echoed the quantitative results, reporting substantially higher rates of applying skills in real life. Several participants described adapting techniques in variations or customized to their particular circumstances such as P26 (treatment) who stated, "*I wasn't necessarily using the breathing techniques that were on the app, but I was doing something similar*" while others described increased awareness of breath (P5), direct application during presentations (P30, P17, P7, P40), or more general daily use (P10, P22, P19). Treatment participants applied skills during public speaking events that they had during the study span, such as:

> "*I did use one in real life recently when I was giving a presentation for my internship...I actually took a minute to stop and breathe just because I was being* *asked a ton of questions all at once and I was struggling to keep up. It was like a team of eight and I was the one presenting, so I took a minute.*" (*P*30, *treatment*)

"*I had two public speaking opportunities during this* [*research study*] *span. There were times I was like, Oh, maybe I can take a second, especially at the start...a big point of the study that I took away was that I did use these techniques almost in real time...I was subconsciously at a poster presentation and I got a little riled up. Let me breathe for a second. So that was cool to see it play out in real time.*" (*P*17, *treatment*)

"*I actually did some breathing because I was super stressed. Public speaking is a big deal for me*" (*P*40, *treatment*)

as well as in non-speaking related engagements:

"*Sometimes, if I'm in a conversation and I'm just listening, I'll randomly just start doing* [3-3-3]." (P22, treatment)

"*Just random times, in the morning, when I reached the office, maybe at night. I would just do box breathing. Or if I'm just sitting idle, I do the muscle relaxation. Those wouldn't signify a stressful moment. It was just because I used the app and got used to the techniques, I was just like, Okay, why not just do the technique more?*" (P10, treatment)

"*When I'm just walking down to work, just breathing. For sort of mentally resetting before work or after work,* [*box breathing*] *has helped a lot*" (P19, treatment)

"*I woke up and saw a rat. The first thing I did was take deep breaths. Because I was terrified. I wouldn't say it was exactly like in the practice system but it was very close. I took a couple of deep breaths, calmed myself down. It was like a minute of complete silence and breathing. And it really helped calm me down*" (P30, treatment)

Surprisingly, despite the lack of social simulation in the intervention, many control participants actually noted that the skills were closely tied to the scenario of public speaking in the headset; as a result, it felt difficult to transfer skills beyond the intervention. For example, P6 stated, "*it's like, closely associated to a task for me, you know, like I have to do this before public speaking*" while P9 said, "*I think I kind of separate the headset from the rest of my life. I didn't really incorporate the two together, and just thought of it as a separate thing.*" Similarly, P16 stated, "*I don't connect it to the real life yet*". Other control participants stated that they were not able to transfer the skills to real-world stress situations due to the, counter-intuitively, stress they felt in those moments. For example, P38 stated, "*I feel like this week has been really stressful...it's the time that I need stress relief the most, but then I neglected during these times*".

In terms of barriers to enacting the skills in their real lives, though, both treatment and control participants described challenges of time and concern over appearing awkward in front of audiences. As P12 (control) explained, "*Real life, you're not gonna be given that chance to try to calm yourself down, Somebody's gonna look at you like, what are you doing*?". As a result, our findings echo (Fang et al., 2025) in that the social norms were still taboo for participants to feel consistently comfortable in using for self-soothing reasons; as we discussed in Section 5.2.1, though, treatment participants expressed that the intervention helped reframe some social dimensions. It is possible that this is unique to the public speaking application, and other scenarios such as interpersonal conflict with close relationships or self-soothing for general stress could allow people to take pauses more liberally. However, we found that several participants found "workarounds" for applying techniques even within the public atmosphere of public speaking, such as shortening the visible time spent on techniques and using skills before/after speaking rather than integrated within. For example, P10 (treatment) described, "*In a real life situation where I am giving a talk, I don't think I would take* 10--20 *seconds off to do box breathing...But I definitely do think that I'll be more confident in asking people, Hey, can I take a few seconds off...I would be pretty comfortable using it, but for how long I use it, I probably keep it to a minimum*".

## 7.5 Discussion

Our study provides empirical evidence that realistic social simulation in augmented reality can meaningfully shape how people learn and apply self-soothing skills for public speaking. Across measures of usage, stress, physiological response, and transfer, the intervention demonstrated both opportunities and limitations on how social simulations matter in the design of embodied mental health technology.

### *Opportunities in VR/AR Simulation for Self-Care Skills*

First and foremost, our findings demonstrate that embedding self-care training within a realistic social stressor environment improved several dimensions of mental health skill training: **transfer of self-care skills**, **physiological resilience and recovery** for real-world stressors, and **depth of engagement** with the system itself. Quantitative evidence supported that participants who used the AR system with a social simulation used a greater level of relevant self-soothing skills in their day-to-day life, and qualitative results showed that this transfer of skills even extended beyond the particular application of public speaking. When participants underwent a mock public speaking task in-person with our research team, physiological outcomes provided further support for the benefits of contextualized training as they experienced slower heart rate increases during stressful speaking tasks and more efficient recovery when practicing calming techniques. As a result, we find that social simulations can benefit the application and generalization of self-care skills.

Additionally, we found qualitative support that participants who practice skills using social simulations were more likely to reframe how they anticipated approaching stress in the future; interestingly, participants who practiced with the social simulation (in this case, public speaking) did not seem to be limit the scope of applying skills to only public speaking, but rather expressed many different generalizations of these skills to other aspects of daily life. For HCI, this highlights a design opportunity: contrary to intuition, using an “anchor scenario” not only trains coping in that domain but may also scaffold broader transfer}, and perhaps even more so than the practice of a mental health skill in a generalized but isolated virtual environment. Designers may wish to select application contexts that are common, evocative, and relatable, enabling users to feel the usefulness of skills but also abstract strategies into their broader lives.

One key challenge our study surfaces is how to design interventions that deliberately resist awe or spectacle. While fantastical XR environments are undoubtedly effective for engagement and can also lead to profound mental health effects (Miller, Stepanova, Desnoyers-Stewart, Adhikari, Kitson, Pennefather, Quesnel, Brauns, Friedl-Werner, & Stahn, 2023), our findings and other prior work (Fang et al., 2025; Nunes et al., 2015b; Nunes & Fitzpatrick, 2018) suggest that technology for mundane self-care can help people create tangible change in their everyday well-being. Designing for the mundane may raise the issue of engagement; although our quantitative findings found the social simulation to lead to longer time of use compared to no social simulation, indeed a few participants in our study suggested various elements of gamification (e.g., P5 who stated they wanted it to be "*more like a video game*", P16 who suggested "*reward incentives*") for the system that they felt would help them continue use — arguably, features that may also run antithetical to the concept of designing for the practical and mundane everyday. One promising design direction for future research may be the incorporation of "reflective" incentives. Rather than points, badges, or streaks, interventions can highlight meaningful progress (e.g., showing users how their stress response has changed over time, prompting continued short reflections on users' lived challenges) or lightweight check-ins that align with mundane rhythms (before a class, during a commute, after a meeting) so that practice feels embedded in daily life rather than as an extra task. Of course, other options can create varied experiences to avoid user boredom like customization features (e.g., changing audience sizes, social dynamics) or difficulty levels, which still preserve relevance.

### *Social Barriers in Adoption of Self-Care Skills from AR Simulation*

While participants found value in the AR system, they also articulated barriers that complicate the adoption of self-soothing in everyday contexts. A central theme was the role of social pressure}. While participants in both control and treatment expressed that they felt comfortable doing the techniques in private moments and felt capable of executing the techniques to high proficiency, several participants described that taking any amount of pause to enact self-soothing would be perceived as awkward, disruptive, or socially unacceptable. As we mentioned, several participants thus created lighterweight versions of skills specifically to "fit into" social norms. As a result, we echo the implications of previous work in the domain of social simulation for self-care (Fang et al., 2025) in that the effectiveness of training people for mental health skills is not only a matter of ability or learning gain, but also broadly shaped by norms of interaction. Our qualitative results highlight that social stigma and contextual constraints remained reasons many participants hesitated to transfer skills to their real-life events. While offering a social simulation for treatment participants seemed to spark some reframing around social factors of stress, our intervention and study were not intensive nor lengthy enough for participants to likely become significantly more comfortable with deviating from social norms.

There are two key directions that we suggest for consideration in HCI technology for mental health: working within social norms or challenging social norms. For example, one way to work within social norms may be to design future self-care and simulation tools to assume the reality that one cannot conduct full-scale self-soothing in social settings and so to instead center micro-practices — short, subtle actions that can be enacted quickly and without drawing social attention. Participants consistently highlighted the difficulty of taking long pauses or engaging in extended visible exercises during public speaking or conversation. Designing for less conspicuous practices (e.g., a single deep breath, quick muscle release) can lower the barrier to real-world use. These micro-practices not only reduce the stigma of caring for self in the moment but also make self-soothing techniques easier to integrate fluidly into everyday moments of stress. Another design possibility for addressing this may be through using immersive technology like VR/AR for learning and training of skills, but integrating with wearable devices or mobile notification that provide discreet prompts for triggering the practice of these skills in real life. For example, haptic pulses from a smartwatch or ring could gently remind users but also guide a single breathing cycle without requiring them to close their eyes, gesture visibly, or interrupt conversation. These cues build a bridge between formal practice in AR and real-world enactment, reminding users of the techniques when they are most needed.

However, another direction is designing for the challenge of the social stigma around self-care. For example, one might design features that also focus on incorporating conversational skills for the user to suggest a break for themselves in a difficult moment (e.g., simulating various reactions to a user asking a virtual audience member if they can take a moment). A related example of challenging these assumptions around stigmatizing practices of self-care may also be designing interventions for group dynamics, intended to bring shared self-care practices into social dynamics. Designing for shared experiences such as group breathing exercises at the start of a meeting, classroom, or performance rehearsal can normalize and reframe well-being as communal rather than individual. From these suggestions, we advocate for future self-care technologies meant for training well-being skills to centrally consider in their design the spectrum at which they accept versus challenge the social environment and cultural expectations of enacting well-being practices.

Together, these barriers suggest that adoption of self-soothing techniques and the attempted scaffolds for this through technological means requires more than effective skill training; it also depends on navigating social norms and broader cultural beliefs about the relationship between technology and well-being.

### *Productive Distress in Self-Care Training Technology*

We found that adding social simulation simulation did not have any significant effects on reducing participants' state or anticipated stress measures, but still increased engagement with the application and the use of these skills outside the technological intervention. Our findings thus support that the social simulation in AR simulation is not necessarily effective at reducing stress, yet is still a useful and effective design. As our findings showed, participants reported no significant differences between control and treatment and qualitatively expressed that they could feel uncomfortable from the simulation given that it made them self-aware of shortcomings or challenges.

Historically, self-care work in HCI has been evaluated through the lens of stress reduction as a positive outcome. Applications such as mindfulness tools, relaxation games, and biofeedback systems often measure success by how much they lower users' immediate anxiety, heart rate, or subjective stress. This framing assumes that comfort is the goal and the metric of progress. However, our work and those similar (Benford et al., 2012; Fang et al., 2025) may instead promote "productive discomfort"; rather than aiming for soothing experiences, systems may deliberately introduce authentic stress to catalyze growth, self-awareness, and skill development (Benford et al., 2012; Halbert & Nathan, 2015). As Halbert and Nathan describe in their work on designing for discomfort, designs can spawn critical reflection and transformative learning from the experience of distress (Halbert & Nathan, 2015). Although designing uncomfortable experiences is not new in HCI, the most prevalent applications have been where maximizing comfort is obviously inappropriate and explicitly counterproductive — addressing deep disparity and colonialist history (Das & Semaan, 2022; Dosono & Semaan, 2020), engaging in debate (Halbert & Nathan, 2015), and sparking entertainment through physical or psychological discomfort (Benford et al., 2012). Our work suggests that even in the field of mental health and self-care — a field marked by the particular desire for comfort and calm — designing with discomfort and distress in mind (and even as a goal) can also be beneficial for long-term well-being}. We show that, even in interventions meant to soothe, discomfort can serve a vital role in offering a safe environment to practice response. We note that participants in our study explicitly mentioned that they did not necessarily feel reduced stress towards public speaking, but rather felt more capable of dealing with the inevitable distress that they expected to have. Participants' interviews illustrated how discomfort led to growth as they learned to normalize pauses, admit when they lacked answers, and confront judgment with less fear. These outcomes represent a shift away from stress reduction and toward stress literacy (Varlow et al., 2009).

# *Chapter 8*
# Discussion

In this dissertation, I have shown that virtual simulation provides a safer, controlled, and empirically grounded method for improving digital mental health, through two applications – agent-based simulation for large-scale communities (Part 1) and training individual-level mental health skills (Part 2). Through both system-level and individual-level study, my work shows how simulation enables responsible experimentation in domains where failure has high human cost.

## 8.1 Thesis Contributions

Over the course of five research studies mentioned in this dissertation, my specific contributions have been as follows.

In Part 1, I contributed:

1. **Empirical understanding for the effectiveness of online peer support (Chapter 3):** I provided large-scale quantitative evidence that participation in online mental health communities can produce small but significant improvements in depression and anxiety, particularly for users with severe symptoms.
2. **Understanding of the algorithmic gaps between current functionalities in online peer support technology and the matching needs of users (Chapter 4):** Through interviews and behavioral-log analysis, I revealed user needs and preferences for matching in OMHCs, highlighting the importance of gender, age, and experience-based compatibility and the necessity of maintaining transparency and user choice.
3. **Agent-based simulation for weighing trade-offs for algorithmic design for mental health communities (Chapter 5):** I built and validated a simulation model that allows researchers to test alternative matching policies without endangering active users, and provide design implications and trade-offs (e.g., topic-based matching as an effective and privacy-preserving proxy, general population versus vulnerable population outcomes for demographic-based matching) surfaced from applying this simulation to the world's largest online peer support platform .

In Part 2, I contributed:

4. **The introduction of immersive simulation for learning self-care, and empirical qualitative understanding of users' design needs for this technology (Chapter 6):** Using created 24 prototypes of immersive simulation with VR/AR, I conducted prototype-driven interviews to examine how people perceive benefit and drawbacks from simulated self-soothing environments, informing design principles for realistic yet psychologically safe exposure.

5. **Quantitative evidence of the benefits of simulation for mental health skill training (Chapter 7):** In a two-week, 44-user experiment study, I demonstrated that training mental health skills leads to greater physiological handling of stress and propensity for applying the trained skills in non-intervention situations.

Together, these studies establish that mixed-method, empirical evidence can guide the creation of virtual simulation as an experimentation method for improving digital mental health.

## 8.2 HCI Methods for Designing Mental Health Technology

In this thesis, I examine a range of HCI methods for identifying users' mental health–related needs, designing technological interventions to address those needs, and evaluating their impacts. Across chapters, I have shown the complementary interaction between qualitative methods (interviews, prototype-driven probes) for identifying users' desiderata and unmet needs for HCI intervention for personal mental health (Chapters 4, 6), large-scale quantitative analysis techniques based in machine learning, natural language processing, and data science for identifying large-scale community gaps (Chapter 4) and identifying behavioral changes (Chapter 7), and how prototype and system development can both explore novel technological possibilities that meet users' needs (Chapter 6) and create full-fidelity interventions for in-the-wild study (Chapter 7).

Taken together, this methodological pipeline is a response to the nature of HCI for mental health being inherently complex, and unable to be addressed through a single methodological lens. Rather, this work emphasizes an integrated approach: using qualitative and quantitative methods to ground design in empirically identified user needs, leveraging simulation-based and system-oriented development to translate these needs into concrete interventions, and evaluating these interventions through a combination of quantitative outcomes (e.g., behavioral change, self-report measures) and qualitative inquiry (e.g., user perspectives, usability, and whether post-use experiences align with initially articulated needs and preferences).

However, a consequence of this integrated approach and a recurring theme throughout this dissertation is the tension between strongly articulated qualitative needs and the relatively small or inconsistent effect sizes observed in quantitative evaluations. Across studies, users clearly express specific desires, preferences, and beliefs about what would support their mental well-being. Yet, when these designs are evaluated empirically, measurable changes in behavior or outcomes are often small to moderate (or non-significant). While on the surface this may represent a failure or shortcoming of either method, what HCI work has shown time-and-time again – and indeed, relevant to my work in digital mental health as well – is that it reflects the inherent complexity of mental health needs and behavior change. Expressed desire is not always straightforward to evaluate using empirical or quantitative measures, in part because users' perspectives on an intervention do not always map onto existing metrics. For instance, frequency of use is often treated as a proxy for usefulness or value; however, many technologies are useful or meaningful even when they are used infrequently. Similarly, an intervention may support reflection, reassurance, or skill-building in ways that are difficult to capture through short-term behavioral logs or standardized self-report measures. Moreover, a range of contextual and psychological factors (e.g., stigma, cognitive and attentional burden, real-world constraints, slowness of skill acquisition) shape how and whether interventions produce observable effects. These factors can attenuate or delay measurable change, even when users report positive subjective experiences. As such, small effect sizes do not necessarily indicate a lack of value, but rather highlight limitations in how we are able to evaluate in this domain. At the same time, this thesis acknowledges that users are not always able to clearly or fully articulate their needs and preferences. For example, when asked whether they like a prototype, participants may express general approval without being able to specify what aspects they find valuable, how they envision using it over time, or what conditions would actually motivate sustained engagement. Disentangling these nuances is a methodological challenge and requires careful refinement of interview techniques, workshop formats, and study protocols. I also wish to acknowledge the importance of evaluating mental health interventions in in-the-wild settings where researchers can observe how users integrate new HCI technology into their everyday lives beyond the artificial conditions of lab environments, where participants may hesitate to express resistance and/or do not encounter the practical barriers that shape real-world use.

A core responsibility of HCI researchers working in mental health, therefore, is to synthesize qualitative and quantitative evidence to form a holistic understanding of intervention effectiveness and usefulness. This requires situating evaluation results within the specific technological context of the intervention, the social and cultural context of the community it is intended to serve, and the relationship between users' qualitatively expressed needs and their observed behaviors and outcomes once an intervention is deployed. As a result, I hope that my thesis serves as an advocate for the integrated process of qualitative, quantitative, design, and development work necessary to obtain a well-rounded picture of HCI for mental health, and the growth of researchers (or teams of researchers) who can represent in all dimensions. Importantly, I argue that designing with this mismatch, rather than attempting to eliminate it, is a central challenge and contribution of HCI for mental health.

## 8.3 **Why Virtual Simulation?:** Considerations for Designing Mental Health Technology

Mental health is a uniquely sensitive and complex context for technological design. Throughout this dissertation, I have emphasized that simulation is an important method for mental health primarily due to the vulnerability of the targeted community; failures in mental health interventions do not simply lead to poor usability metrics but instead can damage trust in care systems and lead to exacerbated health concerns. Thus, simulation is a vital method for developing mental health technology in a safe environment without further harm for already vulnerable individuals.

However, throughout the five studies that I've conducted as part of this dissertation, the considerations that make simulation – and especially *human-centered* development of simulation – important emerged further than just vulnerability; many domains are sensitive, so what makes mental health an area of special need for simulation approaches? I argue that there are four additional constraints that the work of this dissertation surfaces beyond (1) *vulnerability* that make research in mental health of particular consideration – (2) *the need for experimentation and personalization*, (3) *importance of mixed-method outcomes*, and (4) *the inaccessibility of traditional care*. I review each of these below, along with how this dissertation has surfaced these considerations and how simulation addresses these considerations.

### *Consideration 1: Vulnerability*

As I have reviewed, deployment and evaluation of new intervention in the mental health technology space should be done with extreme care as even minor design missteps can have emotional or behavioral consequences. Yet, this sensitivity also limits the degree to which researchers can ethically test new ideas or expose users to realistic but potentially distressing scenarios. Simulation offers a distinct methodological benefit in this regard by providing safety and agency for users through enabling us to explore "what-if" questions that empirical work cannot reveal alone.

In this work, I have shown that from a *researcher* and *community designer* perspective (Part 1), simulation allows us to model new algorithmic intervention and test design hypotheses, in order to evaluate the emergent outcomes that empirical work alone does not always surface. Our work did so in the context of the world's largest peer support platform, a service in which hundreds of thousands of people rely on for well-being support for distressing issues anywhere from depression and anxiety, to relationship advice, to immediate crisis support. Therefore, the typical A/B testing framework or iterative design process of deploying then evaluating then iterating again carries significant risk to these hundreds of thousands of active users. In my work, I collaborated closely with the leaders of the platform to bridge the gap of meeting users' needs for matching while experimenting with different decision possibilities for this relatively unknown space using virtual simulation means. In the end, our work uncovered several trade-offs as reviewed in Chapters 4 and 5, and led to an open-source framework for simulation that other community designers can use as well as informing our collaborators' platform which currently has more advanced self-selection matching using topic.

I have also shown that from a *user* perspective, simulations provide a sense of psychological safety and agency allowing individuals to encounter challenging situations within the virtual world that protects the user from the consequences that make traditional treatments like exposure therapy controversial from real-world trauma and risk. In this way, simulations not only protect against potential harm but also introduce a critical sense of agency and control that is often missing in traditional mental health interventions. Many evidence-based treatments, such as exposure therapy, operate by intentionally eliciting discomfort or anxiety to facilitate lowered reaction or cognitive restructuring. While these methods can be effective, they are also emotionally demanding, difficult to personalize, and frequently constrained by ethical and logistical considerations particularly for early-stage interventions outside of clinical settings. Indeed, these acts of real-world "training" have come under scrutiny for many decades due to psychological safety risks (Gola et al., 2016). By contrast, simulation environments allow individuals to approach challenging experiences and even practice new ways of dealing with this distress (as in our work) on their own terms. This capacity to self-regulate during exposure and face exposure in virtual form makes simulation a unique medium that can support emotional safety while still eliciting needed engagement with discomfort.

In this sense, simulation allows researchers to study complex, high-stakes outcomes through human-centered design intervention. Across both Part 1 and Part 2, my work demonstrates that simulation enables research in contexts where direct experimentation would be irresponsible or infeasible. Therefore, simulation is valuable for a domain as sensitive, consequential, and variable as mental health.

### *Consideration 2: Need for experimentation*

Different from other health domains, I argue that mental health requires experimentation. We see this in traditional mental healthcare, from the norm of trying different therapists before settling on a match, to the typical process of adjusting medications like anti-depressants until side effects emerge and the right balance is found, to the use of methods like exposure therapy that rely on trial-based learning. In nearly all other health domains, treatment is – while perhaps not easy and having different trade-offs – usually a clear prescription or a navigation of at most a couple tried-and-true methods. Mental health, instead, is a domain full of uncertainty. What works for one person works wildly differently for another, and effectiveness can depend on lifestyle, demographics, and contexts that can change day-to-day. As a result, experimentation to find what works is a fundamental property of the mental health domain. Human affective and cognitive systems are dynamic, context-dependent, and deeply personal. As a result, *personalization* is important for mental health technology as interventions need to accommodate individual differences while also being robust enough to produce reliable benefits.

I argue in this thesis that simulation offers a useful methodological solution to these challenges. By creating controlled yet realistic replications of real-world environments (e.g., social interactions, physical environments, realistic outcomes), simulation enables iterative experimentation at the level of research design but also user experience. In Part 1, I demonstrated how agent-based simulations can test alternative system-level interventions (e.g., different matching algorithms) without exposing real users to unproven and potentially harmful conditions. These simulations allow researchers to observe emergent patterns of support, engagement, and well-being at scale. In Part 2, immersive simulations extend this principle to the individual level. Participants can experiment with coping strategies, self-regulation techniques, and stressful encounters in a safe, repeatable environment that provides immediate feedback. Overall, I argue that virtual simulation offers a safer alternative to the traditional therapeutic notion of "trial and error", and allows both researchers and users to explore design and treatment outcomes in ethical and sustainable ways.

Through these studies, I argue that the future of mental health technology should be designed with experimentation as an important need for users in mind. In this dissertation, I've provided examples through my own work on how we can create systems that help users and researchers alike navigate the inevitable uncertainty of their mental states under safe and controlled environments.

### *Consideration 3: Perception as an important evaluation of care*

In mental health, perception is not a secondary component of intervention efficacy; instead, it defines successful care. Unlike most other health domains, where outcomes can be directly measured through well-established physiological or clinical indicators, mental health improvement relies not only on measurable physical correlates (e.g., behavioral change, sleep quality, or physiological responses) but also, critically, on how individuals feel and perceive the intervention itself. People do not continue to use a mental health intervention unless they also perceive it to be useful, and this perception can be independent of whether it is useful on physical health symptoms associated with mental health. Whether an intervention feels supportive, trustworthy, or stigmatizing often determines its effectiveness as much as its clinical content. For example, even a perfectly designed algorithmic intervention may fail if users perceive it to take away their agency or to seem untrustworthy, while a less effective technique may yield positive outcomes by comparison if it fosters a sense of safety or self-efficacy.

To meet this consideration, I argue that the human-centered design component of simulation that has been presented in this work is what matters towards foregrounding perception in both research and practice. Throughout this dissertation, I have shown how simulation can act as both a **technical framework** for experimentation and a **human-centered method** for eliciting and responding to users' felt experiences. In Part 1, for example, the agent-based simulation framework that tests different algorithmic matching policies was grounded in extensive qualitative and quantitative understanding of users' lived experiences within the 7Cups community. We began not by simulating abstract user behavior, but by empirically studying how real users described their relationships with peers, their perceptions of fairness and comfort in being matched, and their unmet needs for autonomy and connection. Only after establishing this empirical foundation did simulation become meaningful; the agent-based model framework extended, rather than replaced, these human insights by exploring the system-level implications of different matching logics that emerged from those perceptions. Similarly, in Part 2, I approached simulation as an embodied and experiential method for examining perception directly. The immersive simulations designed for practicing stress regulation were not built from a priori assumptions of what "should" help users but instead from the patterns of perceived usefulness, realism, and safety that participants themselves articulated when envisioning AR and VR for self-soothing. The final evaluation combined both qualitative feedback—users' reflections on comfort, control, and immersion—and quantitative physiological data, demonstrating that perceptual and physiological measures together produce a more complete account of mental-health outcomes.

Consequently, perception can be viewed as a core dimension of both *design* and *evaluation* of mental health technology, alongside symptom reduction or behavioral change. Designing for perception involves alignment with users' mental models of care. Simulation enables such design by allowing us to test and evaluate human expressions of what they think will work.

### *Consideration 4: Inaccessibility of traditional resources*

Mental health is a domain marked by extreme lack of providers in traditional healthcare systems; globally, the treatment gap remains enormous as millions of individuals cannot access affordable or timely care, particularly in rural, low-income, or marginalized populations. This structural inaccessibility has been one of the primary motivators behind the rise of digital mental health interventions, from peer-support platforms to AI-assisted therapies. Yet accessibility is not only a logistical challenge but also an experiential one as even when services exist, social stigma, scheduling constraints, and cultural barriers commonly deter engagement.

I propose that simulation offers an avenue to address this inaccessibility. Because simulated environments can be deployed on personal devices (e.g., chatbots available on the web, personal AR/VR or lighter-weight mobile devices), they provide situated access that allows people to practice coping skills, rehearse stressful scenarios, or experiment with therapeutic techniques without professional supervision or institutional barriers. In Part 1 of this dissertation, simulation expands accessibility at the system level, enabling platform designers to improve matching algorithms that serve vast online communities. In Part 2, simulation enhances accessibility at the individual level, enabling people to experience evidence-based practices like exposure or breathing training privately and at their own pace.

As we have reviewed throughout this dissertation, accessibility of mental healthcare means not only availability of service but also availability of *safe* contexts for care. For instance, exposure therapy has long been shown to be effective in reducing phobias and anxiety; however, its real-world application often entails psychological discomfort, ethical concerns, and logistical barriers that limit participation. Similarly, youth and other at-risk populations may avoid traditional care resources out of fear of disclosure, perceived failure, or stigma. Simulation in this dissertation has addressed accessibility as an improvement for available services such that they can address the health needs of many (i.e., Part 1 that improves online support communities for meeting provider matching needs for millions) but also the availability of these safe care environments (i.e., Part 2 that provides virtual spaces for people to practice new methods for real-world stressors).

## 8.3 Conclusion

The conceptual implications of this work extend beyond the mental health technology that I have mentioned in this document. Simulation represents a broader methodological shift in HCI towards designing *safe experimental environments*. At its core, this dissertation has been about *proactive and anticipatory care,* as opposed to reactive care or harm mitigation. While reactive systems are vital for responding to crises and minimizing damage, proactive systems aim to prevent harm before it occurs – and proactive design aims to create systems that do not inadvertently cause harm themselves after deployment. The systems that I have contributed in this dissertation showcase the possibilities of allowing researchers to test interventions in parallel before deployment to large-scale communities and promote individual experimentation for new strategies for users without fear of real-world consequence, showcasing philosophy by creating spaces where experimentation and failure are possible while circumventing real-world consequences. I hope to have shown that by offering a mirror of the real-world that can reflect complex social and emotional realities, simulation redefines what responsible innovation can mean in human-computer interaction. In this framework, failure becomes informative rather than dangerous, and iteration becomes a shared process between designers, technologies, and users.

Traditionally, technologies are designed in ways that require users to adapt to the system's constraints and assumptions. Simulation enables a reversal of this paradigm by allowing systems to adapt to people first. Through simulated experimentation, designers can test and refine interventions in anticipation of the complex, lived realities of users before these systems are introduced into the sensitive contexts of human care. This shift reframes design not as a process of optimization for efficiency, but as one of alignment with people's (mental or emotional) needs, and further allows human-computer interaction systems to truly become a mediator between complex human vulnerability and technological care.

---